\documentclass[11pt]{article}
\usepackage{acl}
\usepackage{appendix} 
\usepackage{multicol}
\usepackage{tocloft} 
\usepackage{titletoc}
\usepackage{float}

\usepackage{graphicx} 
\usepackage{xcolor}
\usepackage{colortbl}
\usepackage[most]{tcolorbox}

\usepackage{booktabs}
\usepackage{multirow}
\usepackage{adjustbox}
\usepackage{makecell}
\usepackage{tabularx}
\usepackage{longtable}
\usepackage{array}
\usepackage{subcaption}
\usepackage{wrapfig} 
\usepackage{wrapfig}

\usepackage[patch=none]{microtype}
\usepackage{caption}
\usepackage{stfloats}
\usepackage{lineno}
\usepackage{textcomp}
\usepackage[T1]{fontenc} 
\usepackage{balance}
\usepackage{enumitem} 
\usepackage{times}  
\usepackage[ruled,linesnumbered]{algorithm2e}
\usepackage{algpseudocode}
\usepackage{listings}
\usepackage{amsmath,amsfonts,bm}
\usepackage{amsthm}
\usepackage{verbatim}

\usepackage{pifont}
\usepackage{amssymb} 

\SetKwInput{KwInput}{Input}

\SetKwInput{KwOutput}{Output}

\SetKwComment{Block}{$\triangleright$\ }{}

\SetKwComment{Comment}{//  }{}

\DontPrintSemicolon

\SetKwIF{If}{ElseIf}{Else}{if}{}{else if}{else}{end if}%

\def\HiLi{\leavevmode\rlap{\hbox to \hsize{\color{gray!20}\leaders\hrule height .8\baselineskip depth .5ex\hfill}}}

\definecolor{darkblue}{rgb}{0, 0, 0.5}

\hypersetup{
  colorlinks=true,   % Enable colored links
  citecolor=darkblue, % Set citation color
  linkcolor=darkblue, % Set link color
  urlcolor=darkblue   % Set URL color
}

\theoremstyle{plain}

\theoremstyle{definition}

\theoremstyle{remark}

\def\eqref#1{equation~\ref{#1}}
\def\1{\bm{1}}

\DeclareMathAlphabet{\mathsfit}{\encodingdefault}{\sfdefault}{m}{sl}
\SetMathAlphabet{\mathsfit}{bold}{\encodingdefault}{\sfdefault}{bx}{n}
\newcommand{\tens}[1]{\bm{\mathsfit{#1}}}

\def\tP{{\tens{P}}}

\newcommand{\sigmoid}{\sigma}

\title{PyFi: Toward Pyramid-like Financial Image Understanding for VLMs via Adversarial Agents}

\author{
  Yuqun Zhang$^{1\ast}$ \quad \quad Yuxuan Zhao$^{12\ast}$ \quad \quad  Sijia Chen$^{1\dagger}$ \\
  $^{1}$ The Hong Kong University of Science and Technology (Guangzhou) \\
  $^{2}$ Yantai Research Institute, Harbin Engineering University
  \\
  {\tt\small research@yuqunzhang.com, yx.zhao129@gmail.com, sijiachen@hkust-gz.edu.cn}
}

\begin{document}
\maketitle

\renewcommand{\thefootnote}{\fnsymbol{footnote}}
\footnotetext{$^\ast$ Equal contribution.}
\footnotetext{$^\dagger$ Corresponding author.}

\setcounter{footnote}{0}
\renewcommand{\thefootnote}{\arabic{footnote}}

\begin{abstract}

This paper proposes \emph{PyFi}, a novel framework for \underline{py}ramid-like \underline{fi}nancial image understanding that enables vision language models (VLMs) to reason through question chains in a progressive, simple-to-complex manner. At the core of PyFi is \textbf{\emph{PyFi}-600K}, a dataset comprising 600K financial question-answer pairs organized into a reasoning pyramid: questions at the base require only basic perception, while those toward the apex demand increasing levels of capability in financial visual understanding and expertise. This data is scalable because it is synthesized without human annotations, using \emph{PyFi}-adv, a multi-agent \underline{adv}ersarial mechanism under the Monte Carlo Tree Search (MCTS) paradigm, in which, for each image, a challenger agent competes with a solver agent by generating question chains that progressively probe deeper capability levels in financial visual reasoning. Leveraging this dataset, we present fine-grained, hierarchical, and comprehensive evaluations of advanced VLMs in the financial domain. Moreover, fine-tuning Qwen2.5-VL-3B and Qwen2.5-VL-7B on the pyramid-structured question chains enables these models to answer complex financial questions by decomposing them into sub-questions with gradually increasing reasoning demands, yielding average accuracy improvements of 19.52\% and 8.06\%, respectively, on the dataset.

\end{abstract}

%% Main content of the paper

\section{Introduction}
\label{sec:intro}

Vision language models (VLMs) are capable of step-by-step reasoning, presented as Chain-of-Thought (CoT) \cite{CoT-neurips22}, for image understanding in question answering (Q\&A) across various scientific domains~\citep{MMStar-neurips24,MMMU-cvpr24,Visco-cvpr25}, driven by the development of domain-specific benchmarks~\citep{VisNumBench-iccv25,Mvmath-cvpr25,MMMU-cvpr24,Codevision-arxiv25}, step-wise annotations~\citep{MathShepherd-acl24,o1-coder-arxiv24,mutualreasoning-arxiv24}, and domain-tailored VLMs~\citep{FinGPT-ijcai23,open-finllms-arxiv24,finlmm-r1-arxiv25}. In contrast, advancements in the financial domain remain limited. This is because, in finance image understanding, the creation of CoT reasoning datasets, which are highly necessary for the evaluation and fine-tuning of VLMs, requires a high level of expertise and domain knowledge and is therefore scarce.

\begin{figure*}[t]
  \includegraphics[width=\textwidth]{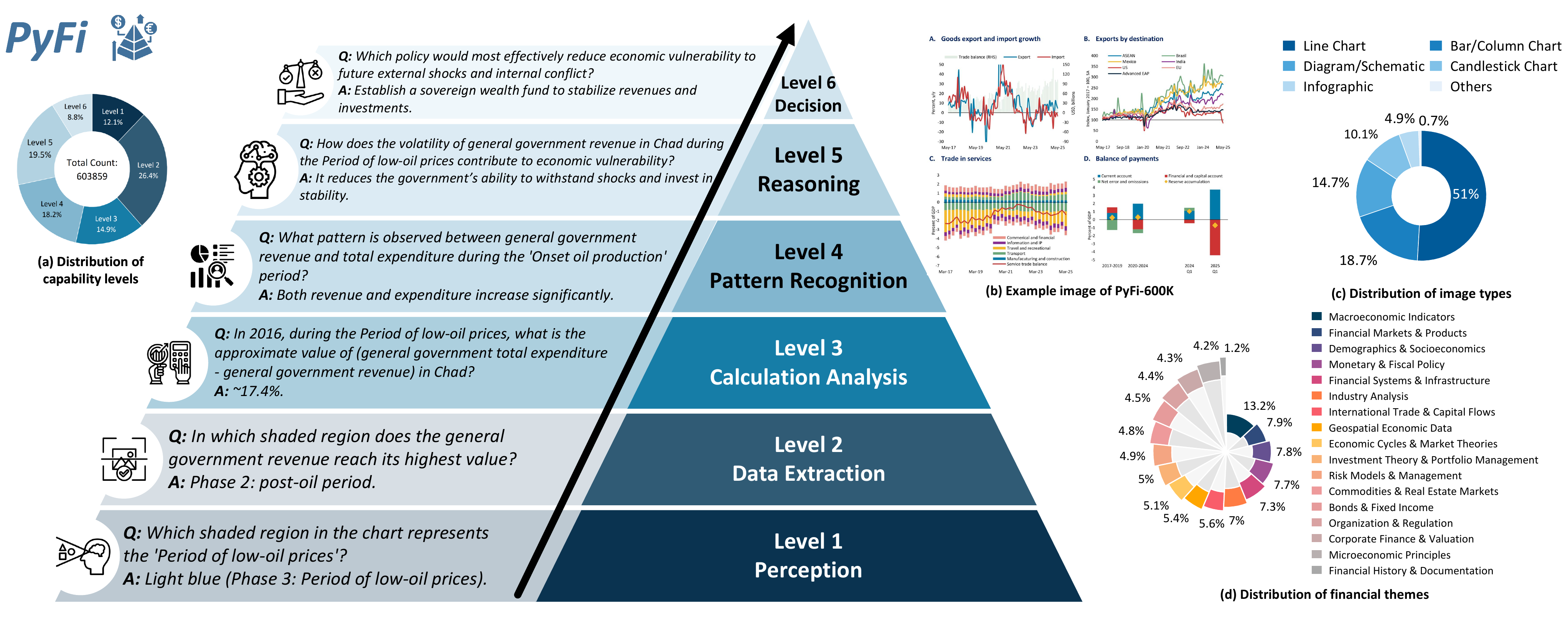}
   \caption{Overview of the \textbf{\emph{PyFi}-600K} dataset within our \textit{PyFi} framework. The dataset is structured as a pyramid comprising 6 capability levels, 11 financial image types, and 17 financial themes.}
  \label{fig:introduction}
  \vspace{-4mm}
\end{figure*}

To address these issues, previous literature~\citep{FinQA-emnlp21} collects Q\&A samples from financial documents manually, a process that is prohibitively time-consuming and difficult to scale. As a result, subsequent works propose datasets by merging extensive existing financial datasets~\citep{FLUE-emnlp22,Finben-neurips24}. These repurposed datasets may introduce accumuated biases and inconsistencies inherited from the original sources. More recently, benchmarks such as \citep{Mme-finance-mm25,FCMR-acl25,FinMME-acl25,BizFinBench-arxiv25,CFBenchmark-MM-arxiv25,VisFinEval-arxiv25} leverage trained VLMs to generate Q\&A pairs augmented with CoTs~\citep{FAMMA-arxiv25}. However, these samples may contain factual errors or hallucinations, which can distort VLM evaluation and fine-tuning and thereby undermine their reliability in real-world financial applications, where precision and expert-level understanding are essential.

Therefore, this paper proposes a framework named \textit{PyFi}, which supports comprehensive evaluation and advanced fine-tuning of VLMs through a \underline{py}ramid-like hierarchy for \underline{fi}nancial image understanding. Specifically, \textit{PyFi} includes a synthesized dataset, \textbf{\textit{PyFi-600K}} shown in Figure~\ref{fig:introduction}, comprising 600 thousand (K) samples with step-wise annotations that provide expert-level guidance on how to progressively solve complex financial problems across \(17\) categories. The construction of this dataset is enabled by \emph{PyFi}-adv, a multi-agent \underline{adv}ersarial mechanism designed to automatically synthesize and refine samples using expert-level financial knowledge.

\textbf{\textit{PyFi-600K}} offers three key benefits for the interpretability and fine-tuning of VLMs in finance. First, the samples are organized in a pyramid structure, where the required image understanding capability increases progressively --- from basic visual perception at the bottom level to complex financial decision-making at the top. This hierarchy, with capability levels ranging from one to six, enables comprehensive evaluation of VLMs across multiple aspects and supports ranking the capabilities required by VLMs in different financial themes. Second, samples across different levels are interconnected through \textit{question chains}, demonstrating how questions at higher levels can be addressed step-by-step by leveraging information and reasoning from lower-level Q\&A samples. Finally, it supports process supervision \citep{PRM800K-iclr23}, as each sample is assigned a reward score that reflects the correctness of its reasoning process in solving the financial problem.

The construction of such a scalable dataset is enabled by our proposed \emph{PyFi}-adv, a multi-agent \underline{adv}ersarial mechanism in which a challenger agent progressively synthesizes increasingly difficult questions, while a solver agent generates answers by reasoning with financial knowledge. The two agents compete within a Monte Carlo Tree Search (MCTS) framework: the challenger begins with simple questions at level 1, either selecting existing ones or generating new ones, and presents them to the solver for response. Based on the answer, the challenger then formulates a more challenging question in the next round. This iterative process continues until the information accumulated suffices to address the decision-making task at level 6.

Our extensive experiments with \textbf{\textit{PyFi-600K}} reveal that 15 well-known VLMs struggle to answer complex financial questions, with average accuracy dropping from 71.80\% at level 1 to 32.95\% at level 6. With \textit{question chains}, we observe that addressing a Level-6 financial decision-making question typically requires correctly answering an average of \(10.48\) sub-questions from lower levels. In particular, our results highlight that errors in calculation analysis are a primary cause of VLMs' failure in final financial decision-making. More importantly, fine-tuning Qwen-VL models on \textbf{\textit{PyFi-600K}} using \textit{question chains} improves accuracy by an average of 13.79\%, with gains reaching up to 19.52\% in the best case.

\section{Related Work}
\label{sec:related} 

\textbf{Financial Benchmark}. The evaluation of Vision-Language Models (VLMs) in the finance domain has seen remarkable progress, from early text-based benchmarks \citep{FinQA-emnlp21,FLUE-emnlp22,Finben-neurips24} to more recent efforts targeting financial image understanding \citep{FCMR-acl25,FinTMMBench-mm25,FinChartBench-arxiv25,CFBenchmark-MM-arxiv25,VisFinEval-arxiv25,Fin-Fact-mm25,FinRAGBench-V-arxiv25,Mme-finance-mm25,FinMME-acl25,FAMMA-arxiv25,FinMR-iclr25}. However, the question-answering datasets of most benchmarks, such as FinMME~\citep{Mme-finance-mm25}, rely on human annotation, making them costly and difficult to scale. Besides, they fail to support comprehensive assessment of VLMs in the financial domain. However, we introduce an automatically synthesized dataset for the fine-grained, hierarchical evaluation of financial visual understanding across diverse image types and financial themes. More related background can be found in Appendix~\ref{sec:related-work-discussion}.

\textbf{Financial Large Models}. Previous efforts primarily centered on text analysis using financial LMs~\citep{FinBERT-car23,Bloomberggpt-arxiv23,FinGPT-ijcai23,PIXIU-neurips23,Finr1-arxiv25}. Recent financial VLMs~\citep{Finvis-gpt-ijcai23,FinTral-acl24,open-finllms-arxiv24,FinTab-LLaVA-kdd25} integrate visual inputs to facilitate step-by-step reasoning, thereby enabling broader financial decision-making.  However, these models lack the expertise and interpretability needed for challenging financial themes, a limitation stemming from the absence of detailed, step-wise problem-solving processes grounded in finance-specific knowledge. In contrast, our dataset contains questions that illustrate how to address financial problems through a progression from simple to complex image understanding and from basic to increasingly deep financial knowledge.

\textbf{Multi-Agent LM Framework}. Recent advances in LMs have enabled Agentic AI to excel in real-world applications, with multiple LM-based agents now collaborating to perceive, learn, reason, and act, supporting intricate tasks such as simulation~\citep{Multi-agent-simulation-arxiv25,Multi-agent-simulator-acl25}, reasoning~\citep{BeTop-neurips24,AgenticReasoning-neurips25,MultiAgentDebate-icml24}, trading~\citep{Hedgeagents-www25,FinCon-neurips24,MountainLion-arxiv25}, and data synthesis~\citep{systematic-simulation-arxiv24,STORM-BORN-acl25,AgentSGEN-arxiv25}. However, to our knowledge, no existing method in finance leverages a multi-agent adversarial approach to synthesize data. In contrast, inspired by Generative Adversarial Networks (GANs) \cite{gan-neurips14}, we present the first method that employs adversarial agents to automatically synthesize hierarchical, fine-grained financial datasets.
\section{Framework}
\label{sec:framework}

\begin{figure*}[t]
    \centering
    \includegraphics[width=\linewidth]{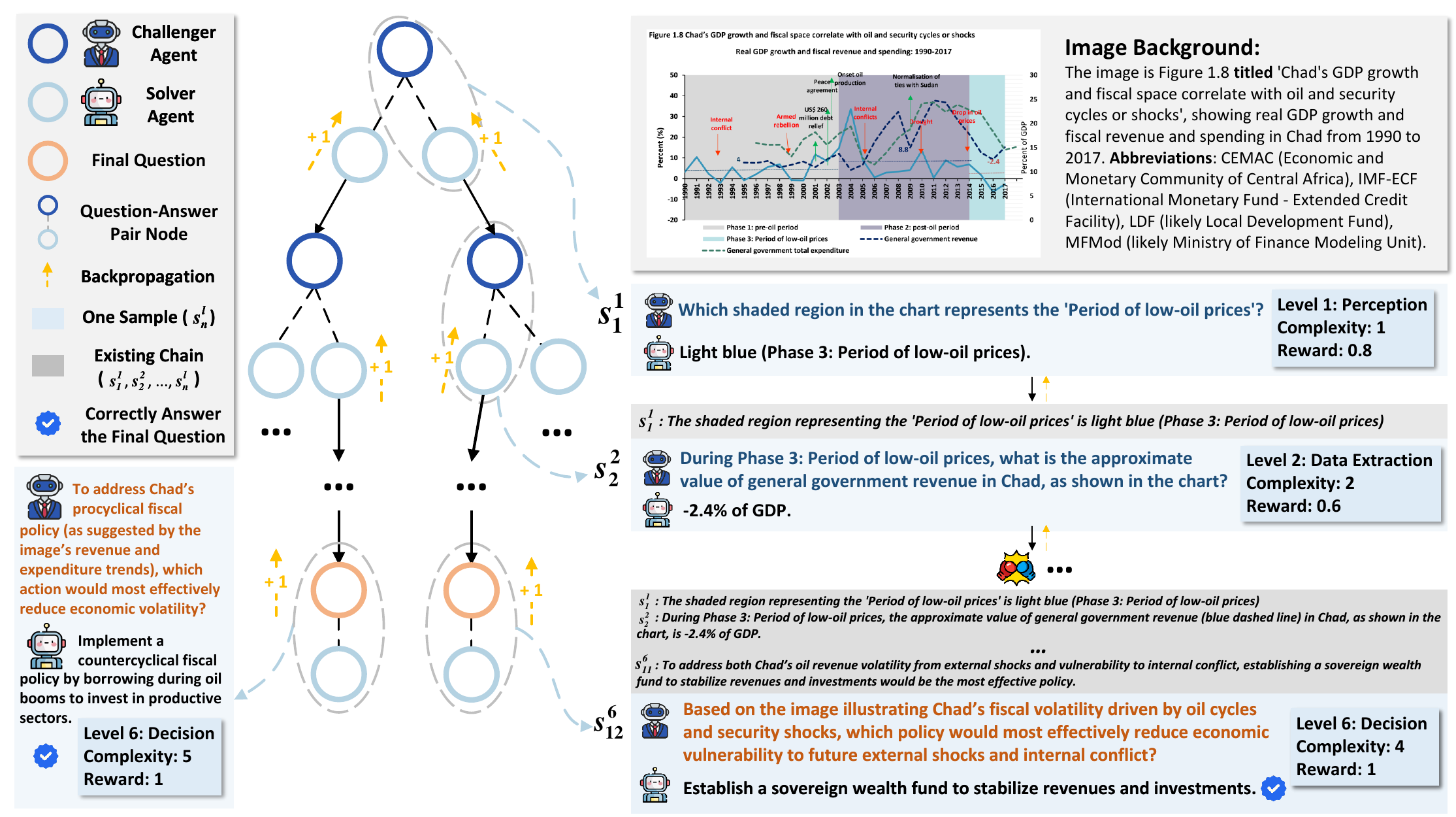}
    \caption{Overview of \textit{PyFi}-adv: a challenger agent competes with a solver agent under the Monte Carlo tree search (MCTS) paradigm to generate question chains that progressively probe deeper capability levels in financial visual reasoning. See the complete chain in Figure~\ref{fig:questionchain} in Appendix.}
    \label{fig:framework}
    \vspace{-4mm}
\end{figure*}

\subsection{Financial Dataset in Pyramid Structure}
\label{subsec:dataset-pyramid}

We release \textbf{\textit{PyFi-600K}}, a large-scale finance dataset comprising 600 thousand (K) samples, each consisting of a question-answer pair, constructed within a pyramid structure that encompasses six levels of finance-specific visual cognition: \textit{Perception} (\textit{PP}), \textit{Data Extraction} (\textit{DE}), \textit{Calculation Analysis} (\textit{CA}), \textit{Pattern Recognition} (\textit{PR}), \textit{Logical Reasoning} (\textit{LR}), \textit{Decision Support} (\textit{DS}). More importantly, \textbf{\textit{PyFi-600K}} introduces \textit{sample chains} that demonstrate how to address higher-level pyramid questions by leveraging knowledge from lower-level samples.

Specifically, we denote each finance sample as \(\bm{s}^l=\left(I, Q, A, r\right)\), where \(r\in \left[0,1\right]\) is the reward reflecting its correctness, \(I\) is the finance image to understand, \(l \in \left[1,6\right]\) indicates the pyramid level to which the question \(Q\) and the answer \(A\) belong. The sample complexity degree is \(C\left(\bm{s}^l\right) \in \left[1,5\right]\). And, as presented in right sub-figure of Figure~\ref{fig:framework}, we define \textit{sample chain} as \(\bm{S}^{l}_{1\dots n}=\left[\bm{s}^1_1,...,\bm{s}^l_n\right]\) where \(n\) is the length and \(n \geq l\) because samples within the chain can come from the same level. Considering the chain, the sequence of questions increases in levels or complexity degrees, as the answer for each one requires more insightful image understanding and financial expertise than the preceding ones. That is, for any \(\bm{s}^{h}_{i}\) and \(\bm{s}^{k}_{j}\) with \(i, j \in [1, \dots, n]\), \(h, k \in [1, \dots, l]\), \(i < j\), we have that \(h \leq k\), and if \(h = k\), then \(C\left(\bm{s}^{h}_{i}\right) \leq C\left(\bm{s}^{k}_{j}\right)\). Subsequently, the reasoning process for answering the question in any sample within the chain depends on its predecessor samples. Specifically, when \(\bm{s}^{l}_{j}\) is addressed by a VLM parameterized by \(\bm{\theta}\), we have \(p_{\bm{\theta}}\left(A|I, Q\right) < p_{\bm{\theta}}\left(A|I, Q, \left\{\bm{s}^{l}_{i}\right\}^j_{i=1}\right)\), meaning that the reasoning information and financial knowledge in answers from predecessor samples enable more reliable question answering for the current sample.

Therefore, compared to existing manual and automatic evaluation dataset construction methods, we observe the following three advantages of the proposed approach for the finance domain: 

\noindent \textbf{Hierarchical Cognition} in the dataset embodies human-level problem solving processes, progressing from basic perception to complex financial decision-making. Samples at each level \(l\) correspond to a specific cognitive capability, which builds upon those from lower levels \(i=1..,l-1\) or complexities. Thus, evaluating VLMs on these hierarchically structured samples enables a comprehensive assessment of their abilities in the finance domain. That is, any financial reasoning performed for questions from higher levels, which require rare, expert-level knowledge, relies on foundational financial questions involving base cognitions such as \textit{CA}. Besides, we can identify how VLMs perform, whether better or worse, at each cognition level during financial reasoning, thereby revealing which level is crucial for which financial questions and to what extent.

\noindent \textbf{Progressive Logical Chain} in the dataset is presented as the \textit{sample chain} \(\bm{S}^{l}_{1\dots n}\) shown in Figure~\ref{fig:framework}, which demonstrates how a challenging financial question at level \(l=6\) is solved by progressively addressing a sequence of simple-to-complex sub-questions. This aligns with financial decision-making logic, where each step builds upon the previous one to form a cumulative and logically coherent reasoning flow. Such a traceable chain enhances the interpretability and justifiability required in the finance domain. Moreover, VLMs fine-tuned on such chains are capable of solving complex financial problems by accumulating knowledge through first addressing a series of simpler problems.

\noindent \textbf{Step-wise Reward Score} \(r\) of each sample reflects its reliability; thus, the scores along a sample chain support process supervision~\citep{PRM800K-iclr23} in reasoning with financial image understanding. Using these scores, we enable the training of process reward models that act as verifiers for any financial VLM.

% The title should be revised. All titles should be in similar form.
\subsection{Automatic Synthesization with Adversarial Agents}

Existing financial datasets rely solely on human annotation, while automated methods from other domains, such as rStar-Math~\citep{rStar-Math-arxiv25} for mathematical reasoning, fail to meet the requirements of finance, where interpretability and reliable reasoning are essential, along with increasing complexity, domain expertise, and hierarchical cognition. We achieve this by proposing a multi-agent \underline{adv}ersarial mechanism, referred to as \emph{PyFi}-adv, which constructs question chains, as illustrated in Figure~\ref{fig:framework}.

Specifically, \emph{PyFi}-adv contains a challenger agent that is supported by an off-and-shelf VLM, denoted as \(\psi\), and similarly, a solver agent \(\phi\). Inspired by the adversarial mechanism in AlphaGo~\citep{AlphaGo-nature}, our two agents compete within an MCTS framework to automatically synthesize high-quality reasoning samples. In the finance domain, this approach substantially outperforms existing automatic data augmentation methods discussed in subsection~\ref{subsec:automatic-data-augmentation}.

\noindent \textbf{Confrontation for Sample Synthesization}, as shown by the Gray oval circle in Figure \ref{fig:framework}, is that when \(\phi\) aims to correctly answer the given question, \(\psi\) checks the answer to propose a more challenging follow-up question to hinder the solver. This is formulated as \(Q_i \sim p_{\psi}\left(Q_i|I, \tP_q, \bm{S}^{l}_{1\dots i-1}\right)\) and thus \(A_{i} \sim p_{\phi}(A_i|I, Q_i, \tP_a, \bm{S}^{l}_{1\dots i-1})\), where \(\tP_q\) and \(\tP_a\) are the agent prompts. 

\noindent \textbf{Exploration for Diversity Guarantee} aims to prevent the adversarial process from converging to a set of questions with identical content. At each step, a Bernoulli sampling is used to determine the \(\psi\) whether to \emph{explore} a new question path or \emph{exploit} an existing one. For exploitation, the UCT formula is employed to select the most promising prior question. 

\noindent \textbf{Expansion for Hierarchical Questions} is naturally achieved through the adversarial agents: once \(\psi\) performs a \emph{question expansion}, it generates a subsequent, more complex question based on the accumulated \(\bm{S}^{l}_{1\dots i}\). Agent \(\phi\) then produces an answer \(A_{i+1}\), forming a new sample \(\bm{s}^l_{i+1}\). Thus, a hierarchical question structure is built through repeated adversarial interactions that continue until either the top-level (\(l=6\)) question is addressed or a predefined depth limit is reached.

\noindent \textbf{Backpropagation for Reward Assignment} adjusts the success rate of each sample by backpropagating the comparison between the answer of the top-level question and the ground truth down to individual samples in the chain, as indicated by the orange dotted line in Figure~\ref{fig:framework}. Upon convergence of the MCTS, the resulting success rate is used as the reward score, providing a correctness label that measures the reliability of the synthesized dataset.

%% ~~~~~~~~~~~~~~~~~~~~~~~~~~~~~~~~~~~~~~
%% ~~~~ Scalable Applications in Financial Field ~~~~
%% ~~~~~~~~~~~~~~~~~~~~~~~~~~~~~~~~~~~~~~

% \subsection{Scalable Applications in Financial Field}

% Our PyFi-600K generated using PyFi-adv, integrating the benefits of hierarchical cognition, progressive logical chains, and step-wise reward scores, creates immediate applications for financial AI development. This pyramid-like structured dataset serves as exceptional training material for fine-tuning financial VLMs, where models learn to navigate increasingly complex reasoning paths $\bm{S}^{l}_{1\dots n}$ with immediate validation reward signals. This training paradigm enables VLMs to develop sophisticated interpretive skills essential for financial tasks such as multi-step chart analysis and risk assessment.

% Beyond model enhancement, the framework can be used to establish a comprehensive evaluation dataset. The inherent hierarchy allows for assessing reasoning capabilities across cognitive levels, while the chained structure tests logical coherence, and the reward mechanism $r_i$ provides granular performance metrics. This integrated approach offers advanced insights into model capabilities, moving beyond simple accuracy measures to evaluate the quality of financial reasoning processes.
\section{Experiments}
\label{sec:Experiments}

\subsection{Setup}
\label{subsec:exp-setup}

\begin{table*}[t!]
\centering
\caption{Evaluation of 15 Pre-trained VLMs on \textbf{\emph{PyFi}-600K} across 
6 capability levels and 5 complexity degrees, conducted on the \textbf{Evaluation Dataset after Exclusion}. This dataset is derived from an initial random selection of 1000 instances from PyFi-600K, followed by an additional filtering step to \textbf{address potential data leakage}. The final test set consists of 301 samples. For simplicity, we only present the results from complexity degrees 3, 4, and 5. The full model names listed in the "Model" column are: GPT-4.1, InternVL3-38B, Claude-opus-4-1-20250805, DeepSeek-VL2, Hunyuan-Large-Vision, ERNIE-4.5-turbo-vl, Moonshot-V1-128k-Vision-Preview, Moonshot-V1-32k-Vision-Preview, Moonshot-V1-8k-Vision-Preview, Qwen3-VL-Plus, Qwen2.5-VL-72B-Instruct, Qwen2.5-VL-32B-Instruct, Qwen2.5-VL-7B-Instruct, and Qwen2.5-VL-3B-Instruct.}
\label{tab:vlms-capability}
\begin{adjustbox}{max width=\linewidth}
\begin{tabular}{l c *{10}{>{\centering\arraybackslash}p{1.3cm}}}
\toprule
\textbf{Model} & 
\textbf{Overall} & 
\textbf{\textit{PP}} & 
\textbf{\textit{DE}} & 
\textbf{\textit{CA}} & 
\textbf{\textit{PR}} & 
\textbf{\textit{LR}} & 
\textbf{\textit{DS}} & 
\textbf{\(3\)} &
\textbf{\(4\)} &
\textbf{\(5\)} \\

\midrule
\hline
GPT-4.1 \citep{gpt4.1} & 52.99 & 86.84 & 47.64 & 40.62 & 57.14 & 54.84 & 38.46 & 52.63 & 46.55 & 53.85 \\
InternVL3 \citep{Internvl3-arxiv25} & 52.91 & 76.32 & 56.37 & 44.53 & 43.88 & 53.23 & 30.77 & 41.78 & 44.83 & 67.31 \\
Claude-opus \citep{anthropic2024claude3} & 64.70 & 80.92 & 62.74 & \textbf{66.02} & 58.16 & 72.58 & 32.69 & 66.12 & 55.60 & 73.08 \\
GLM-4.5V \citep{glm4.5v-arxiv25} & \underline{\textbf{74.75}} & 89.47 & \underline{\textbf{78.07}} & 65.23 & \underline{\textbf{75.51}} & \underline{\textbf{75.81}} & 46.15 & \underline{\textbf{72.37}} & \underline{\textbf{64.22}} & 73.08 \\
DS-VL2 \citep{deepseekvl2-arxiv24} & 45.18 & 76.32 & 44.58 & 36.33 & 42.35 & 42.74 & 19.23 & 34.54 & 42.24 & 46.15 \\
Hunyuan-LV \citep{Hunyuan-arxiv24} & 59.72 & \underline{\textbf{90.79}} & 58.96 & 36.72 & 63.27 & 69.35 & \underline{\textbf{51.92}} & 50.00 & 57.76 & \underline{\textbf{80.77}} \\
ER-4.5-TVL \citep{ERNIE4.5-arxiv25} & 34.47 & 34.87 & 33.96 & 36.72 & 25.00 & 45.16 & 36.54 & 38.82 & 25.43 & 65.38 \\
MV1-128k \citep{kimi-vl-arxiv25} & 54.57 & 80.92 & 61.08 & 39.06 & 41.84 & 58.87 & 38.46  & 46.71 & 43.97 & 57.69 \\
MV1-32k \citep{kimi-vl-arxiv25} & 54.40 & 80.26 & 61.56 & 37.11 & 40.31 & 62.90 & 38.46  & 45.72 & 43.97 & 61.54 \\
MV1-8k \citep{kimi-vl-arxiv25} & 54.90 & 80.26 & 61.79 & 39.84 & 42.35 & 58.06 & 38.46  & 48.03 & 43.97 & 55.77 \\
Q3-VL-Plus \citep{Qwen3-arxiv25} & 51.00 & 77.63 & 54.25 & 37.50 & 41.84 & 57.26 & 32.69  & 45.72 & 37.50 & 65.38 \\
Q2.5-VL-72B \citep{qwen2.5vl-arxiv25} & 48.84 & 70.39 & 59.91 & 33.20 & 35.71 & 46.77 & 26.92  & 36.51 & 31.03 & 57.69 \\
Q2.5-VL-32B \citep{qwen2.5vl-arxiv25} & 43.19 & 60.53 & 45.28 & 37.50 & 36.73 & 42.74 & 28.85  & 39.14 & 30.60 & 59.62 \\
Q2.5-VL-7B \citep{qwen2.5vl-arxiv25} & 37.87 & 54.61 & 49.53 & 34.77 & 18.37 & 24.19 & 15.38  & 31.91 & 17.67 & 38.46 \\
Q2.5-VL-3B \citep{qwen2.5vl-arxiv25} & 20.51 & 36.84 & 15.57 & 21.48 & 16.84 & 21.77 & 19.23  & 23.03 & 18.97 & 30.77 \\

\bottomrule
\end{tabular}
\end{adjustbox}
\vspace{-4mm}
\end{table*}

\noindent \textbf{Datasets}. \emph{PyFi}-600K dataset consists 603,859 samples and 62,660 sample chains with reward scores of 1. For evaluation, we proportionally sample 1,000 samples from each capability level to construct the test set across all cases. See detail in Appendix~\ref{sec:data-details}.

\noindent \textbf{Data Leakage Mitigation}. To ensure a fair evaluation, we systematically identify and exclude samples that exhibit data leakage. Specifically, any question that can be correctly answered by more than half of the evaluated models without the image is considered leaked. This reduces our test set from 1,000 to \emph{301} samples. Detailed analysis and justification are provided in Appendix~\ref{sec:dataleakage}.

\noindent \textbf{Models}. For evaluation, we use 15 VLMs as listed in Table~\ref{tab:vlms-capability}. For SFT, we fine-tune Qwen2.5-VL-3B-Instruct and Qwen2.5-VL-7B-Instruct on sample chains to obtain \emph{PyFi-QwenVL-3B} and \emph{PyFi-QwenVL-7B}, respectively. 

\noindent \textbf{SFT}. We fine-tuned the models on approximately 47K sample chains using two dataset variants: one with chain-of-thought (CoT) annotations (w/ CoT) and one without (w/o CoT). In the CoT variant, each question chain is converted into a CoT reasoning sequence, where each reasoning step corresponds to a sub-question and its answer. Thus, a chain with  \(n\) samples yields a CoT sequence with \(n\) reasoning steps. In the non-CoT variant, only the question and answer from the final sample in the chain are used. We use the AdamW optimizer with a learning rate of \(1.0 \times 10^{-4}\), cosine learning rate scheduling, and a warmup ratio of 0.1. Training is conducted for one epoch with an effective batch size of 8. For parameter-efficient fine-tuning, we apply LoRA with full-module adaptation. All experiments are carried out on four NVIDIA RTX 5090 GPUs.

\noindent \textbf{Evaluation}. We prompt all models to place the solution within \texttt{\textbackslash boxed\(\left\{\right\}\)}. For all inference runs, we use a temperature of 0.1 and a top-p value of 1.0. Accuracy is reported as the evaluation metric.

% The title should be revised.
\subsection{Main Results}
\label{subsec:main-results}

\noindent \textbf{Evaluations of \(15\) Pre-trained VLMs} are presented in Table~\ref{tab:vlms-capability}. In summary, we observe that as the capability level and complexity of financial image understanding increase, the accuracy of VLMs gradually declines. Specifically, the average accuracy across all models decreases from 71.80\% at level 1 (Perception) to 32.95\% at level 6 (Decision Support). This indicates that while current VLMs can handle basic visual perception in the financial image understanding, they struggle with complex financial questions that require higher-level cognitive abilities. Notably, GLM-4.5V achieves the highest overall accuracy of 74.75\%, demonstrating its strong performance across various capability levels. However, even this model's accuracy drops to 46.15\% at level 6, highlighting the challenges in financial decision-making. Most models, including GPT-4.1 and InternVL3-38B, achieve over 70\% accuracy at level 1 but fall below 40\% at level 6, resulting in a relatively low average overall accuracy of 50\%. 

\begin{figure}[t]
    \centering
    \includegraphics[width=\linewidth]{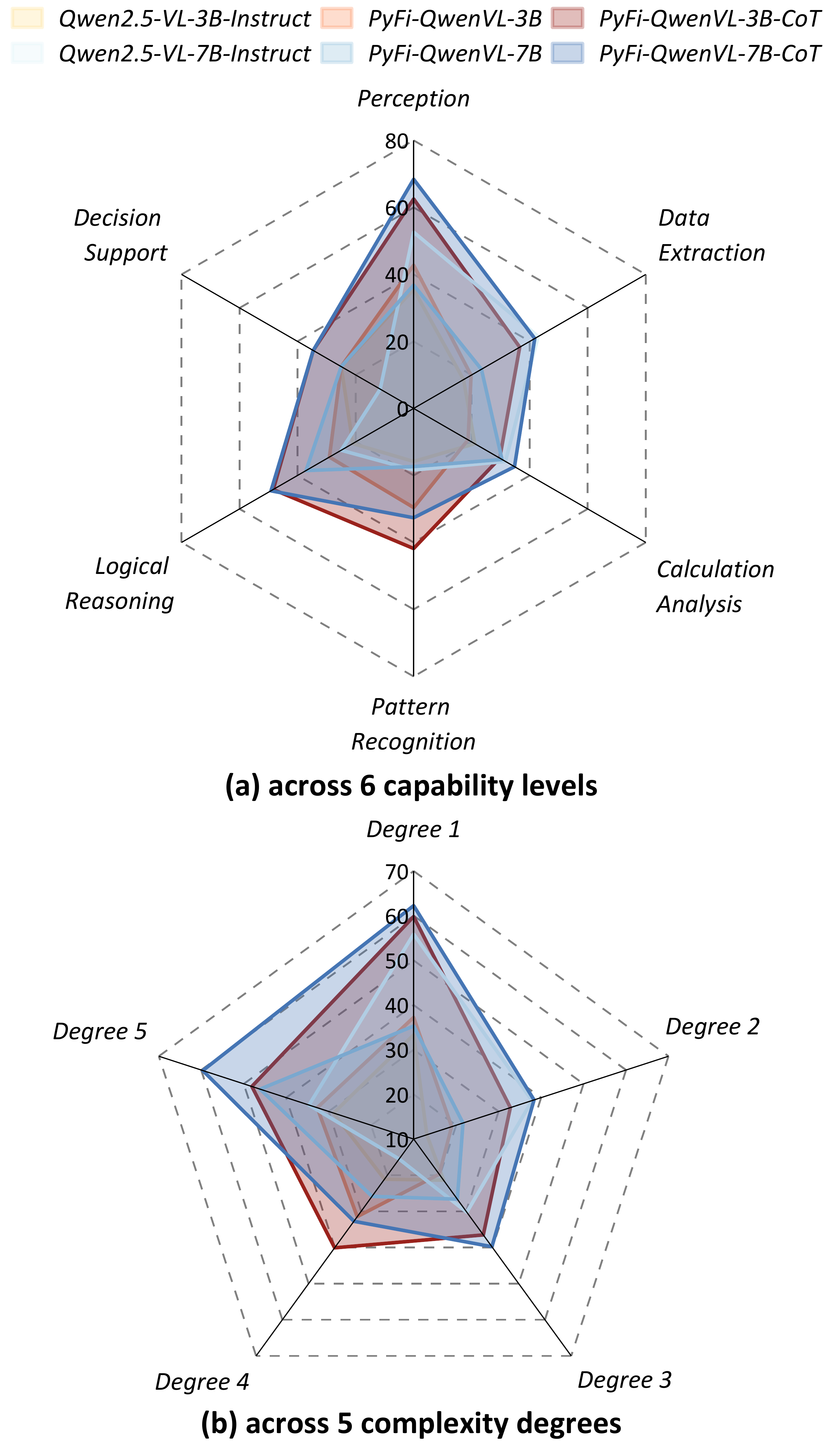}
    \caption{Comparison between Qwen2.5-VL models and ours ( \emph{PyFi-QwenVL-3B} and \emph{PyFi-QwenVL-7B}).}
    \label{fig:evaluations}
    \vspace{-4mm}
\end{figure}

\noindent \textbf{Evaluations of Fine-tuned Qwen-VL models} are presented in Figure~\ref{fig:evaluations}. In summary, with \textit{question chains} arranged in a pyramid manner, the models, especially the smaller ones, show a significant increase in accuracy and gain the ability to reason progressively, moving from low-level to high-level questions until they reach the solution, as shown in Figure 5. Specifically, both our \emph{PyFi-} models exhibit substantial accuracy gains. \emph{PyFi-CoT} related models improves the 3B model by 19.52\% points and the 7B model by 8.06\% points over their local baselines. The gain is notably larger for the smaller 3B model. Notably, on higher-level financial questions, such as Decision Support (DS), \emph{PyFi-CoT} ones improve the 3B model by 9.62\% and the 7B model by 23.08\%. The substantially larger gain for the 7B VLM suggests that larger VLMs benefit more from question-chain fine-tuning in challenging financial scenarios. Notably, in Pattern Recognition (PR) and Logical Reasoning (LR), \emph{PyFi-QwenVL-3B} even surpasses the accuracy of the much larger 72B VLM, as shown in Table~\ref{tab:vlms-capability}. Building on the detailed comparison in Figure~\ref{fig:experiments_finetuning}, we argue that sample-chain fine-tuning equips VLMs with the ability to solve complex financial problems by iteratively addressing a sequence of self-generated sub-questions of increasing capability levels, thereby significantly improving interpretability and reliability.

\begin{figure*}[t]
    \centering
    \includegraphics[width=\linewidth]{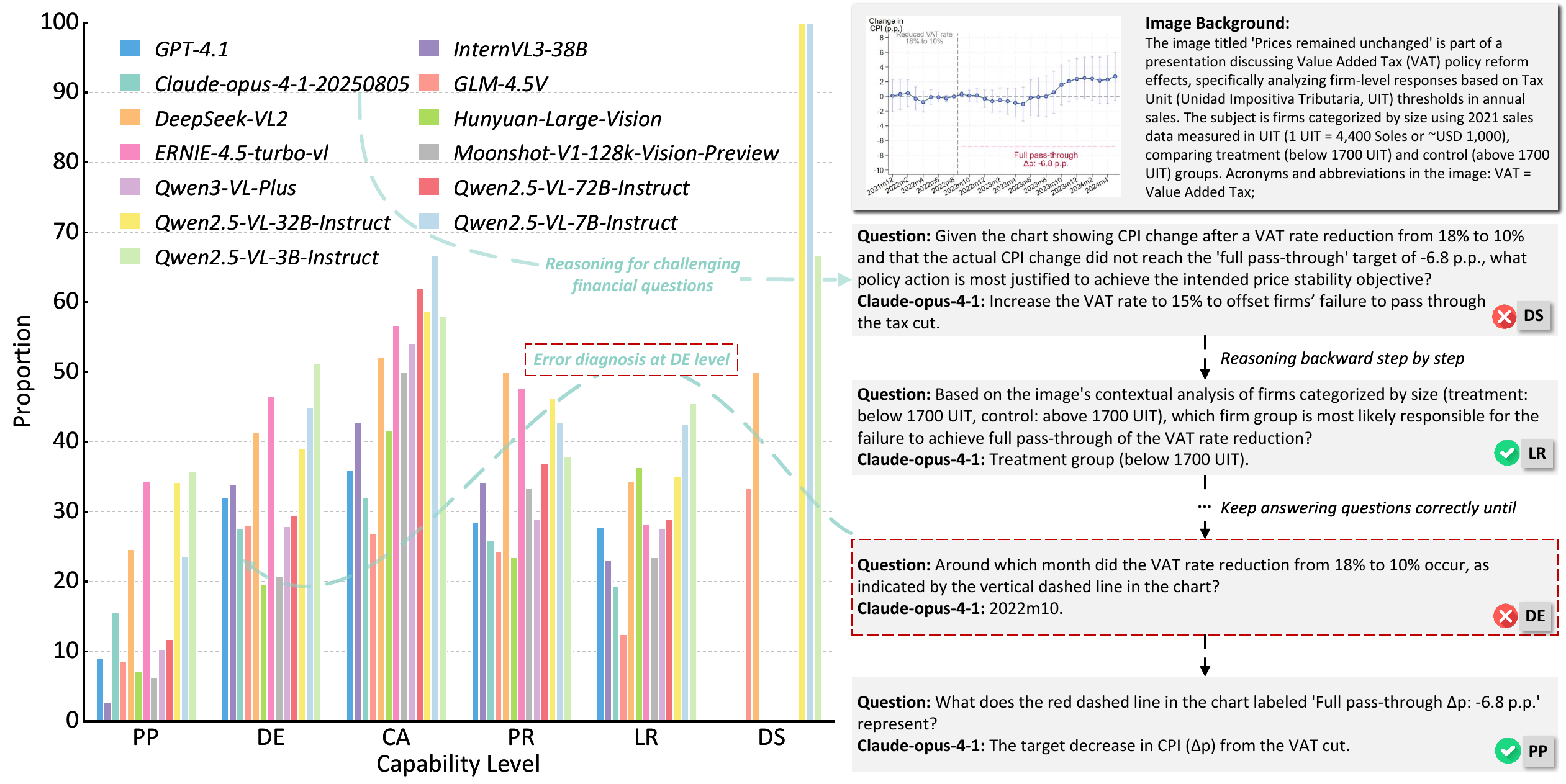}
    \caption{Evaluation of error proportions across capability levels leading to incorrect financial decisions. As illustrated in the right sub-figure, the \textit{sample chain} enables error tracing: when Claude-Opus-4-1 fails to answer the Level 6 financial decision-making question, we can \textbf{trace back} to identify the specific lower-level sub-questions that were answered incorrectly, thereby pinpointing the exact level at which the error originated.}
    \label{fig:chain_error_analysis}
    \vspace{-4mm}
\end{figure*}

\subsection{Main Benefits and Insights}
\label{subsec:analysis}

\noindent \textbf{Evaluations} in Table~\ref{tab:vlms-capability} and Figure~\ref{fig:evaluations} enable a fine-grained analysis of how --- and to what extent --- VLMs meet the varying capability requirements of financial image understanding across different levels of complexity. The phenomenon of decreasing accuracy as the level increases in Table~\ref{tab:vlms-capability} indicates that existing VLMs can only meet the most basic image understanding in finance and are difficult to be used for complex financial decision-making. For instance, while models like GLM-4.5V perform well on basic perception questions (level 1), their accuracy drops considerably on decision support questions (level 6). This trend is consistent across most models, suggesting that greater efforts are needed to enhance VLMs' capabilities in the finance domain. Thus, the results in Figure~\ref{fig:evaluations} demonstrate the importance of our question-chain fine-tuning: Qwen models learn to make accurate financial decisions by first addressing a series of sub-questions in a progression from simple to complex, a process aligned with human cognition. %Specifically, \emph{PyFi-CoT} ones achieve significant accuracy gains at higher capability levels, indicating that in financial problem-solving with VLMs, accumulating information from simpler questions facilitates more effective reasoning for accurate decision-making, a process aligned with human cognition.

\noindent \textbf{Tracing back from Decision Support to Lower-level Q\&As}, as illustrated in Figure~\ref{fig:chain_error_analysis}, significantly improves the interpretability of VLMs in financial decision-making. In Figure~\ref{fig:chain_error_analysis}, we first collect Level-6 samples that VLMs fail to answer correctly and analyze their \textit{question chains} by evaluating VLMs on the questions at each lower level in the chain. In the finance domain, addressing a Level-6 financial decision-making question typically requires correctly answering sub-questions from lower levels. Thus, when an error occurs, as indicated by the red cross in the right sub-figure, \textbf{\emph{PyFi}-600K} enables us to trace the problem-solving chain from \textit{DS} to \textit{LR} and eventually to \textit{PP}, thereby diagnosing how an error in the \textit{DE} leads to the final mistake. Such a tracing-back approach is crucial for understanding the limitations of VLMs in financial decision-making, as it reveals at which capability level a failure occurs.

\noindent \textbf{Identifying the Most Frequent Failure Capability Levels}, as presented in the left part of Figure~\ref{fig:chain_error_analysis}, reveals that the key to improving VLMs' performance in the financial domain lies in strengthening their capabilities at the CA (Calculation Analysis) level. Specifically, the pyramid-like structure and the \textit{question chain} of \textbf{\textit{PyFi-600K}} facilitates detailed error analysis, allowing us to identify specific levels where VLMs struggle. That is, most VLMs, even the advanced Claude-opus-4-1-20250805 and Qwen3-VL-Plus, fails to answer more than 40\% of the question from CA, revealing that the calculation analysis is a critical bottleneck for them. More importantly, by comparing the error proportions between six capability levels, we can prioritize which capabilities to enhance first. For instance, since CA has the highest error proportion, focusing on improving calculation analysis skills in VLMs could lead to the most significant overall performance gains in financial decision-making tasks. In addition, this comparison also shows that, in the finance domain, analyzing numerical calculations from images is one of the most critical capabilities for accurate decision-making, even more important than logical reasoning, aligning with conclusions drawn by real-world finance experts.

\begin{figure*}[t]
  \includegraphics[width=\textwidth]{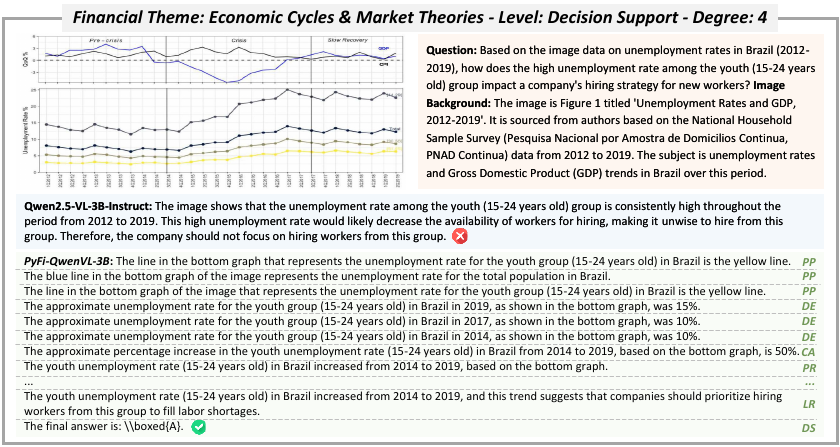}
   \caption{Qualitative analysis of level-wise COT by comparing the Qwen2.5-VL-3B-Instruct and \emph{PyFi-QwenVL-3B} in the challenging financial decision-making.}
  \label{fig:experiments_finetuning}
  \vspace{-4mm}
\end{figure*}

\noindent \textbf{Addressing Financial Decision-Making with Level-wise CoT}, as depicted in Figure~\ref{fig:experiments_finetuning}, is not only reliable but also highly interpretable. Specifically, we compare the responses of Qwen2.5-VL-3B-Instruct and \emph{PyFi-QwenVL-3B} --- the latter fine-tuned on \textit{question chains}--- when addressing the financial problem. The Qwen2.5-VL-3B-Instruct produces a short and disorganized reasoning process, reflecting its limited financial reasoning capability. In contrast, \emph{PyFi-QwenVL-3B} follows a clear, structured, and logical reasoning flow, which is essential for expert-level performance in finance. For instance, our model first addresses basic \textit{PP} questions, then proceeds to \textit{DE} and \textit{CA}, and finally tackles high-level \textit{LR} and \textit{DS}. This level-wise CoT not only improves answer reliability but also provides a transparent and traceable reasoning path that users can easily follow and verify, making it especially valuable in financial contexts where stakes are high. Moreover, experts can inspect this professional decision-making process and provide detailed and direct feedback to further refine the model.

\noindent \textbf{Providing Process Supervision}, in the form of a reward score for each sample in our \textbf{\textit{PyFi-600K}}, enables research into training reliable verifier models, known as process reward models (PRMs), to facilitate checking the correctness of each step generated by financial VLMs. Even though, due to space limitations, this paper does not perform such training, we use the reward scores to filter out low-quality samples during both evaluation and fine-tuning.

\noindent \textbf{Mitigating Hallucinations}. As shown in Figure~\ref{fig:experiments_finetuning} and Figure~\ref{fig:chain_error_analysis}, our support of Level-wise CoT in training effectively mitigates model hallucination by enforcing a grounded reasoning process: high-level conclusions (e.g., DS) are derived strictly from verified low-level findings (e.g., PP and DE), rather than from ungrounded speculation. This structure ensures traceability and verifiability, as every reasoning step can be explicitly traced back to concrete evidence in preceding steps.

\section{Conclusion}
\label{sec:conclusion}
In this paper, we have proposed a framework named \emph{PyFi}, containing the dataset \textbf{\emph{PyFi}-600K} and the multi-agent adversarial mechanism \emph{PyFi}-adv, to facilitate \underline{py}ramid-like \underline{fi}nancial image understanding for VLMs. With the Q\&A samples organized in a pyramid structure, \textbf{\emph{PyFi}-600K} enables comprehensive evaluation of VLMs across \(6\) capability levels, \(17\) financial themes and \(11\) image types. Our novel level-wise \textit{question chain} supports the interpretability of the performance of VLMs in financial decision-making by 1) showing importance ranking of capability level in accurate decisions and 2) enabling a back-track of which errors in lower levels led to the decision failure. More importantly, with \emph{PyFi}-adv, the dataset is scalable without human annotations, meaning the size and quality can be improved continuously. After performing extensive experiments on \(15\) VLMs and especially fine-tuning on Qwen models with \textit{question chains}, we have verified these benefits of the \emph{PyFi} by showing detailed, hierarchical, and fine-grained results, exposing important insights such as Calculation Analysis (\textit{CA}) is the most crucial for correct decision-making and fine-tuned small VLMs gain \(19.52\%\) accuracy improvement without compromising the model's existing capabilities on general tasks.

\bibliography{main}

\clearpage
\newpage

\begin{appendices}

\onecolumn
\section*{Appendix Contents}
\addcontentsline{toc}{section}{Appendix}

\renewcommand{\cftsecfont}{\normalfont}
\renewcommand{\cftsecpagefont}{\normalfont} 
\renewcommand{\cftsecleader}{\cftdotfill{\cftdotsep}}

\setcounter{tocdepth}{2}

\startcontents[appendices]
\printcontents[appendices]{}{1}{}

\twocolumn

\section{Related Work Discussion}
\label{sec:related-work-discussion}

% The part needs to write a comprehensive data discussion that specifically describes why finance is necessary and highlights why our dataset is urgent and important in finance+AI.

\subsection{VLM Reasoning}

Vision-language models (VLMs) have achieved impressive performance on complex tasks by leveraging step-by-step reasoning, decomposing intricate problems into simpler subtasks and solving them sequentially to reach a final solution~\citep{seed1.5vl-arxiv25,kimi-vl-arxiv25,qwen2.5vl-arxiv25,glm4.5v-arxiv25,Gemma3-arxiv25,Internvl3.5-arxiv25}. Benchmarking reasoning ability is essential for guiding future advancements. In contrast to earlier methods that evaluated reasoning holistically via a single scalar score, recent work has shifted toward fine-grained assessment, enabling step-by-step validation of the chain-of-thought (CoT) process~\citep{Visco-cvpr25,MME-arxiv23,MMDocBench-arxiv24,MMStar-neurips24,Mmbench-eccv24,MMMU-cvpr24,RMultiplex200K-iccv25}. This paradigm has seen significant traction in domains like mathematical reasoning~\citep{Mathvision-neurips24,Mathvista-iclr24,Mathverse-eccv24,VisNumBench-iccv25,MathOPEval-arxiv25,MATPBench-arxiv25,Mvmath-cvpr25,Wemath-arxiv24} and code generation~\citep{Designbench-arxiv25,Interaction2Code-arxiv24,Codevision-arxiv25,WebMMU-emnlp25,Mmcode-emnlp24,Plot2code-arxiv24}, where outputs are inherently verifiable and intermediate steps can be automatically annotated.

The success in these domains stems from the inherent verifiability of their outputs, which enables automatic construction of process-annotated datasets for fine-tuning large language models (LLMs). Two types of verifiers are commonly employed: the Outcome Reward Model (ORM), which assigns a scalar reward to the entire solution, and the Process Reward Model (PRM)~\citep{StepPRM-arxiv25,lessonsPRM-arxiv25,Visualprm-arxiv25,AgentPRM-arxiv25,Rewardbench-arxiv24}, which evaluates correctness at each step of the reasoning path. PRM's step-wise scoring enables precise error localization, facilitating more targeted feedback.

However, this progress has yet to be effectively transferred to vertical domains such as finance. Reasoning in finance typically involves multistep interpretation of charts, tables, and contextual narratives, requiring not only accurate understanding of financial visuals and deep domain knowledge, but also often yielding open-ended answers that are inherently difficult to verify precisely. This ambiguity impedes the generation of dense, step-wise process rewards essential for PRM training.

\subsection{Financial Benchmarks}

Existing financial benchmarks remain limited: they reduce complex reasoning to single scalar scores, obscure intermediate reasoning steps, and lack support for hierarchical or diagnostic evaluation~\citep{FCMR-acl25,FinTMMBench-mm25,FinChartBench-arxiv25,CFBenchmark-MM-arxiv25,VisFinEval-arxiv25,Fin-Fact-mm25,FinRAGBench-V-arxiv25,Mme-finance-mm25,FinMME-acl25,FAMMA-arxiv25,FinMR-iclr25}. Additionally, current evaluation datasets consist only of isolated question-answer pairs, making it difficult to uncover hidden relationships within the data~\citep{SBP-arxiv25}. Compounding this challenge, these benchmarks rely on costly manual annotations that hinder scalability, and no large-scale synthetic dataset currently exists to train finance-specific preference reward models (PRMs) for visual reasoning tasks.

Specifically regarding scalability, FAMMA-LivePro~\citep{FAMMA-arxiv25}, a manually annotated benchmark, contains 103 questions crafted by senior finance experts. These questions are of high quality and accuracy, and capture the complexity of real-world financial scenarios. However, such manual annotation demands highly qualified experts and is both time-consuming and labor-intensive, making it difficult to scale. While some benchmarks~\citep{Omnieval-arxiv24,Mme-finance-mm25,FinMME-acl25,BizFinBench-arxiv25,CFBenchmark-MM-arxiv25,FinTMMBench-mm25,FinRAGBench-V-arxiv25,FinChartBench-arxiv25,Mmbench-eccv24} explore using large models for automatic data synthesis, they typically prompt these models to directly generate expert-level, complex financial questions. However, due to the difficulty large models face in understanding complex financial visuals, they often rely heavily on provided textual cues, leading to question homogenization and hallucinations. This results in question-answer pairs of inconsistent quality that still require substantial manual verification. On the other hand, most financial benchmarks only release their datasets without disclosing their data augmentation methods, making direct comparison of such methods difficult.

As a result, the finance domain still lacks benchmark datasets capable of supporting hierarchical capability assessment, evaluating step-by-step reasoning, enabling fine-grained reward modeling, and achieving fully scalable data augmentation.

\subsection{Automatic Data Augmentation}
\label{subsec:automatic-data-augmentation}

Existing automated annotation methods fall into two categories. The first includes Math-Shepherd \citep{MathShepherd-acl24}, AutoPSV \citep{AutoPSV}, AutoPRM \citep{AutoPRM}, and AdaptiveStep \citep{AdaptiveStep}, which score or decompose steps of a fixed problem using verifiers or confidence heuristics. While efficient, they are passive and cannot generate new problems or structure them by difficulty. The second category, including OmegaPRM \citep{OmegaPRM}, rStar-Math \citep{rStar-Math-arxiv25}, GroundedPRM \citep{GroundedPRM}, and RMultiplex200K \citep{RMultiplex200K}, leverages MCTS or structured planning to create higher-quality, verifiable paths. Specifically, OmegaPRM employs a divide-and-conquer MCTS algorithm to systematically locate the first incorrect step in a reasoning chain, enabling efficient and high-quality data generation. GroundedPRM also uses MCTS but integrates external validation tools to ensure the fidelity of each step, resulting in a more data-efficient PRM that achieves SOTA performance with far fewer annotations.

While existing methods such as Math-Shepherd, AutoPSV, AutoPRM, and AdaptiveStep effectively annotate steps of a given problem through passive, single-agent verification, they are inherently limited to flat, post-hoc labeling and cannot generate new problems or structure them by difficulty. Even MCTS-based approaches though capable of constructing high-fidelity reasoning paths—focus exclusively on verifying or refining a single solution trajectory. In contrast, our work introduces an active, adversarial two-agent framework embedded within the MCTS architecture: one agent (the questioner) generates financial reasoning problems, while the other (the answerer) solves them, with the questioner progressively increasing difficulty based on the answerer's performance.  

This co-evolutionary process is specifically designed for the financial domain, where reasoning often requires layered knowledge, from basic accounting principles to complex risk modeling, and where errors must be interpretable and traceable. Our method explicitly constructs a hierarchical, interpretable curriculum of financial problems, in which each subsequent question builds upon prerequisite concepts and demands deeper domain understanding. Non-MCTS methods lack the structured search mechanism needed to jointly optimize problem generation, difficulty progression, and financial knowledge dependency. MCTS, by contrast, provides the necessary planning backbone to balance exploration of novel financial scenarios with exploitation of solvable, pedagogically meaningful chains. Thus, MCTS is essential not only for scalability but also for realizing a generative, knowledge-aware annotation process that produces explainable, difficulty-graded financial reasoning data, extending MCTS's role beyond verification to structured curriculum synthesis tailored to domain-specific reasoning.

\subsection{Our Contributions}

To address above shortcomings, we aim to: (1) comprehensively evaluate VLMs across a hierarchy of cognitive levels, from basic perception to expert financial decision-making, enabling granular diagnosis of strengths and weaknesses at each reasoning stage and revealing which foundational cognitions are critical for higher-level financial tasks, and uncover meaningful data relationships; (2) provide traceable, progressive logical chains that mirror real-world financial reasoning, enhancing interpretability, explainability, and justifiability while enabling VLMs to learn complex problem-solving through cumulative, step-by-step knowledge building; (3) introduce step-wise reward scores for each reasoning step, facilitating automated, scalable verification via Process Reward Models and overcoming the limitations of costly manual annotation, and (4) propose an automatic data augmentation method that decomposes the task of generating complex financial question-answer pairs into incremental process, starting from simple visual perception questions and progressively increasing difficulty and financial reasoning depth, all without requiring human verification. Finally, we present a comparison between our \textbf{\emph{PyFi}-600K} dataset and existing evaluation datasets in Table~\ref{app:dataset_comparison}.

\begin{table*}[t]
\centering
\caption{Comparison of \textbf{\emph{PyFi}-600K} with Existing Evaluation Dataset. In Multimodal column, "T" means Text modality, "I" means Image modality. In the Volume column, K is used as the unit.}
\label{app:dataset_comparison}
\begin{adjustbox}{max width=\linewidth}
\begin{tabular}{l c *{6}{>{\centering\arraybackslash}p{1.7cm}}}
\toprule
\textbf{Dataset} & 
\textbf{Volume}  &  
\thead{\textbf{Multimodal}} & 
\thead{\textbf{Automatic}\\ \textbf{Synthesized}} & 
\thead{\textbf{Hierarchical}\\ \textbf{Cognition}} & 
\textbf{CoT} & 
\thead{\textbf{Step-wise}\\ \textbf{Reward}} \\
\hline
\hline
\multicolumn{7}{c}{\textbf{\textit{General Domain}}} \\\hline
MME~\citep{MME-arxiv23} & 2.4  & T \& I &  &  & \checkmark &  \\
MMDocBench~\citep{MMDocBench-arxiv24} & 4.3  & T \& I &  &  & \checkmark & \\
MMStar~\citep{MMStar-neurips24} & 1.5  & T \& I &  &  &  &  \\
MMBench~\citep{Mmbench-eccv24} & 3.2  & T \& I &  & \checkmark &  &  \\
MMMU~\citep{MMMU-cvpr24} & 11.6  & T \& I &  &  &  &  \\ 

\hline
\hline
\multicolumn{7}{c}{\textbf{\textit{Finance Domain}}} \\ \hline
FinQA~\citep{FinQA-emnlp21} &  8.3  & T &  &  & & \\
FLUE~\citep{PIXIU-neurips23} &  26.3  & T &  &  & & \\
FLARE~\citep{PIXIU-neurips23} & 19.2  & T &  &  &  &  \\     
CF-Benchmark~\citep{Cfbenchmark-arxiv23} & 3.9  & T &  &  &  &   \\
BizFinBench~\citep{BizFinBench-arxiv25} & 7.0  & T &  &  & \checkmark &   \\
FinMME~\citep{FinMME-acl25} & 11.1  & T \& I &  & \checkmark &  &   \\     
MME-Finance~\citep{Mme-finance-mm25} & 1.2  & T \& I &  & \checkmark &  &   \\  
FAMMA-Basic~\citep{FAMMA-arxiv25} & 1.9  & T \& I &  &  & \checkmark &   \\

CFBenchmark-MM~\citep{CFBenchmark-MM-arxiv25} & 9.7  & T \& I &  &  &  &   \\
FinMMR~\citep{FinMMR-iccv25} & 4.3  & T \& I &  &  & \checkmark  &   \\
MultiFinBen-Vision~\citep{MultiFinBen-arxiv25} & 21.8  & T \& I &  & \checkmark &  &   \\
\textbf{\emph{PyFi-600K} (Ours)} & 600 & T \& I & \checkmark & \checkmark & \checkmark & \checkmark  \\  

\bottomrule
\end{tabular}
\end{adjustbox}
\end{table*}

\section{Data Details}
\label{sec:data-details}

\subsection{Data Statistics}
\label{subsec:data-statistics}

The complete \textbf{\emph{PyFi}-600K} dataset comprises 603,859 samples, each comprising a financial image paired with a corresponding question-answer (Q\&A) pair, along with essential background information (caption, theme, acronym expansions) of the image and a summary analysis of its context. Each sample is annotated with a capability level and complexity rating. Specifically, Level 1 (Perception) contains 73,006 samples (12.09\% of the total); Level 2 (Data Extraction) includes 159,601 samples (26.43\%); Level 3 (Calculation Analysis) comprises 90,229 samples (14.94\%); Level 4 (Pattern Recognition) has 109,848 samples (18.19\%); Level 5 (Logical Reasoning) accounts for 117,751 samples (19.50\%); and Level 6 (Decision Support) includes 53,424 samples (8.85\%). Notably, Level 2, Data Extraction, is the most prevalent, underscoring its foundational role in complex reasoning tasks. The dataset also includes over 60,000 sample chains, whose lengths vary from 1 to 26 samples, with an average chain length of 11.48. Question lengths range from 29 to 853 characters, with an average length of 142 characters, as shown in Table~\ref{tab:question_and_chain_length}. The full distribution across capability levels and complexity degrees is shown in Table~\ref{tab:combined_distribution}.

\begin{table}
\centering
\caption{Question Length and Question Chain Length Statistics}
\begin{adjustbox}{max width=\linewidth}
\label{tab:question_and_chain_length}
\begin{tabular}{lrr}
\hline
\textbf{Statistic} & \textbf{Question Length (Characters)} & \textbf{Chain Length} \\
\hline
Mean           & 142.11 & 11.48 \\
Median         & 137.00 & 12 \\
Minimum        & 29     & 1 \\
Maximum        & 853    & 26 \\
Std Dev        & 50.54  & 2.55 \\
\hline
\end{tabular}
\end{adjustbox}
\end{table}

\begin{table}[htbp]
\centering
\caption{Distribution of Question Capability Levels and Complexity Degree}
\label{tab:combined_distribution}
\begin{adjustbox}{max width=\linewidth}
\begin{tabular}{cllrr}
\toprule
\multicolumn{2}{c}{\textbf{Classification}} & \multicolumn{1}{c}{\textbf{Type}} & \multicolumn{1}{c}{\textbf{Count}} & \multicolumn{1}{c}{\textbf{Percentage (\%)}} \\
\midrule
\multirow{6}{*}{\begin{tabular}[c]{@{}c@{}}\textbf{Capability}\\ \textbf{Level}\end{tabular}} 
& 1 & Perception & 73,006 & 12.09 \\
& 2 & Data Extraction & 159,601 & 26.43 \\
& 3 & Calculation Analysis & 90,229 & 14.94 \\
& 4 & Pattern Recognition & 109,848 & 18.19 \\
& 5 & Logical Reasoning & 117,751 & 19.50 \\
& 6 & Decision Support & 53,424 & 8.85 \\
\midrule
\multirow{5}{*}{\begin{tabular}[c]{@{}c@{}}\textbf{Complexity}\\ \textbf{Degree}\end{tabular}}
& 1 & --- & 91,143 & 15.09 \\
& 2 & --- & 148,465 & 24.59 \\
& 3 & --- & 165,072 & 27.34 \\
& 4 & --- & 139,654 & 23.13 \\
& 5 & --- & 59,527 & 9.86 \\
\midrule
\multicolumn{3}{l}{\textbf{Total}} & \textbf{603,859} & \textbf{100.00} \\
\bottomrule
\end{tabular}
\end{adjustbox}
\end{table}

The distribution of question chain lengths by segment has shown in Table~\ref{tab:chain_distribution_segmented}. The results indicate that the majority of sample chains (84.47\%) fall within the 11-15 length range, suggesting that most tasks in the dataset require a moderate to substantial number of reasoning steps. Since chain length reflects the number of logically sequential question-answer pairs needed to arrive at a final answer, longer chains correspond to higher task complexity. The scarcity of very short (1-10, 15.11\%) or very long (16-26, 0.42\% combined) chains further implies that the dataset is intentionally curated to focus on multistep reasoning scenarios that are neither trivial nor excessively extended.

\begin{table}[ht]
\centering
\caption{Distribution of question chain lengths by segment}
\label{tab:chain_distribution_segmented}
\begin{tabular}{lrr}
\toprule
\textbf{Length Range} & \textbf{Count} & \textbf{Percentage (\%)} \\
\midrule
1--10   & 9,456  & 15.11 \\
11--15  & 52,863 & 84.47 \\
16--20  & 241    & 0.39 \\
21--26  & 21     & 0.03 \\
\bottomrule
\end{tabular}
\end{table}

\begin{figure}
    \centering
    \includegraphics[width=\linewidth]{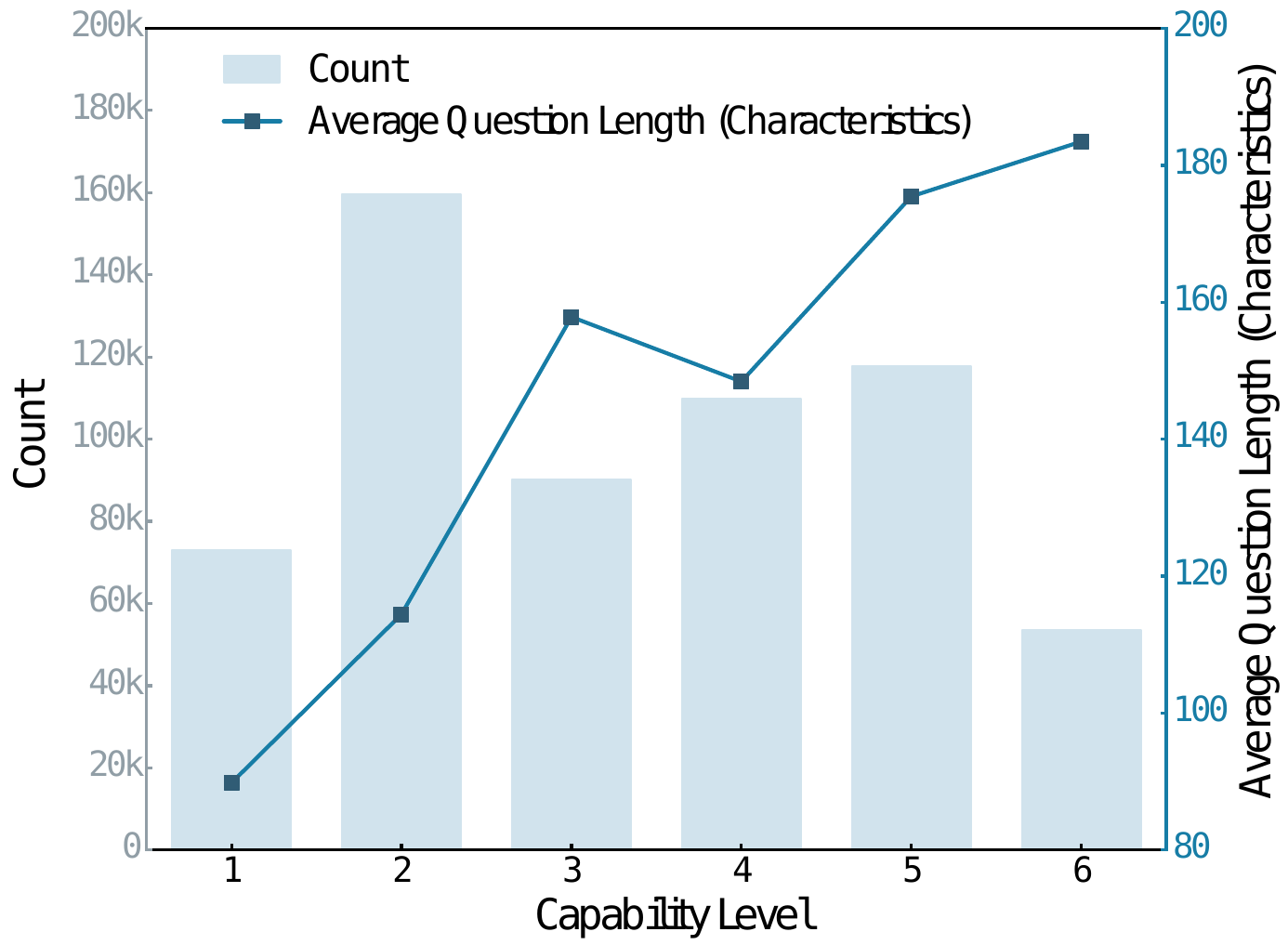}
    \caption{Capability Level with Question Length.}
    \label{fig:capability_level_with_question_length}
\end{figure}

We also observed that, generally, the average question length increases with higher capability levels, indirectly reflecting a progressive rise in task difficulty, as shown in Figure~\ref{fig:capability_level_with_question_length}. However, an unexpected decrease occurs at Level 4 (Pattern Recognition). This may be because pattern recognition tasks often require identifying and interpreting intricate patterns rather than relying on simple textual matching, which could lead to more concise yet semantically dense questions that challenge model performance. 

\begin{figure}
    \centering
    \includegraphics[width=\linewidth]{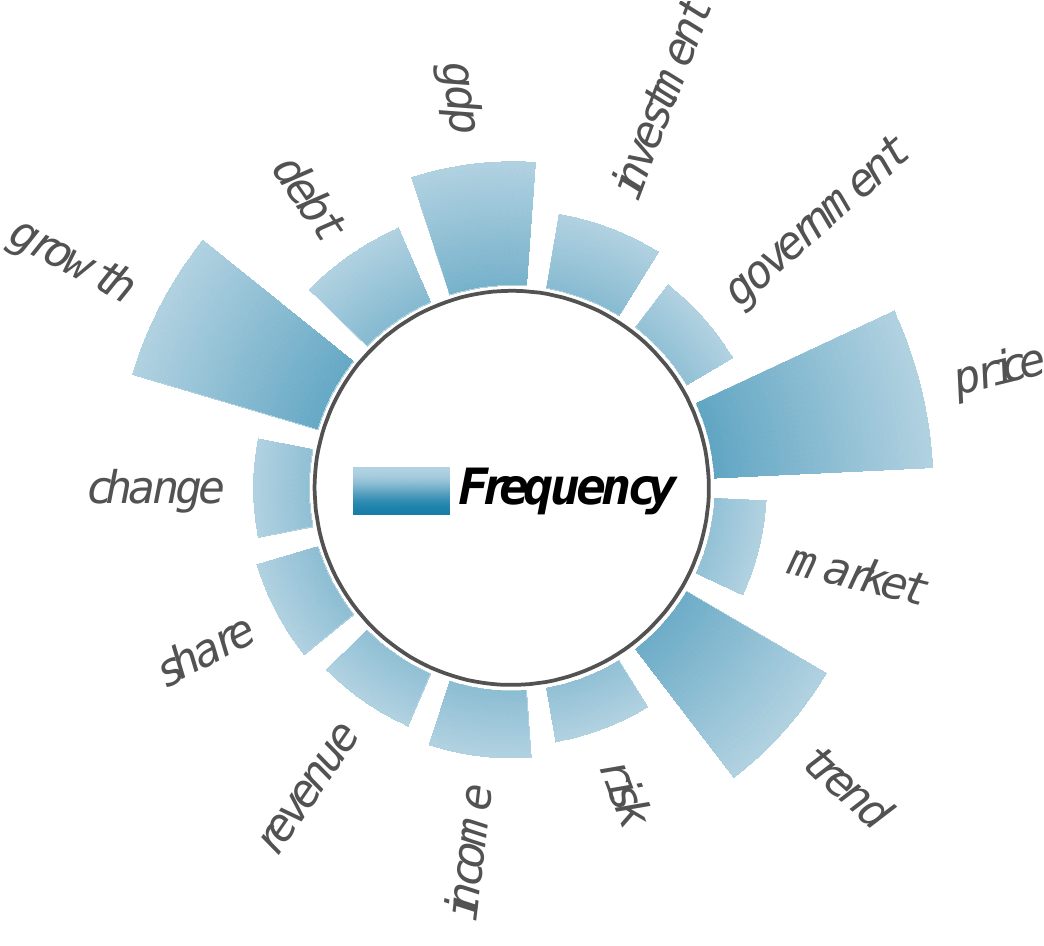}
    \caption{Top most frequent words in the questions.}
    \label{fig:top_words}
\end{figure}

Finally, we analyzed word frequency in the questions, as shown in Table~\ref{tab:top_words} and Figure~\ref{fig:top_words}. Excluding common function words such as prepositions and basic terms like "and" or "where," the most frequently occurring words include "price", "growth", "trend", "GDP", "debt", "investment", "income", "share", "revenue", "change", "risk", "government", and "market". This vocabulary highlights the dataset's rich thematic coverage across diverse domains of finance.

\begin{table}[ht]
\centering
\caption{Top most frequent words in the questions}
\label{tab:top_words}
\begin{adjustbox}{max width=\linewidth}
\begin{tabular}{llrr}
\toprule
\textbf{Rank} & \textbf{Word} & \textbf{Frequency} & \textbf{Percentage (\%)} \\
\midrule
1  & price        & 112,502 & 1.31 \\
2  & growth       & 99,789  & 1.17 \\
3  & trend        & 83,465  & 0.97 \\
4  & gdp          & 63,960  & 0.75 \\
5  & debt         & 42,057  & 0.49 \\
6  & investment   & 38,931  & 0.45 \\
7  & income       & 34,923  & 0.41 \\
8  & share        & 33,044  & 0.39 \\
9 & revenue      & 29,665  & 0.35 \\
10 & change       & 29,019  & 0.34 \\
11 & risk         & 28,486  & 0.33 \\
12 & government   & 27,847  & 0.33 \\
13 & market       & 26,763  & 0.31 \\
\bottomrule
\end{tabular}
\end{adjustbox} 
\end{table}

Before fine-tuning, we analyzed the answer distribution across all samples in the question chains and observed significant category imbalance. As shown in Table~\ref{tab:answer_distribution}, answer option A accounts for 43.39\% of all responses, while B makes up 34.20\%. In contrast, options C and D represent only 15.81\% and 6.08\%, respectively, with options E-J collectively comprising less than 0.52\%. This skew suggests a generation bias, likely introduced during synthetic question creation, where the model tends to place the correct answer in the first two choices. To mitigate this bias and ensure a fair evaluation, we manually rebalanced the answer distribution in a subset of 1,000 test samples so that options A, B, C, and D each appear with approximately equal frequency. During fine-tuning, we then compared model accuracy under the original imbalanced distribution versus the rebalanced version to quantify the impact of answer-position bias on model performance. This controlled comparison helps isolate whether apparent gains in accuracy stem from genuine reasoning ability or merely from exploiting skewed answer patterns.

\begin{table}[ht]
\centering
\caption{Answer distribution in the Question Chain with non-standard responses grouped as "Wrong Format"}
\label{tab:answer_distribution}
\begin{adjustbox}{max width=\linewidth}
\begin{tabular}{lr}
\toprule
\textbf{Answer Category} & \textbf{Percentage (\%)} \\
\midrule
A  & 43.39 \\
B  & 34.20 \\
C  & 15.81 \\
D  & 6.08 \\
E  & 0.44 \\
F  & 0.06 \\
G  & $<$0.01 \\
H  & $<$0.01 \\
I  & $<$0.01 \\
J  & $<$0.01 \\
\midrule
Wrong Format & $<$0.01 \\
\midrule
\textbf{Total} & \textbf{100.00} \\
\bottomrule
\end{tabular}
\end{adjustbox}
\end{table}

\subsection{Instructions}
\label{subsec:instructions}
\paragraph{Data Privacy}
Data is collected from open-source and publicly available resources, including reports, textbooks, and online documents. When using MinerU~\citep{Mineru-arxiv24} for data extraction, we strictly adhere to the copyright and licensing terms of each source. Furthermore, we provide comprehensive citation information for every financial document in our collection, including details such as title, author, and publisher. Most importantly, the dataset is used exclusively for research purposes and involves no financial gain.

\paragraph{Data Format}
All textual data in \textbf{\emph{PyFi}-600K} are formatted in JSON, while images are stored in \texttt{.jpg} format. We verify that each sample contains exactly one image, belongs to a specific level within the pyramid hierarchy, and is part of one or more sample chains.

\paragraph{Data Structure}

The complete dataset is stored in a tree structure represented in JSON format. The root node of the tree is a placeholder with no semantic meaning, while every other node represents a Q\&A pair. Each branch extending from the root to a leaf node constitutes a complete sample chain. Nodes are assigned to different levels of a capability pyramid based on their hierarchical competence. Each node contains the following fields: the question, the answer, an associated image, essential visual context required to answer the question, and a summary/analysis of the contextual information.

\subsection{Data Processing}
\label{subsec:data-processing}

\begin{figure*}[t]
    \centering
    \includegraphics[width=\linewidth]{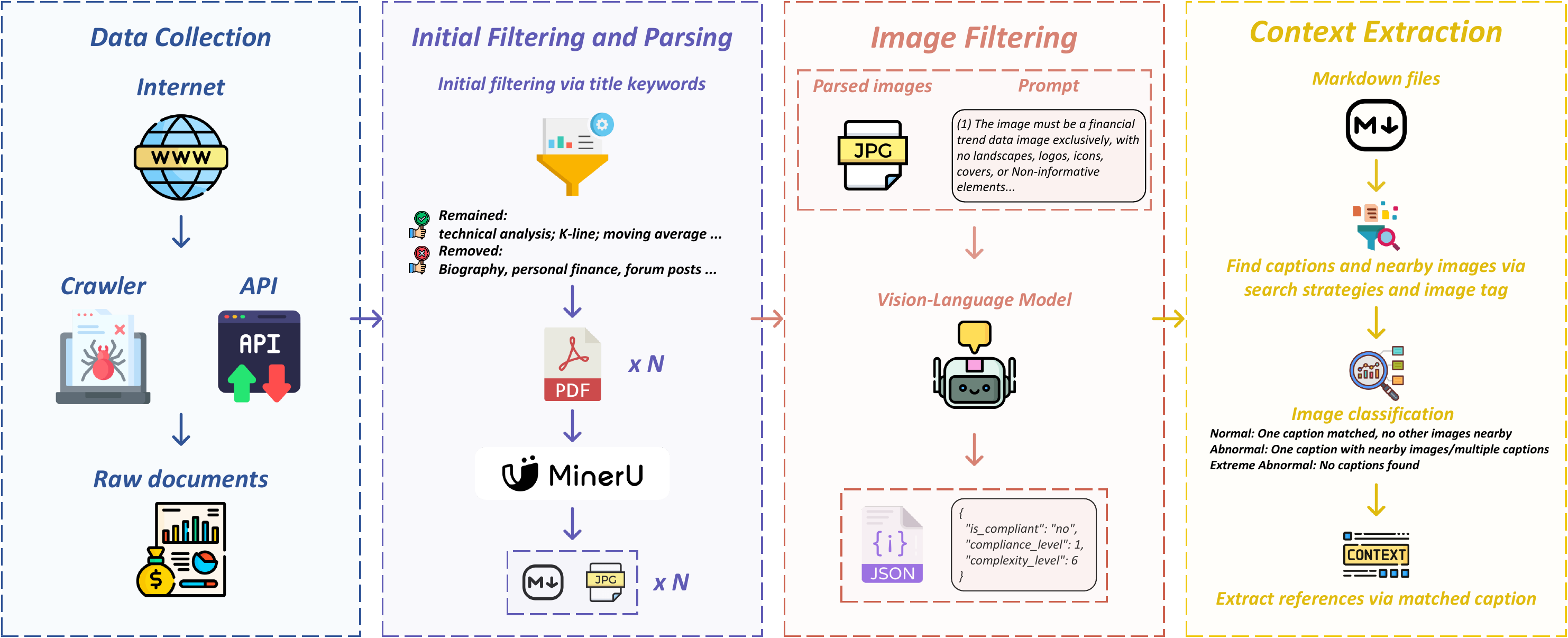}
    \caption{Overview of Data Processing Pipeline.}
    \label{fig:data_process}
\end{figure*}

To construct the final \textbf{\emph{PyFi}-600K} dataset, we adopt a concise multi-stage pipeline: collect public financial documents; parse PDFs to extract images and associated text; evaluate compliance and classify images (normal/abnormal/extremely abnormal), enhancing only normal ones with contextual metadata; categorize images into content themes and chart types; maintain provenance; and sample by relevance and complexity with expert validation. An overview is shown in Figure~\ref{fig:data_process}.

We aggregated over 30,000 public financial PDFs in Chinese and English via web crawling and institutional APIs, then applied technical checks (corruption, password protection, readability) and domain relevance filters with curated inclusion/exclusion keywords to retain about 10,000 documents. We leverage \texttt{MinerU}~\citep{Mineru-arxiv24} to parse validated PDFs and extract images, text blocks, and layout metadata, preserving image–caption associations with page indices and bounding coordinates. Extracted assets were reorganized into a standardized hierarchical directory by document and page (separating images and contextual text) for efficient batch processing and traceability. We screened approximately 820,000 images using a VLM-guided dual 0–10 scoring for compliance (financial visualization criteria) and complexity (visual/data density and structural sophistication), generated per-image JSON summaries, and conducted a targeted manual audit. Images were labeled as normal/abnormal/extremely abnormal by combining caption-match windows around image tags with nearby-image proximity signals; only normal images were enhanced with contextual metadata. We follow the following criteria to select the target images for our study:

\begin{itemize}

    \item Financial trend data chart only; no non-informative elements.
    \item Clear, legible Chinese or English text; unambiguous recognition.
    \item No obvious errors or omissions; follows financial data conventions.
    \item Clearly highlights dynamic financial processes.
    \item Tables/grids are secondary and must not dominate.
    \item Explicit relationships shown via visual connections.
    \item Addresses at least one finance-related issue.
    \item Universally discernible; self-contained composition.

\end{itemize}

\textbf{Context Extraction}. The retained images are classified into three status categories (\textit{normal}, \textit{abnormal} and \textit{extremely abnormal}) based on their contextual clarity within the document. A \textit{normal} image is defined by a clear one-to-one match with a single caption located nearby and no other adjacent images, indicating reliable and unambiguous correspondence. Only images classified as \textit{normal} are enriched with detailed contextual metadata from the surrounding text, providing each with explanatory insights.

An \textit{abnormal} image arises when there is contextual ambiguity, either because the image has other images close to it or because multiple captions appear near it. An \textit{extremely abnormal} image is identified when no caption is found within the search window, signaling a high risk of missing context. This classification ensures that contextual enhancement is applied only to images with a well-defined and correct caption relationship.

\textbf{Topic Classification}. We utilize the VLM to classify each image into one of 17 distinct financial content themes and one of 11 chart types, based on both the visual content of the image and its extracted contextual information.

\vspace{1em}

\textbf{Content Theme and Chart Type Classification}

\vspace{1em}

\textbf{I. Content Theme}

\begin{figure*}[h!]
\centering

% --- Row 1 ---
\begin{minipage}{0.45\linewidth}
    \centering
    \includegraphics[width=\linewidth, height=3.5cm, keepaspectratio]{"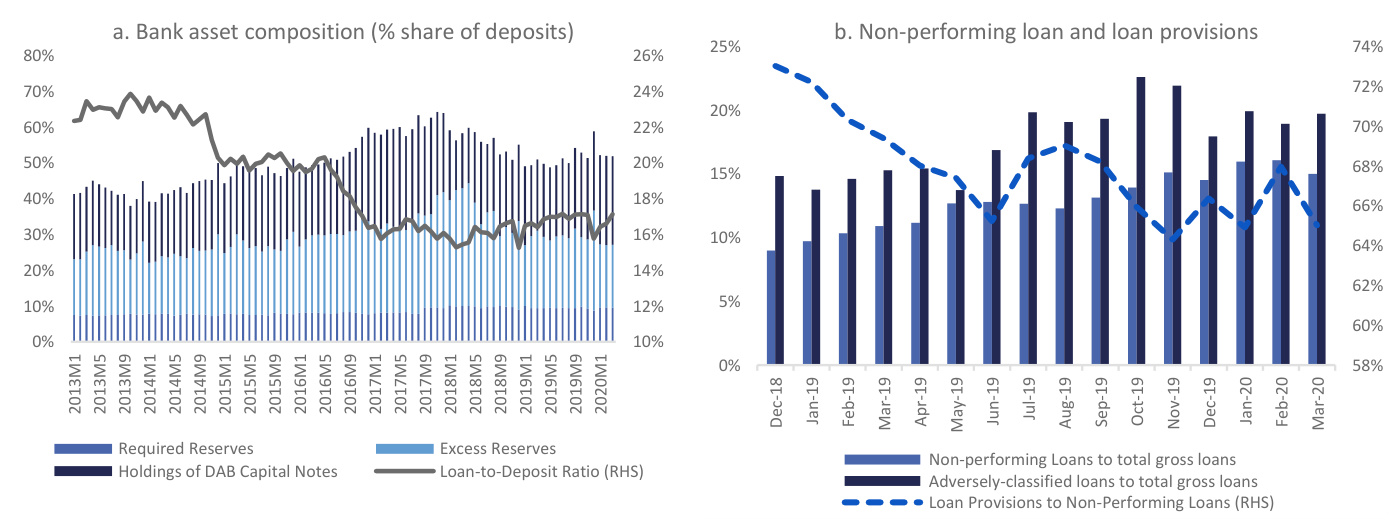"} \\
    \vspace{0.5em}
    Macroeconomic Indicators
\end{minipage}%
\hfill
\begin{minipage}{0.45\linewidth}
    \centering
    \includegraphics[width=\linewidth, height=3.5cm, keepaspectratio]{"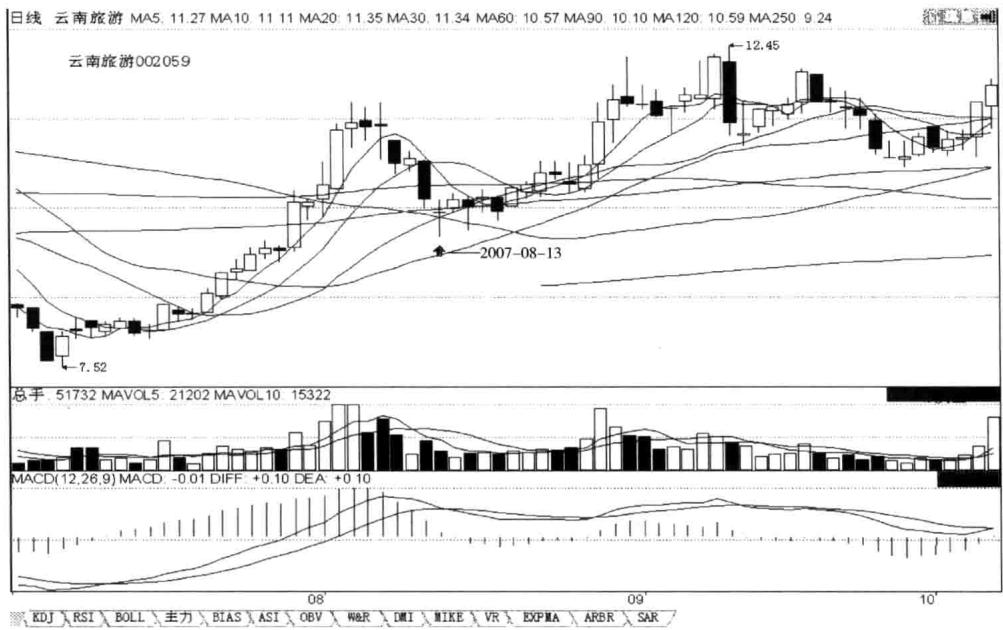"} \\
    \vspace{0.5em}
    Financial Markets \& Products
\end{minipage}

\vspace{1em}

% --- Row 2 ---
\begin{minipage}{0.45\linewidth}
    \centering
    \includegraphics[width=\linewidth, height=4.5cm, keepaspectratio]{"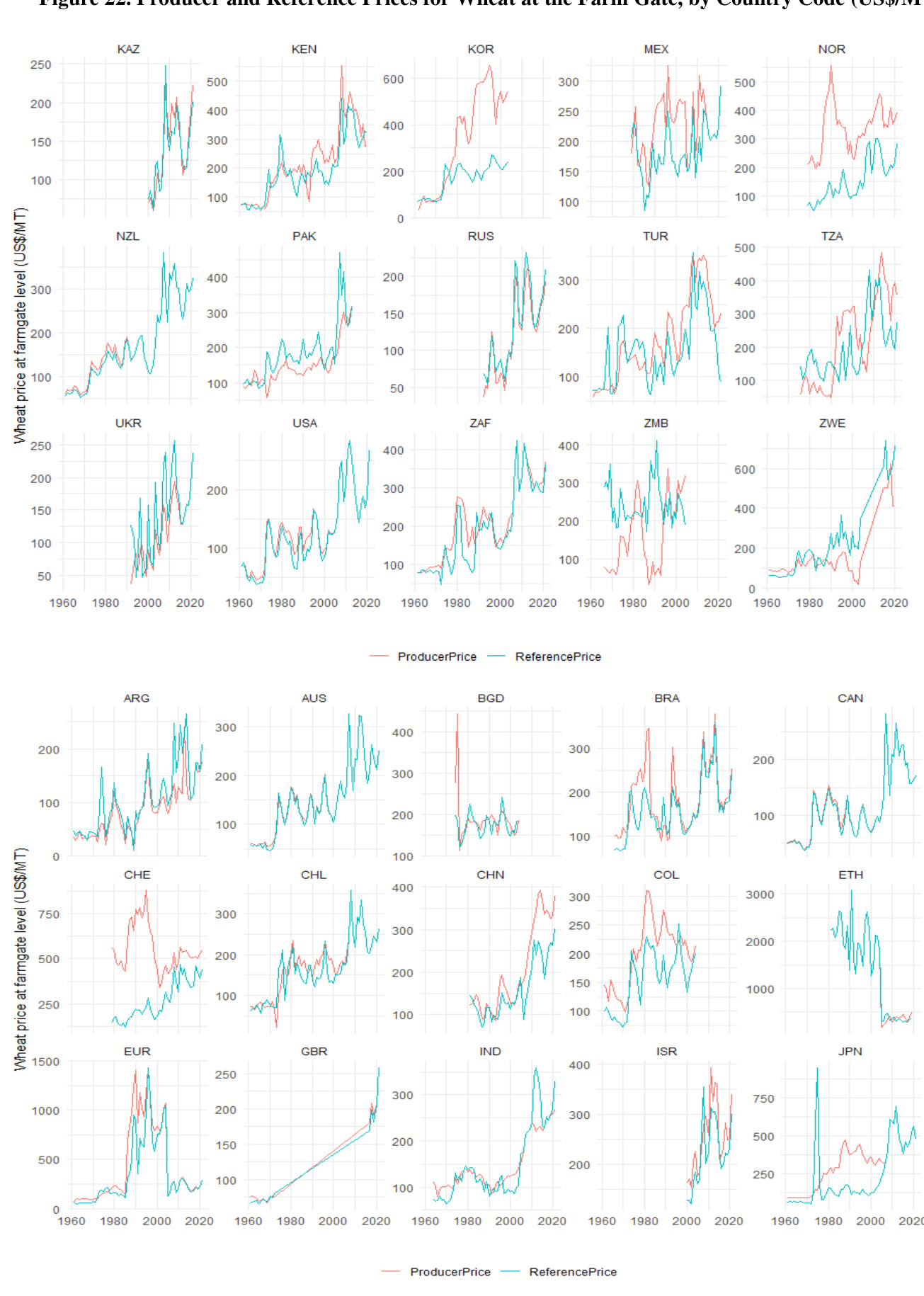"} \\
    \vspace{0.5em}
    Commodities \& Real Estate Markets
\end{minipage}%
\hfill
\begin{minipage}{0.45\linewidth}
    \centering
    \includegraphics[width=\linewidth, height=4.5cm, keepaspectratio]{"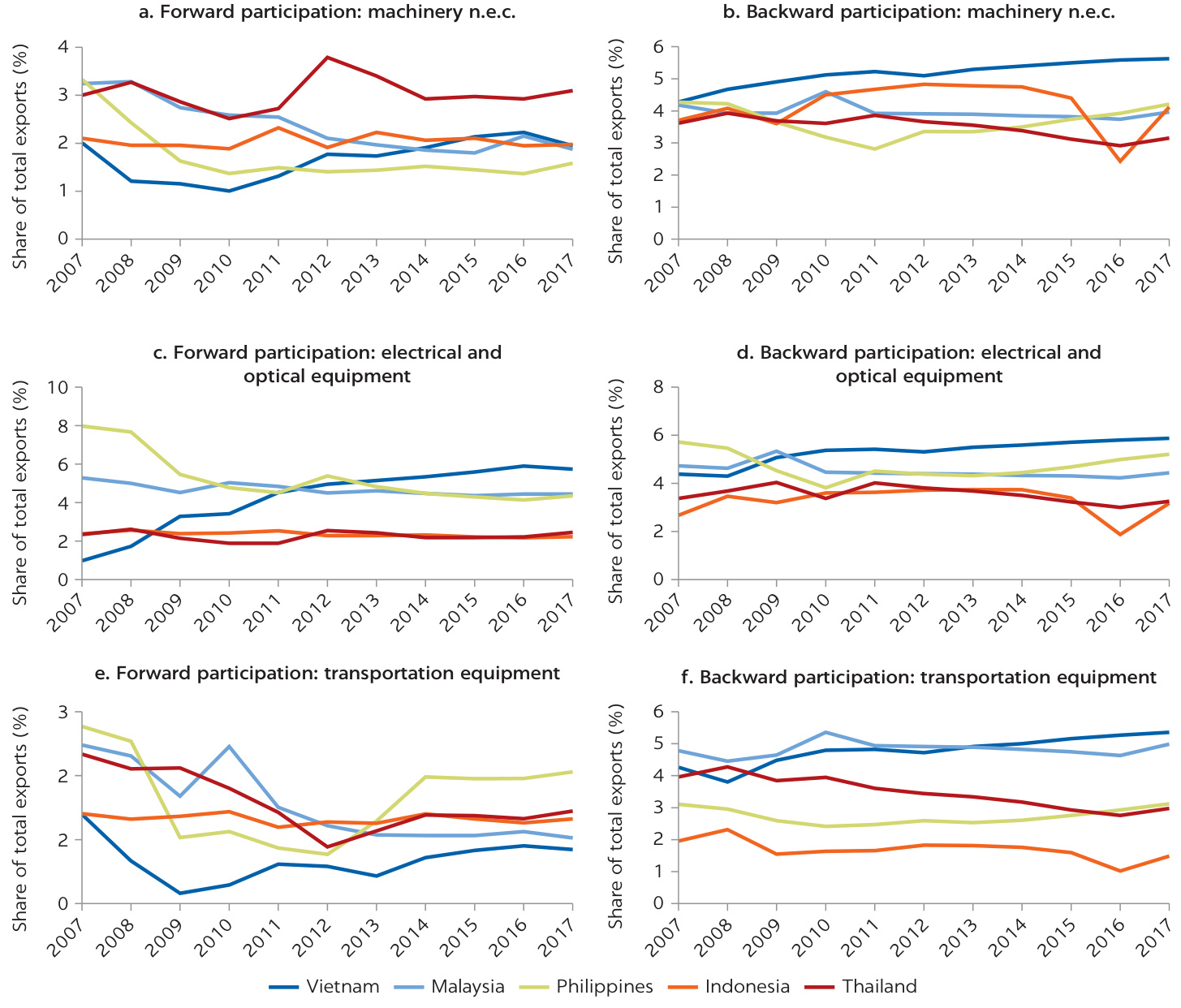"} \\
    \vspace{0.5em}
    Industry Analysis
\end{minipage}

\vspace{1em}

% --- Row 3 ---
\begin{minipage}{0.45\linewidth}
    \centering
    \includegraphics[width=\linewidth, height=3.5cm, keepaspectratio]{"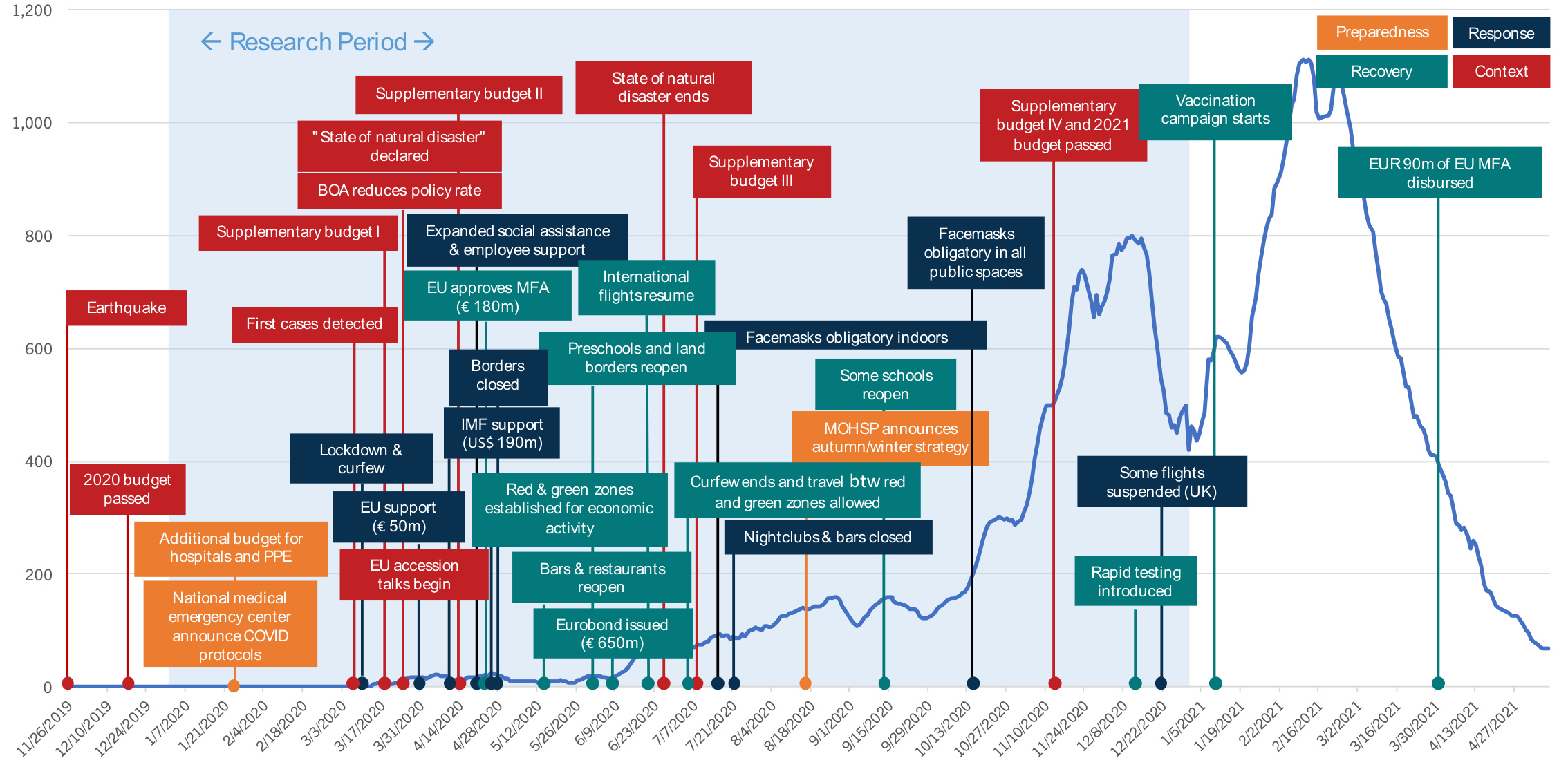"} \\
    \vspace{0.5em}
    Monetary \& Fiscal Policy
\end{minipage}%
\hfill
\begin{minipage}{0.45\linewidth}
    \centering
    \includegraphics[width=\linewidth, height=3.5cm, keepaspectratio]{"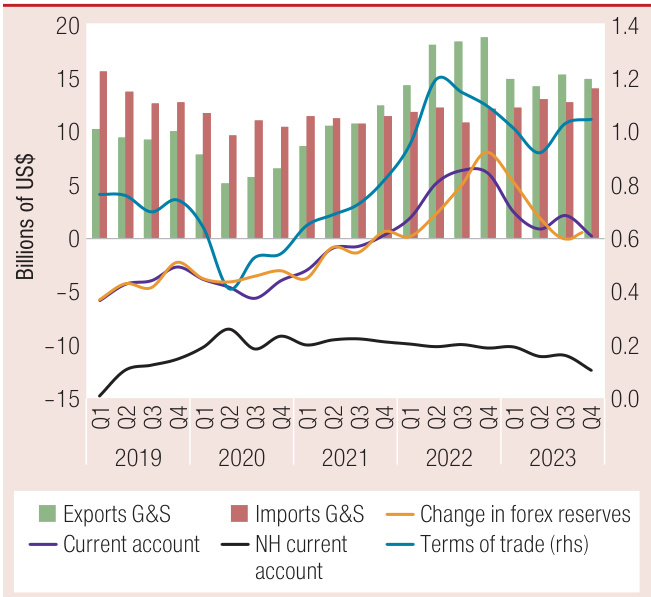"} \\
    \vspace{0.5em}
    International Trade \& Capital Flows
\end{minipage}

\vspace{1em}

% --- Row 4 ---
\begin{minipage}{0.45\linewidth}
    \centering
    \includegraphics[width=\linewidth, height=3.5cm, keepaspectratio]{"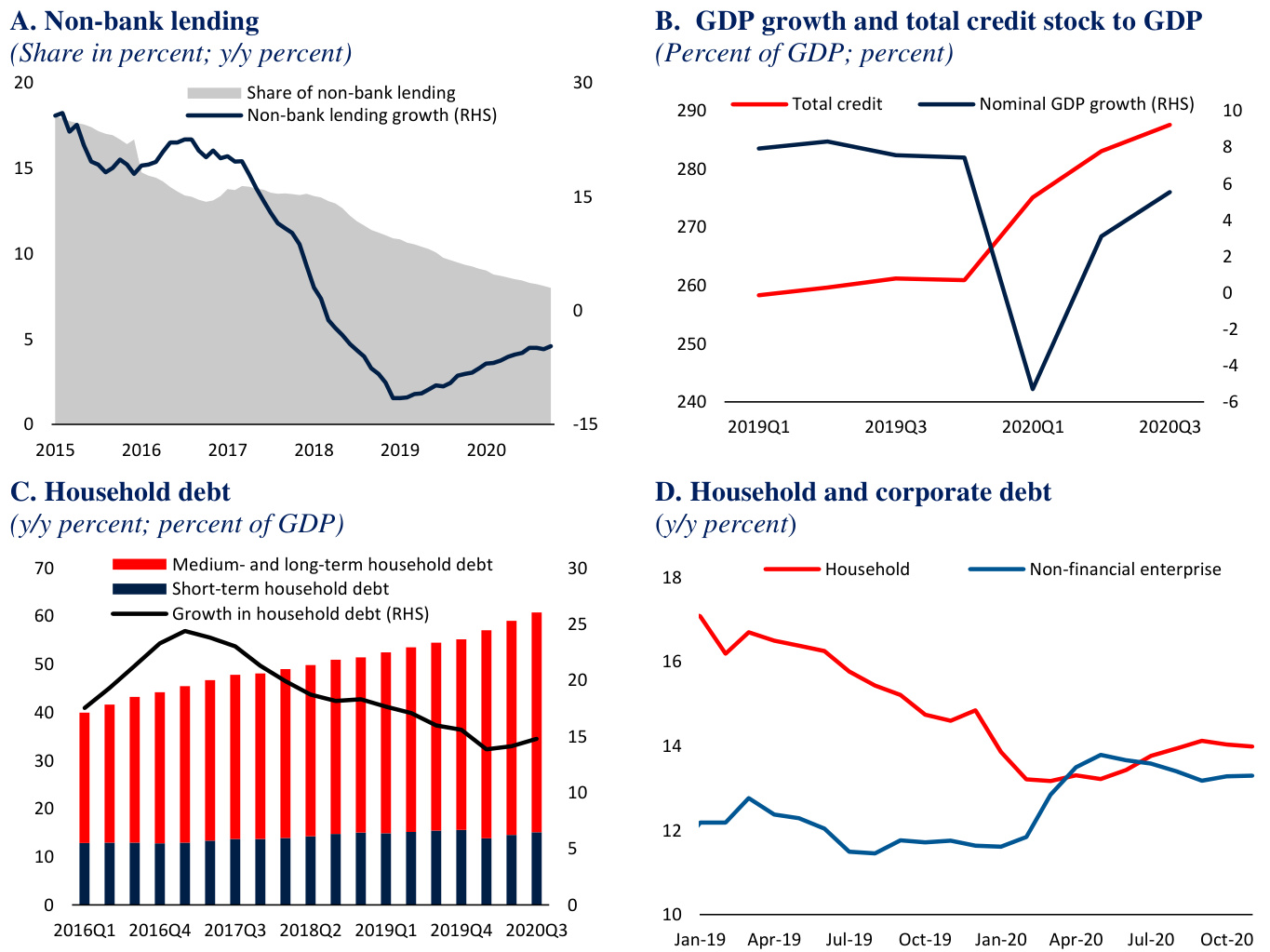"} \\
    \vspace{0.5em}
    Corporate Finance \& Valuation
\end{minipage}%
\hfill
\begin{minipage}{0.45\linewidth}
    \centering
    \includegraphics[width=\linewidth, height=3.5cm, keepaspectratio]{"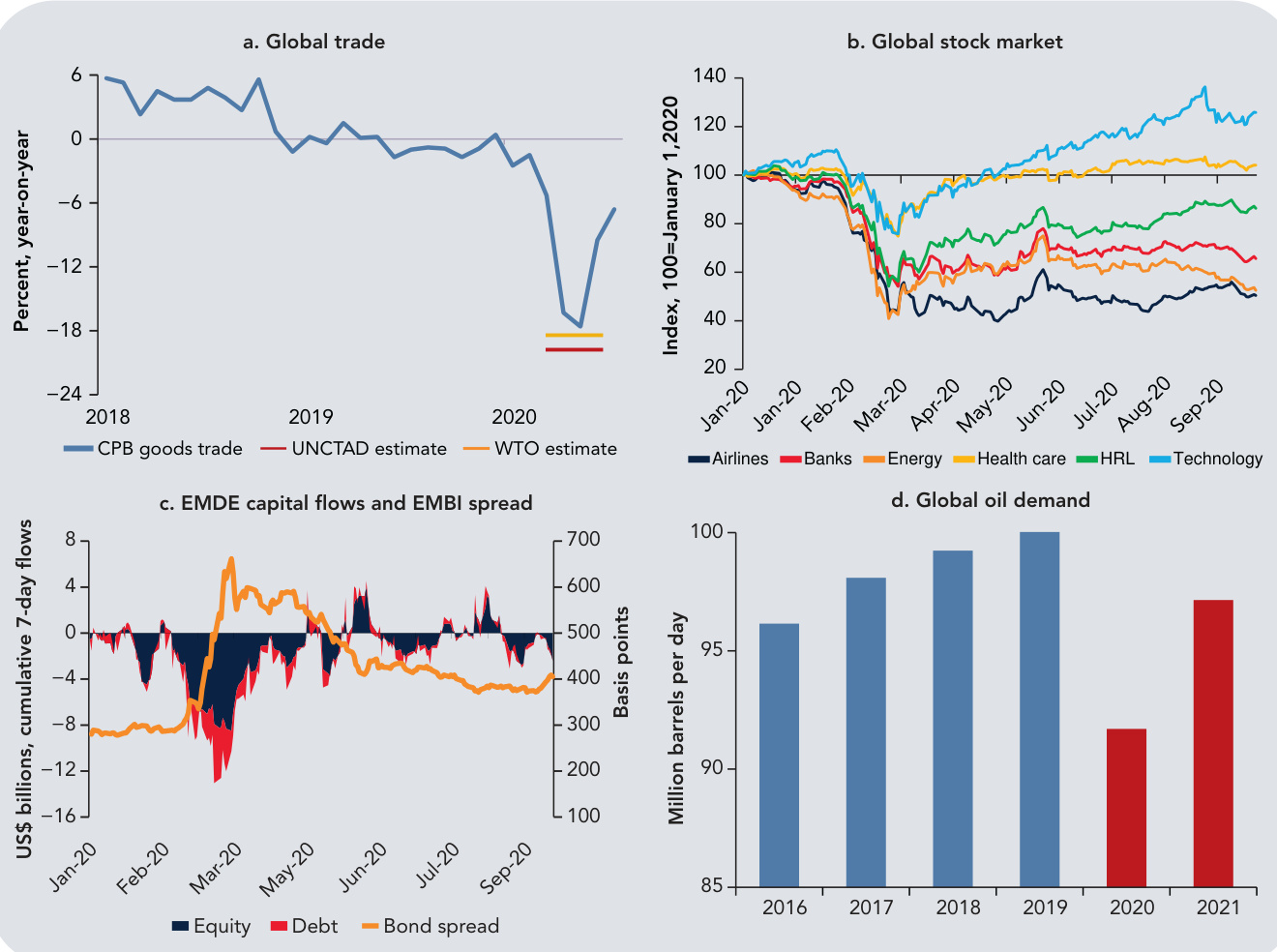"} \\
    \vspace{0.5em}
    Bonds \& Fixed Income
\end{minipage}

\vspace{1em}

% --- Row 5 ---
\begin{minipage}{0.45\linewidth}
    \centering
    \includegraphics[width=\linewidth, height=3.5cm, keepaspectratio]{"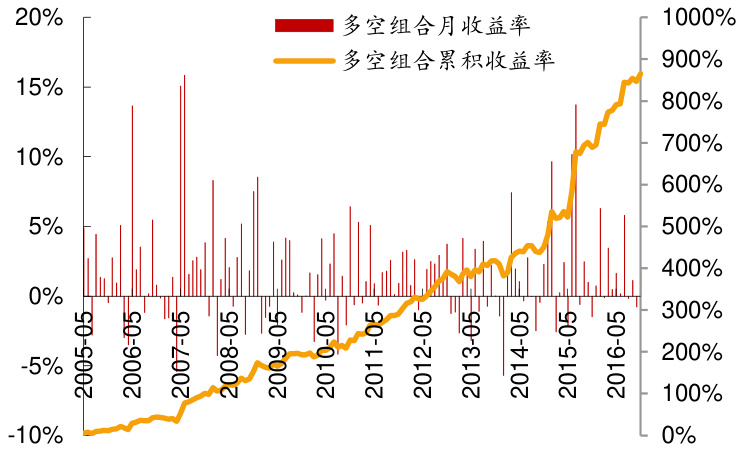"} \\
    \vspace{0.5em}
    Investment Theory \& Portfolio Management
\end{minipage}%
\hfill
\begin{minipage}{0.45\linewidth}
    \centering
    \includegraphics[width=\linewidth, height=3.5cm, keepaspectratio]{"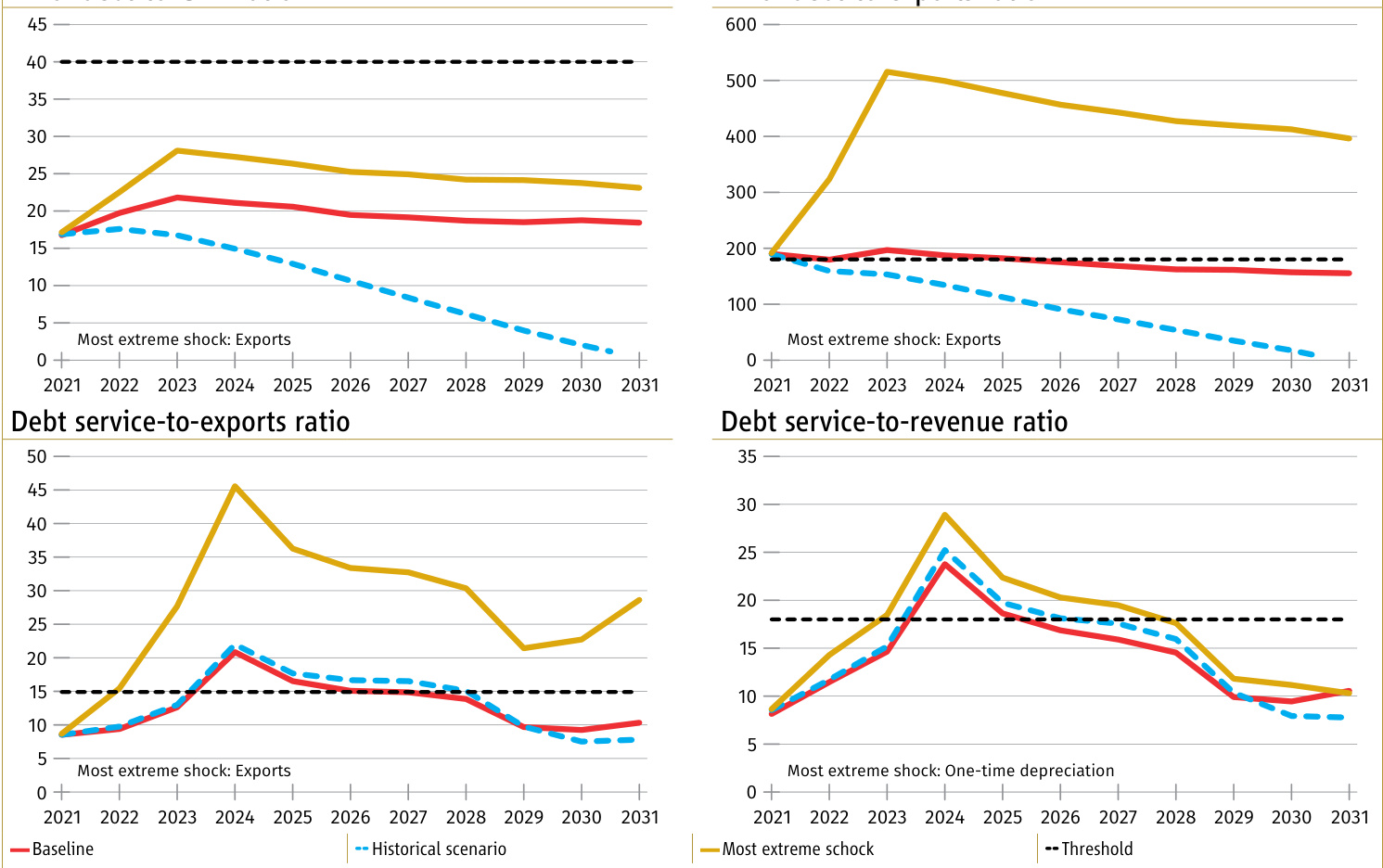"} \\
    \vspace{0.5em}
    Risk Models \& Management
\end{minipage}
\caption{Overview of content themes.}
\label{fig:content_theme_examples}
\end{figure*}

\begin{figure*}[h!]

\vspace{1em}

% --- Row 1 ---
\begin{minipage}{0.45\linewidth}
    \centering
    \includegraphics[width=\linewidth, height=6.5cm, keepaspectratio]{"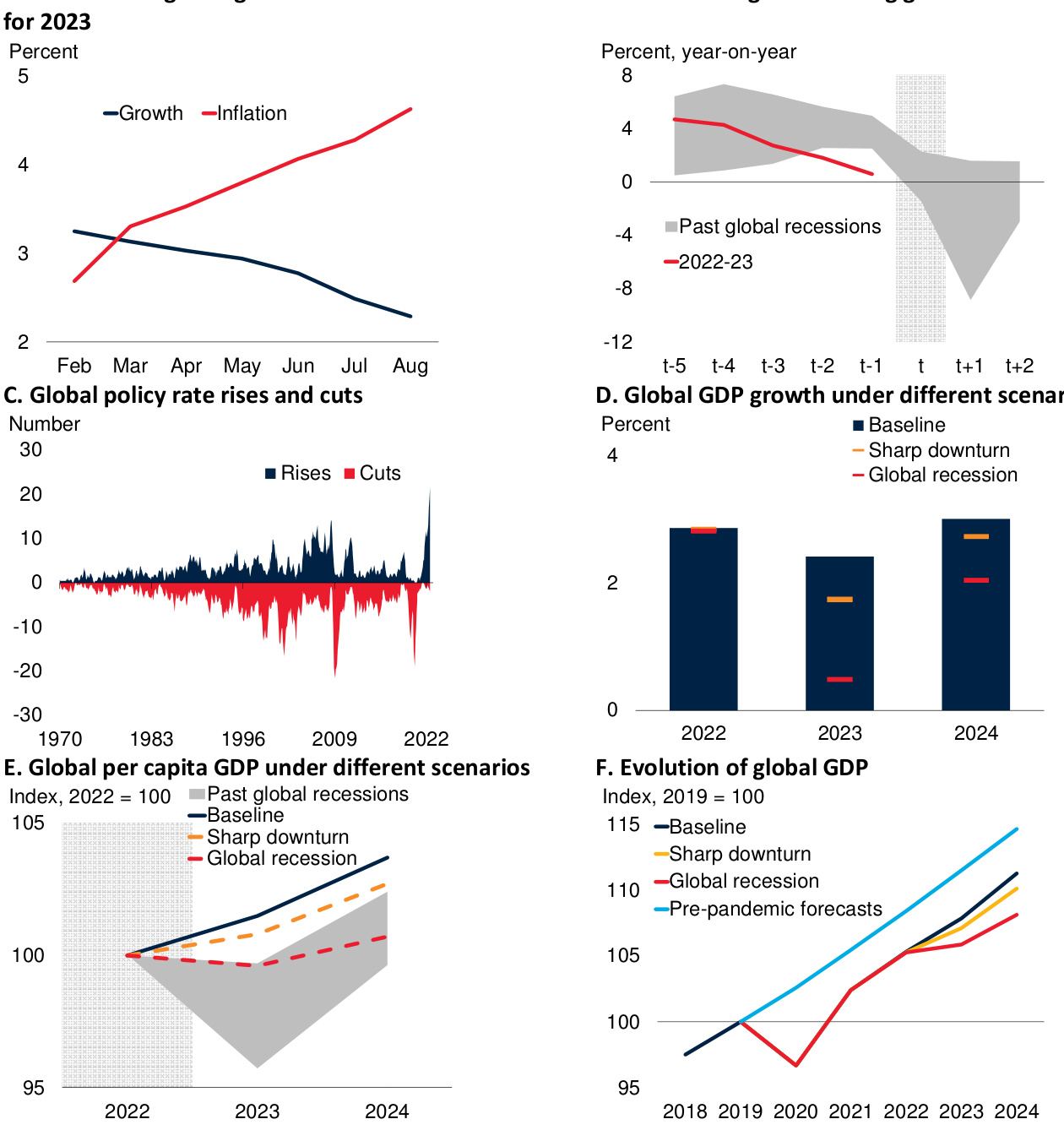"} \\
    \vspace{0.5em}
    Economic Cycles \& Market Theories
\end{minipage}%
\hfill
\begin{minipage}{0.45\linewidth}
    \centering
    \includegraphics[width=\linewidth, height=6.5cm, keepaspectratio]{"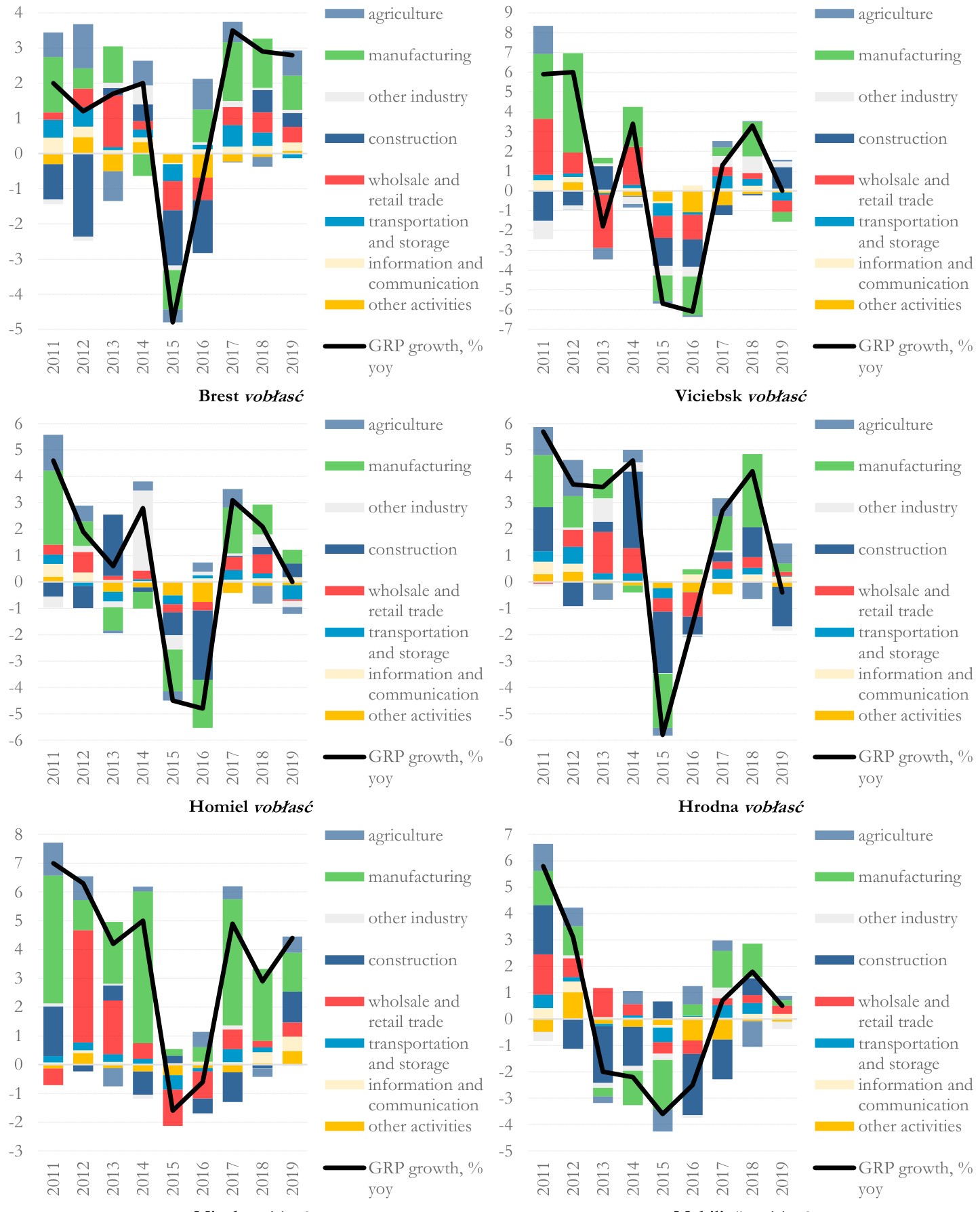"} \\
    \vspace{0.5em}
    Geospatial Economic Data
\end{minipage}

\vspace{1em}

% --- Row 2 ---
\begin{minipage}{0.45\linewidth}
    \centering
    \includegraphics[width=\linewidth, height=4.5cm, keepaspectratio]{"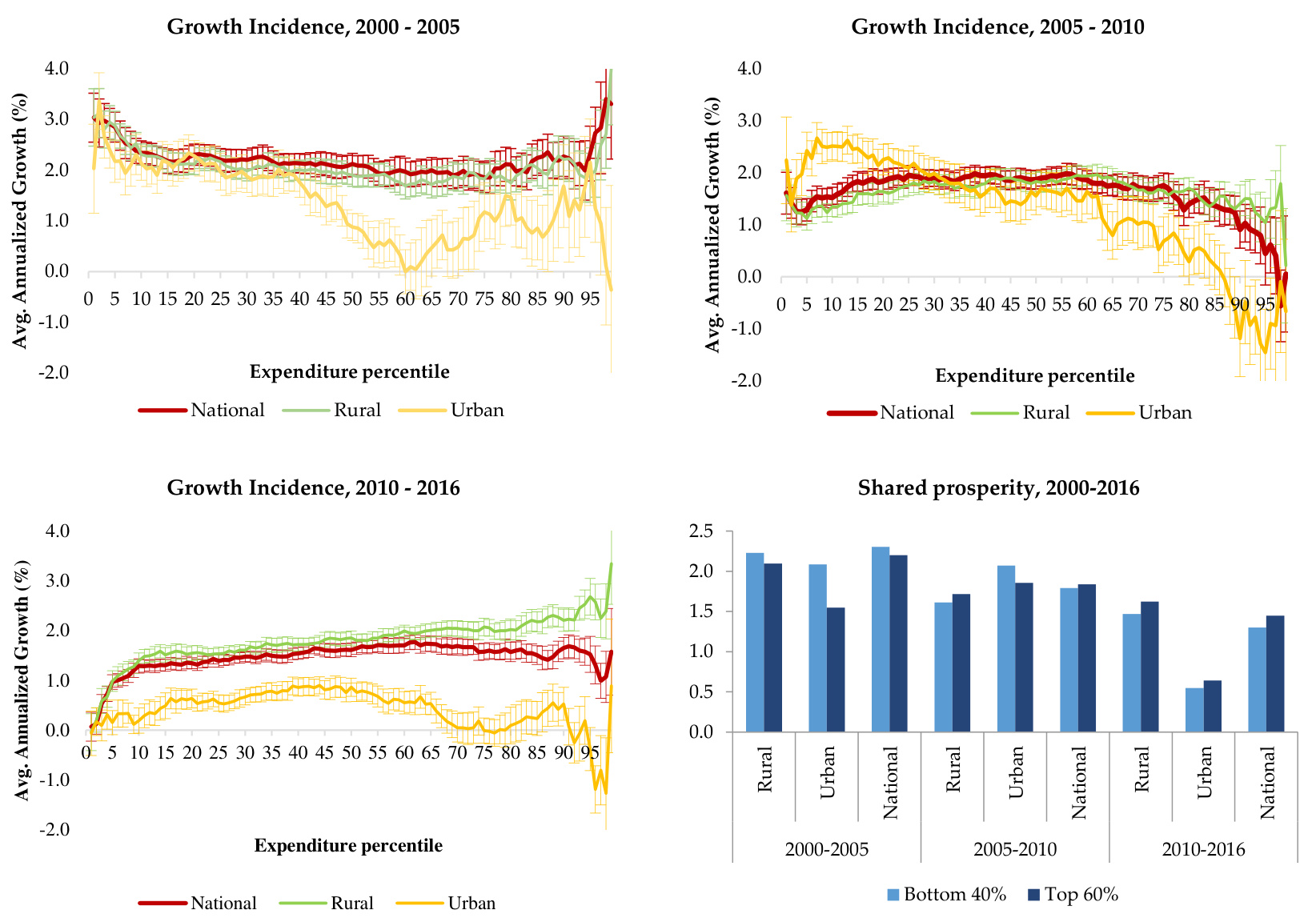"} \\
    \vspace{0.5em}
    Demographics \& Socioeconomics
\end{minipage}%
\hfill
\begin{minipage}{0.45\linewidth}
    \centering
    \includegraphics[width=\linewidth, height=4.5cm, keepaspectratio]{"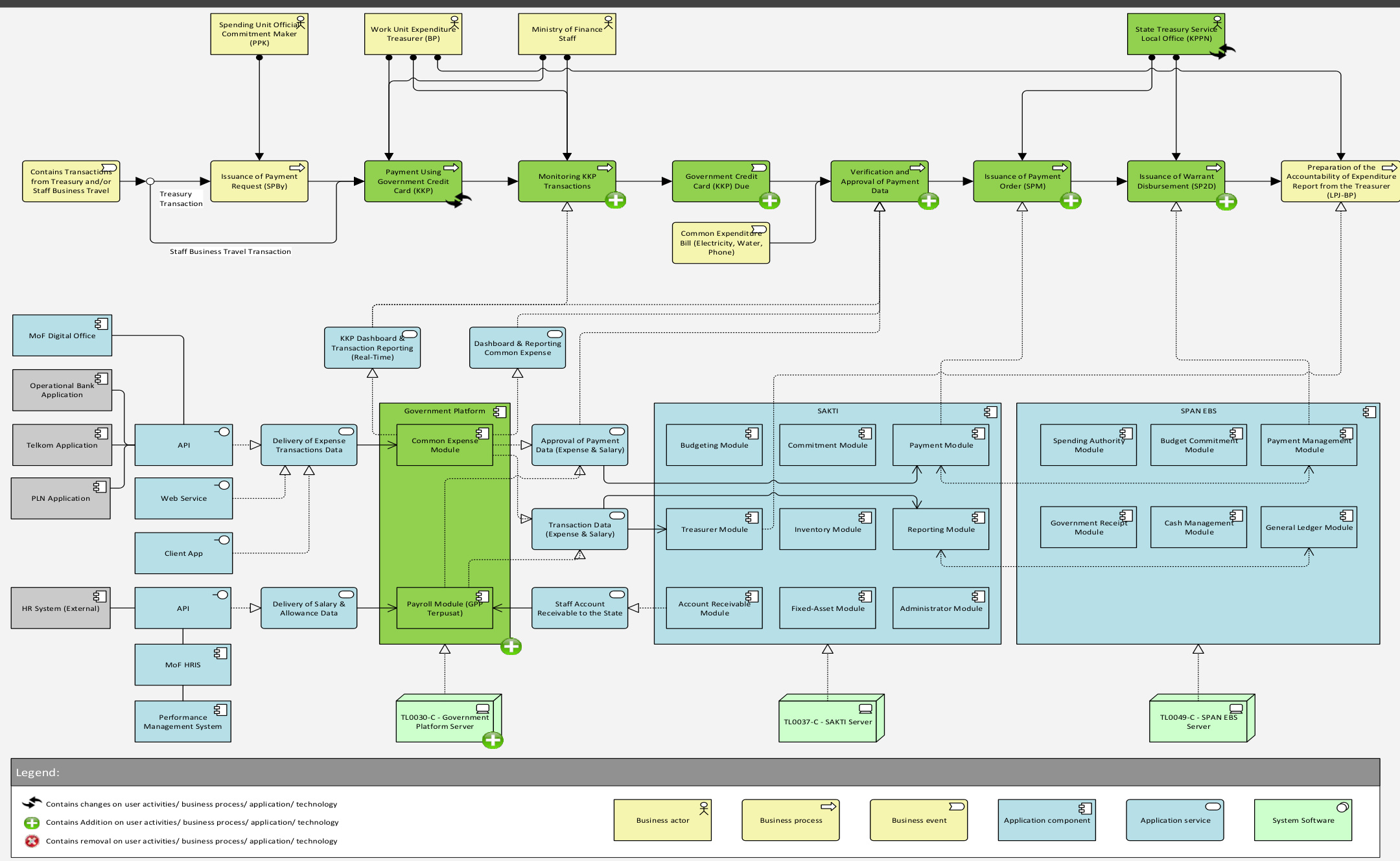"} \\
    \vspace{0.5em}
    Financial Systems \& Infrastructure
\end{minipage}

\vspace{1em}

% --- Row 3 ---
\begin{minipage}{0.45\linewidth}
    \centering
    \includegraphics[width=\linewidth, height=4.5cm, keepaspectratio]{"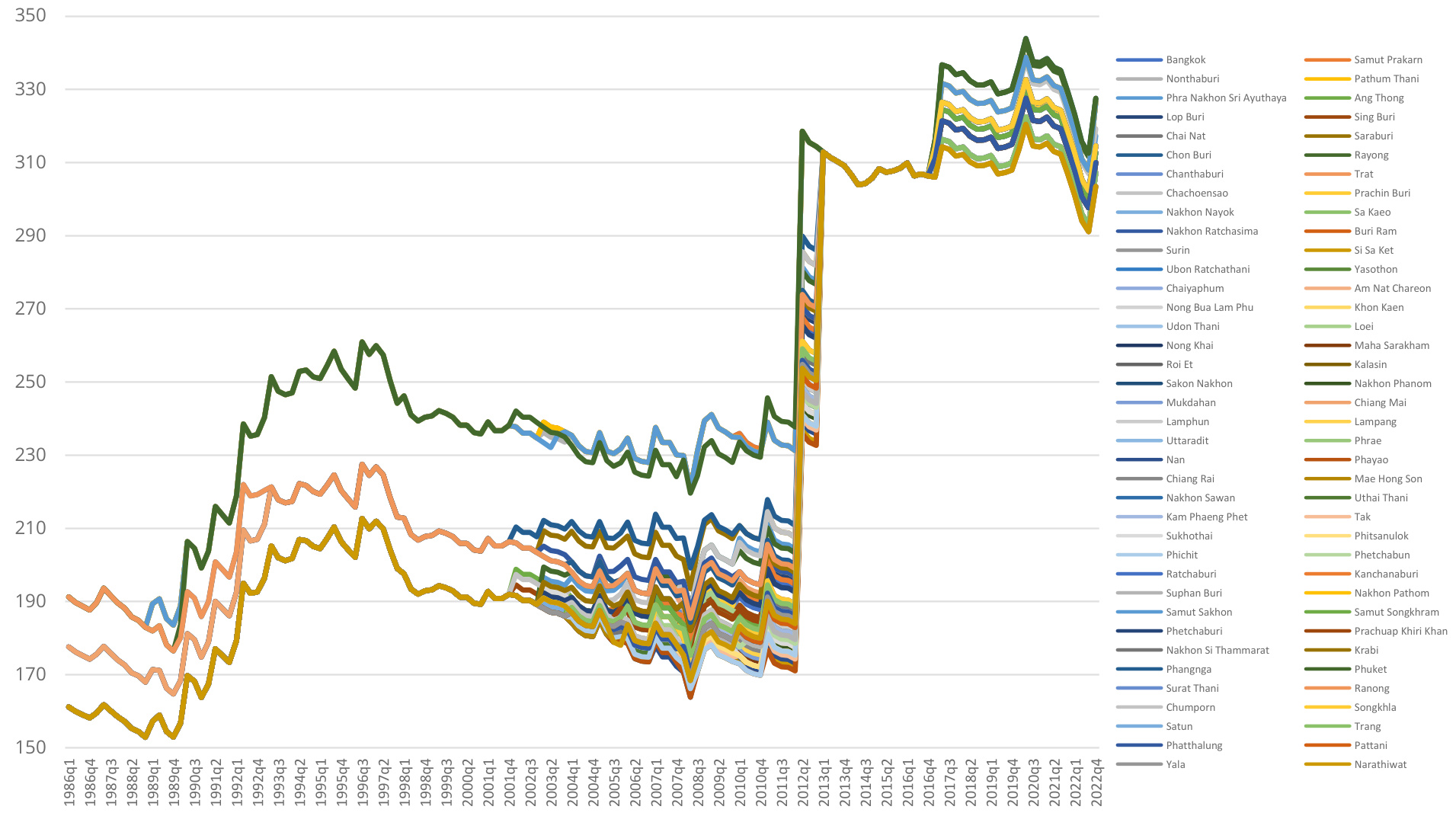"} \\
    \vspace{0.5em}
    Microeconomic Principles
\end{minipage}%
\hfill
\begin{minipage}{0.45\linewidth}
    \centering
    \includegraphics[width=\linewidth, height=4.5cm, keepaspectratio]{"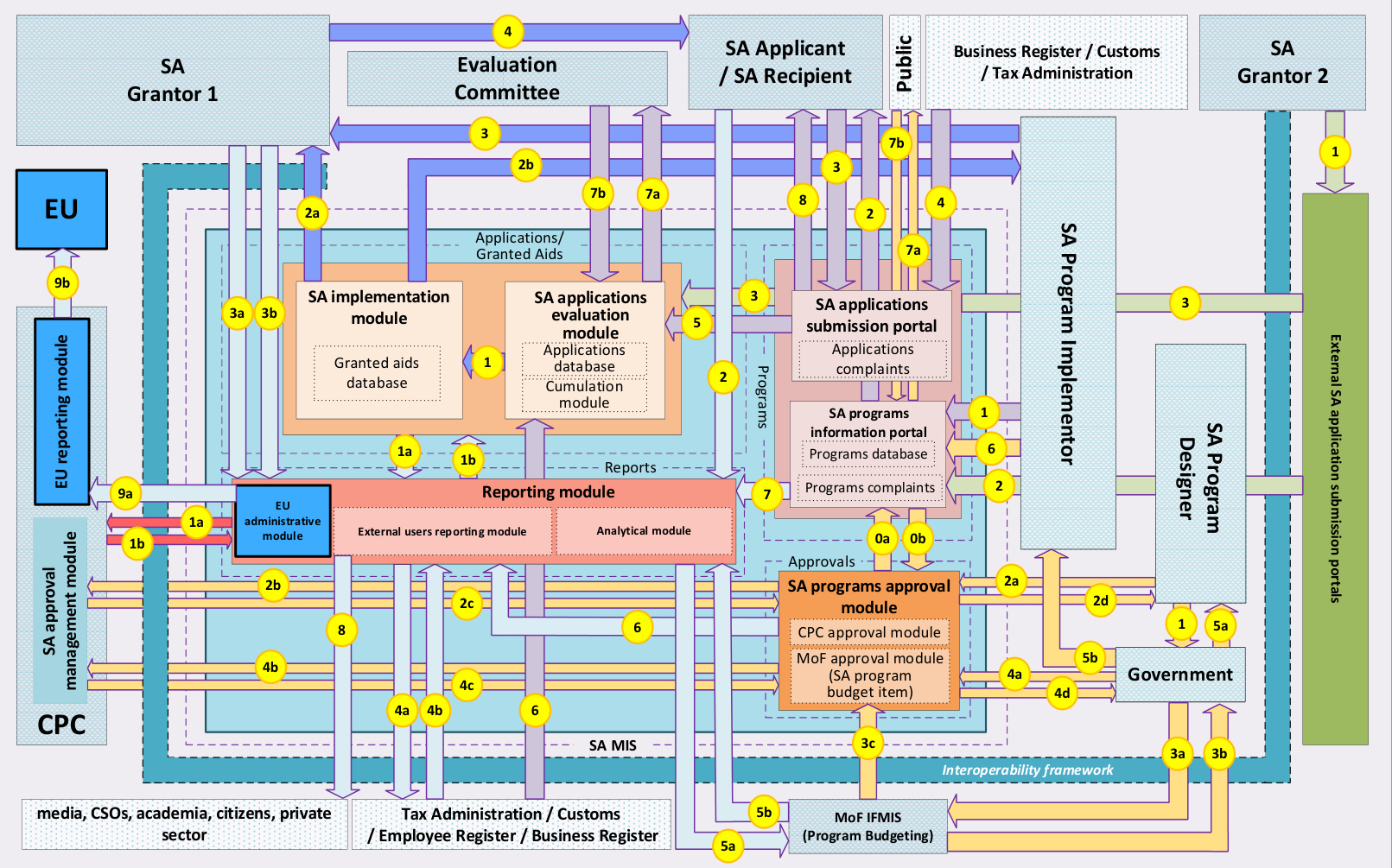"} \\
    \vspace{0.5em}
    Organization \& Regulation
\end{minipage}

\vspace{1em}

% --- Row 4 (Single Item) ---
\begin{minipage}{0.45\linewidth}
    \centering
    \includegraphics[width=\linewidth, height=5cm, keepaspectratio]{"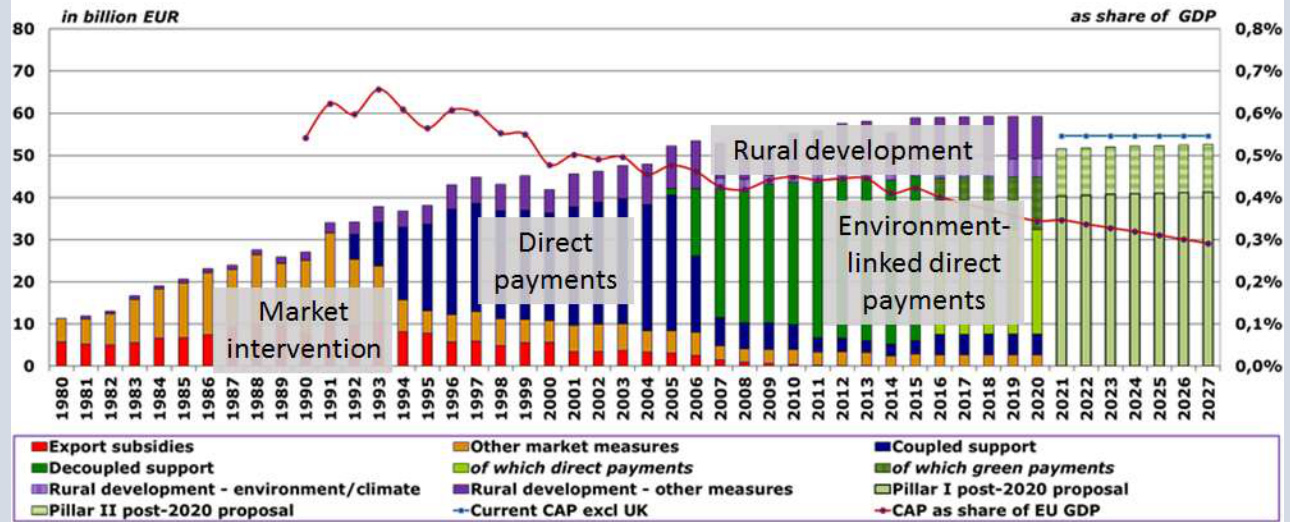"} \\
    \vspace{0.5em}
    Financial History \& Documentation
\end{minipage}
\caption{Overview of content themes (continued).}
\label{fig:content_theme_examples_cont}
\end{figure*}

This dimension describes the core financial subject of the image, including \textit{Macroeconomic Indicators}, \textit{Financial Markets \& Products}, \textit{Commodities \& Real Estate Markets}, \textit{Bonds \& Fixed Income}, \textit{Monetary \& Fiscal Policy}, \textit{International Trade \& Capital Flows}, \textit{Corporate Finance \& Valuation}, \textit{Industry Analysis}, \textit{Investment Theory \& Portfolio Management}, \textit{Risk Models \& Management}, \textit{Economic Cycles \& Market Theories}, \textit{Microeconomic Principles}, \textit{Demographics \& Socioeconomics}, \textit{Financial Systems \& Infrastructure}, \textit{Geospatial Economic Data}, \textit{Organization \& Regulation}, and \textit{Financial History \& Documentation}, as shown in Figure~\ref{fig:content_theme_examples} and Figure~\ref{fig:content_theme_examples_cont}.

\textbf{II. Chart Type}

\begin{figure*}[h!]
\centering

% --- Row 1 ---
\begin{minipage}{0.45\linewidth}
    \centering
    \includegraphics[width=\linewidth, height=5cm, keepaspectratio]{"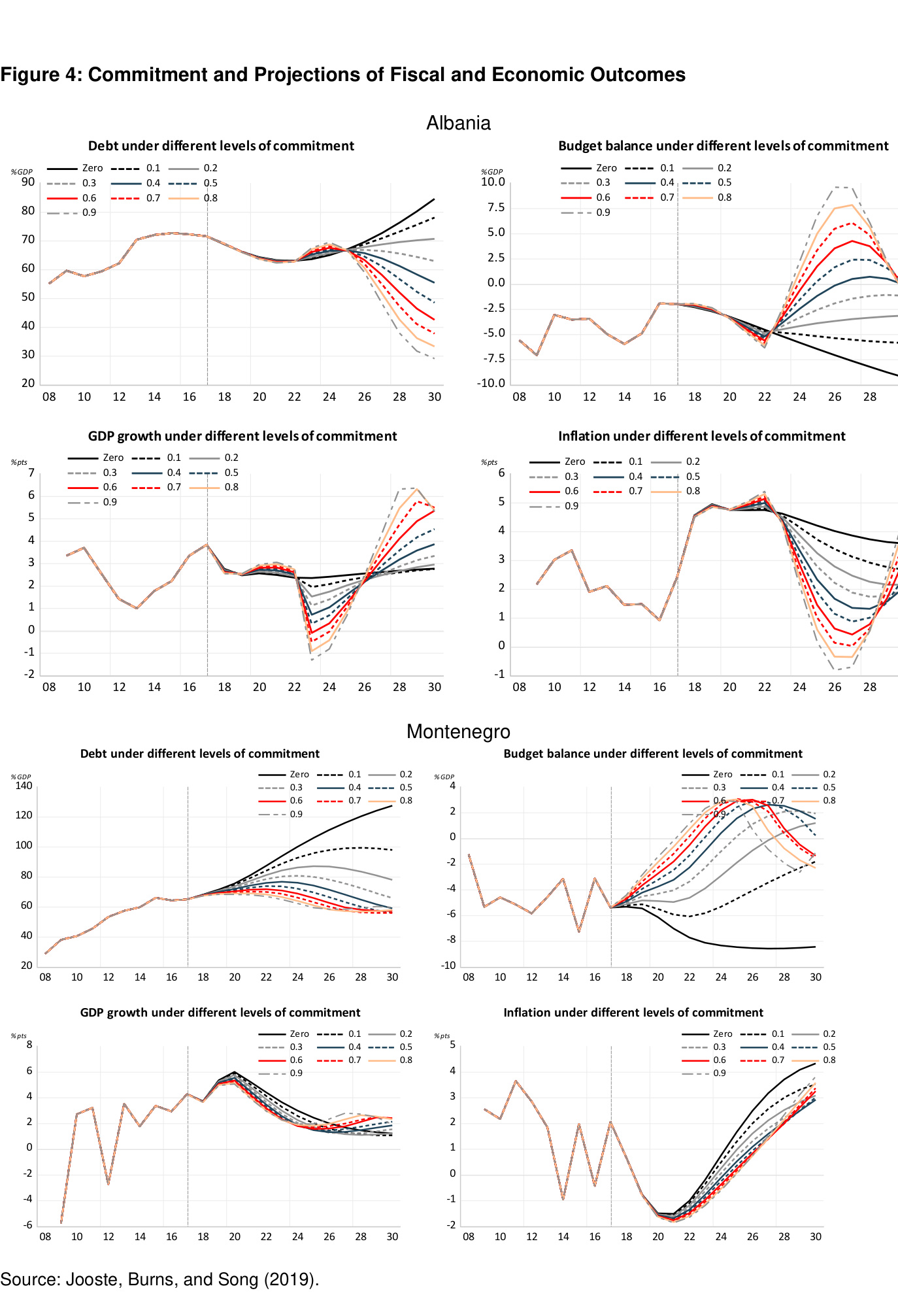"} \\
    \vspace{0.5em}
    Line Chart
\end{minipage}%
\hfill
\begin{minipage}{0.45\linewidth}
    \centering
    \includegraphics[width=\linewidth, height=5cm, keepaspectratio]{"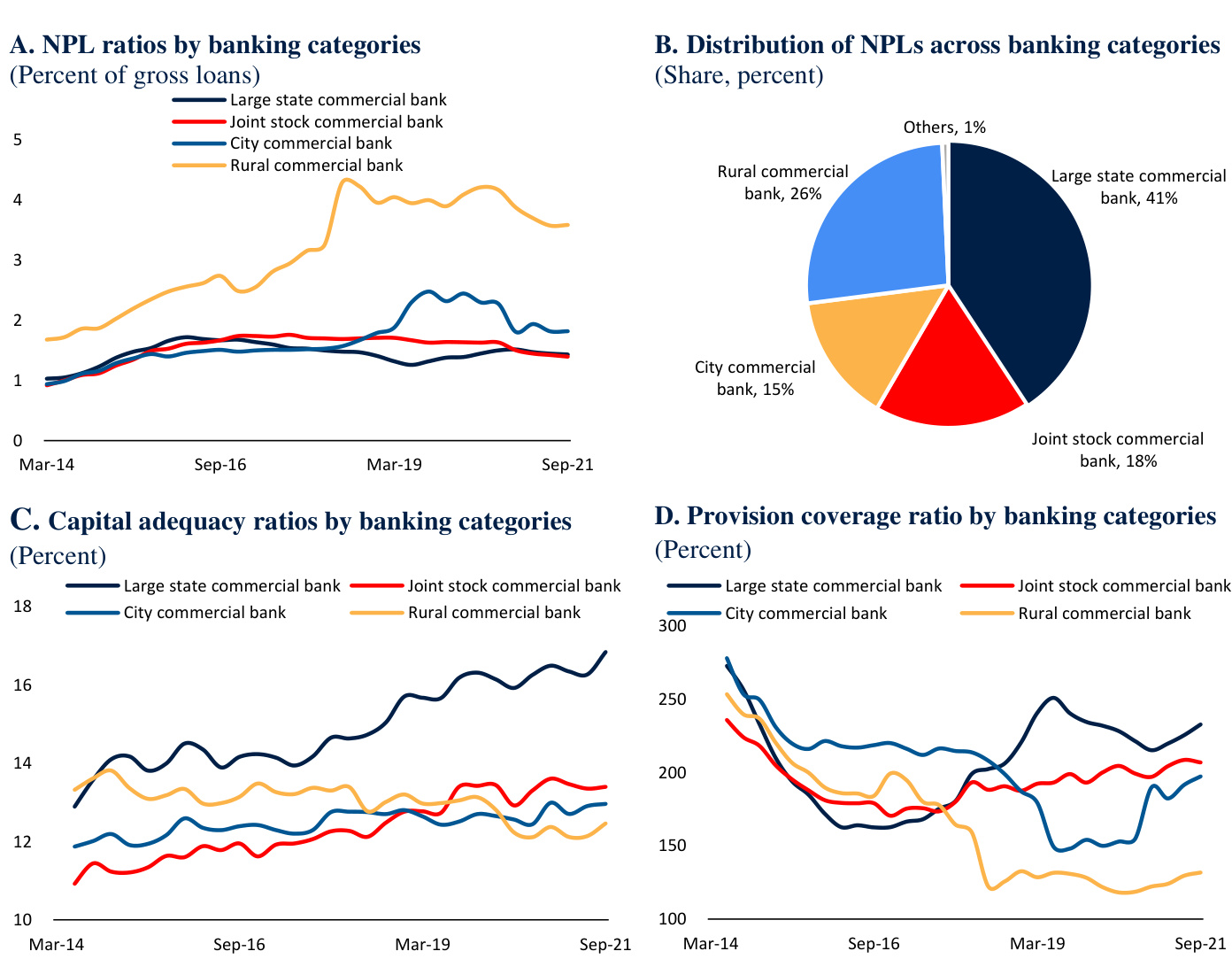"} \\
    \vspace{0.5em}
    Pie Chart \& Donut Chart
\end{minipage}
\vspace{1em}

% --- Row 2 ---
\begin{minipage}{0.45\linewidth}
    \centering
    \includegraphics[width=\linewidth, height=3.5cm, keepaspectratio]{"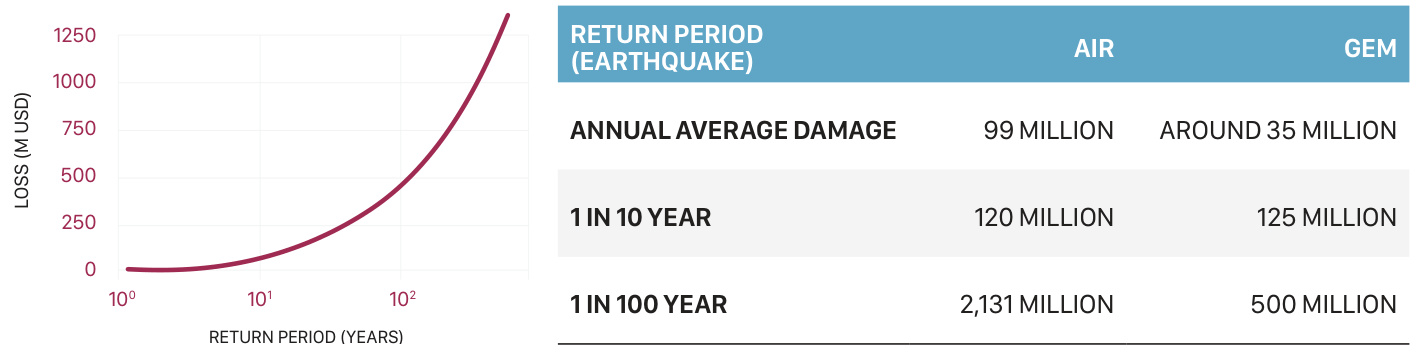"} \\
    \vspace{0.5em}
    Table
\end{minipage}%
\hfill
\begin{minipage}{0.45\linewidth}
    \centering
    \includegraphics[width=\linewidth, height=3.5cm, keepaspectratio]{"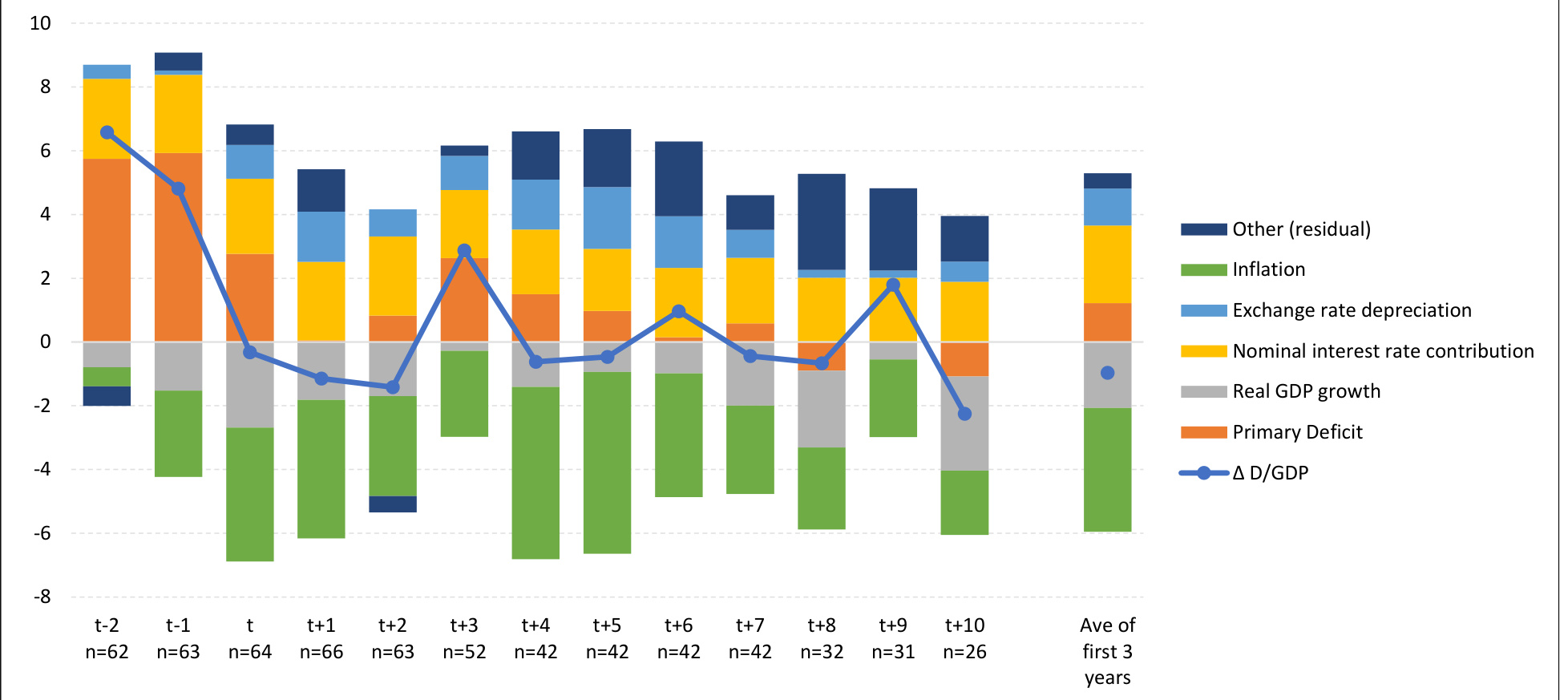"} \\
    \vspace{0.5em}
    Bar Chart \& Column Chart
\end{minipage}
\vspace{1em}

% --- Row 3 ---
\begin{minipage}{0.45\linewidth}
    \centering
    \includegraphics[width=\linewidth, height=4cm, keepaspectratio]{"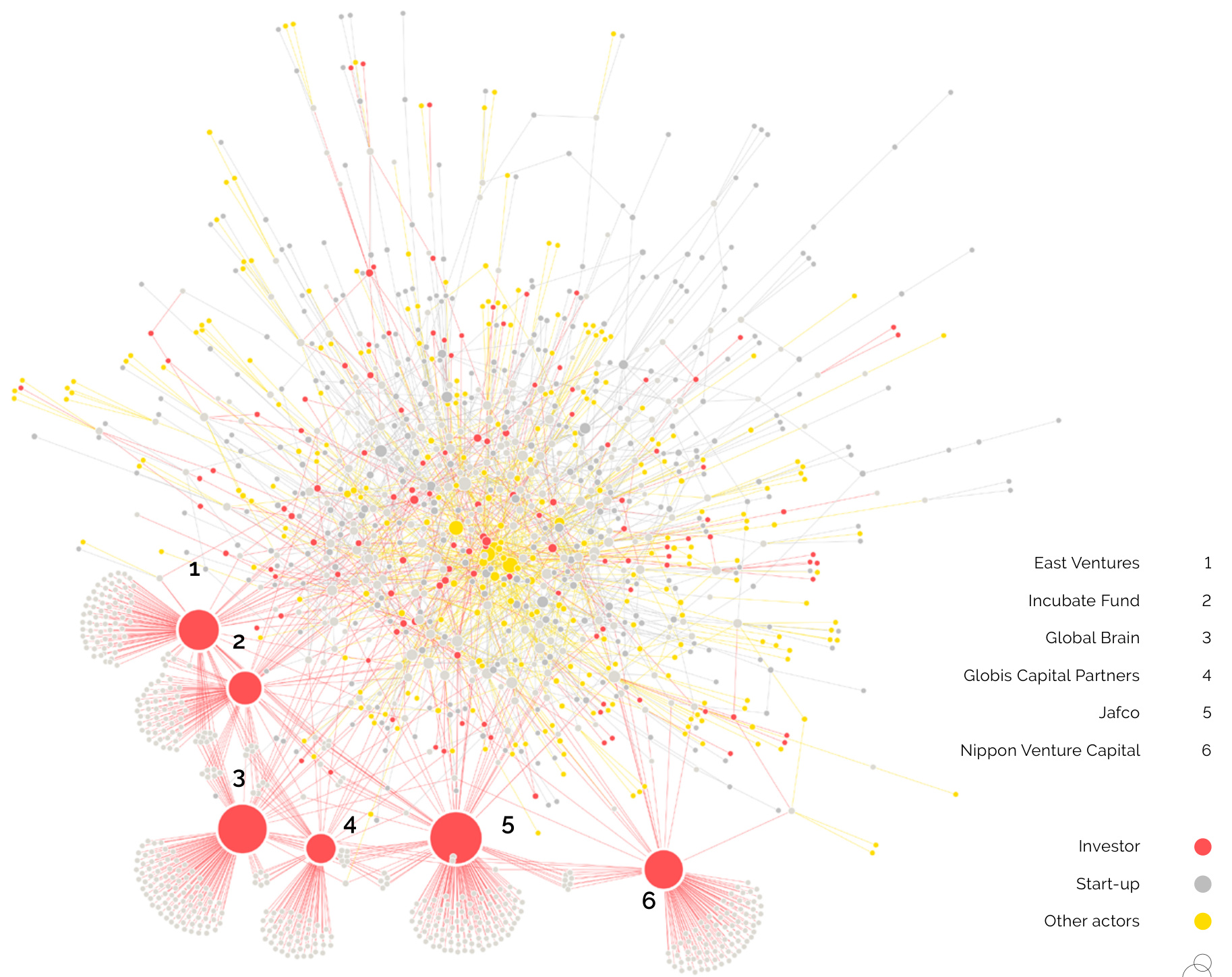"} \\
    \vspace{0.5em}
    Scatter Plot \& Bubble Chart
\end{minipage}%
\hfill
\begin{minipage}{0.45\linewidth}
    \centering
    \includegraphics[width=\linewidth, height=4cm, keepaspectratio]{"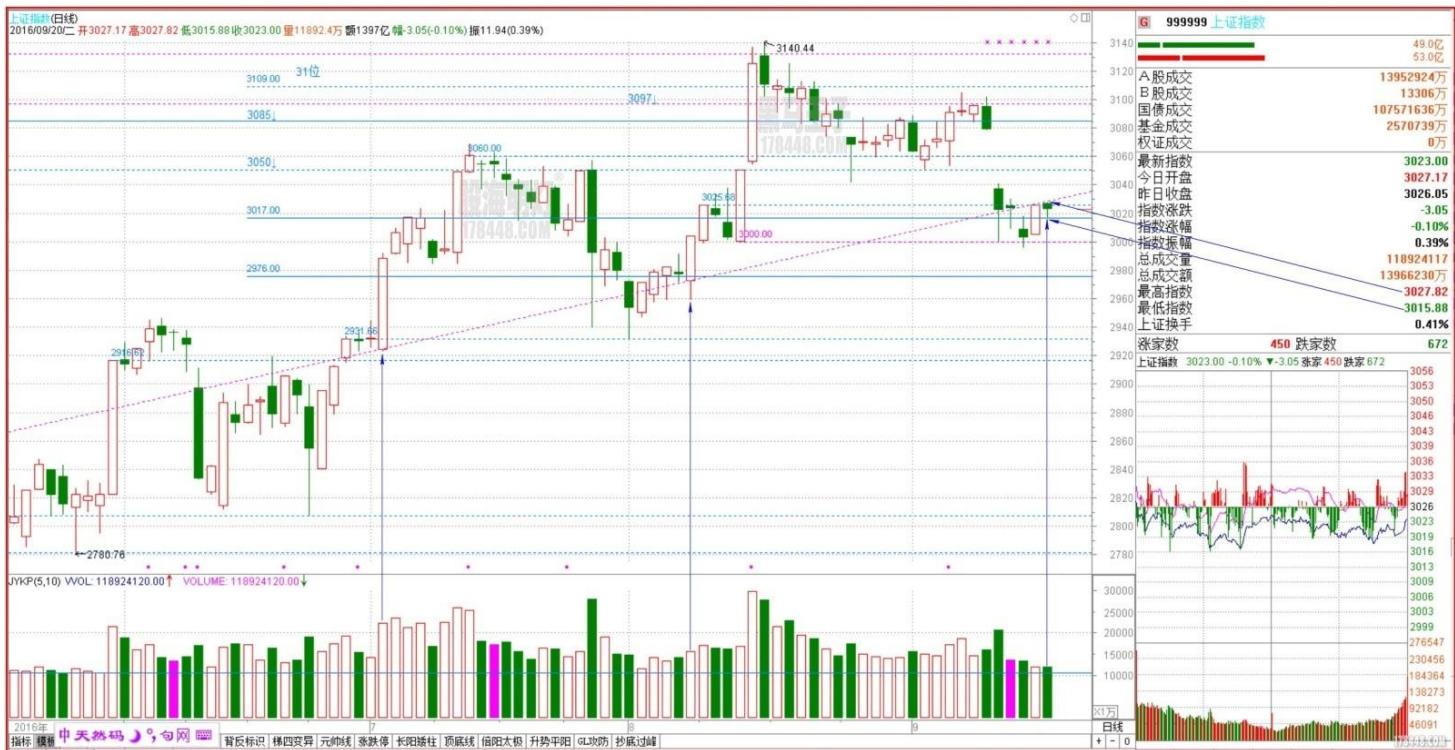"} \\
    \vspace{0.5em}
    Candlestick Chart
\end{minipage}
\vspace{1em}

% --- Row 4 (Single Item) ---
\begin{minipage}{0.45\linewidth}
    \centering
    \includegraphics[width=\linewidth, height=6cm, keepaspectratio]{"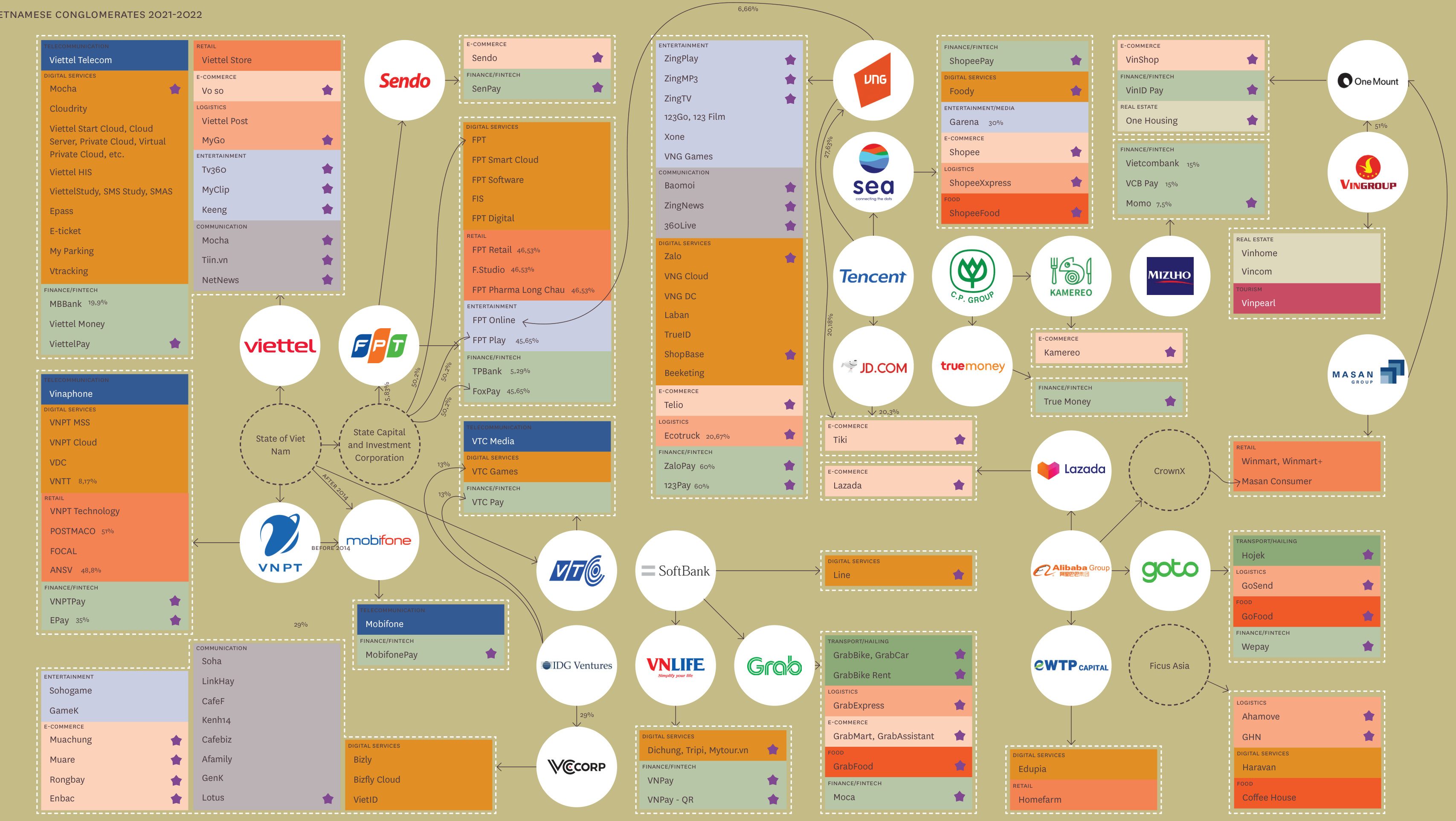"} \\
    \vspace{0.5em}
    Diagram
\end{minipage}

\centering
\caption{Overview of chart types. In images used for generating question-answer pairs, Radar Charts, Heatmaps, and Photographs are rarely used.}
\label{fig:chart_type_examples}

\end{figure*}

The chart types including \textit{Line Chart}, \textit{Bar Chart \& Column Chart}, \textit{Pie Chart \& Donut Chart}, \textit{Scatter Plot \& Bubble Chart}, \textit{Table}, \textit{Diagram \& Schematic}, \textit{Radar Chart}, \textit{Heatmap}, \textit{Candlestick Chart}, \textit{Photograph}, and \textit{Infographic}, as shown in Figure~\ref{fig:chart_type_examples}.

\vspace{1em}

\vspace{1em}

\textbf{Reference Info Extraction}. We employ a LLM to extract bibliographic references and citation metadata from financial and academic documents. The model parses footnote, endnotes, and reference sections to identify cited works and extract key fields, including author(s), publication venue, year, and DOI (when available), while mapping each citation to its contextual use within the text.

\textbf{Data Sampling}. The image sampling strategy is structured around four key dimensions: complexity degree, compliance level, content theme, and chart type. The compliance level (ranging from 0 for non-compliant to 10 for fully compliant) reflects the degree to which the image satisfies the specified financial data visualization criteria. The complexity degree (from 0 for very simple to 10 for very complex) is assigned based on visual intricacy, data density, and structural sophistication. Each image is tagged with at least one of 17 content themes and at least one of 11 image types.

From an initial pool of \textbf{820,816} images, we applied a two-stage filtering process to construct a comprehensive financial evaluation dataset. First, we retained only images with compliance levels 9 and 10, indicating full or near-full adherence to the specified financial data visualization criteria. Non-financial images were excluded.

From the compliant subset, we selected the top 20,000 images with the highest complexity degree, reflecting visual intricacy and informational density. We further refined the set by including only image types relevant to financial data analysis: Line Chart, Bar Chart / Column Chart, Diagram / Schematic, Candlestick Chart / OHLC Chart, and Infographic. Content themes were balanced across categories such as macroeconomic indicators, financial markets, risk management, and economic cycles to ensure thematic diversity.

After automated filtering, a manual review was conducted to exclude any remaining unsuitable images, such as those with poor readability, non-compliant elements, or ambiguous content. The final curated dataset consists of \textbf{3,092} images. The resulting dataset supports rigorous computational analysis of financial visualizations with high compliance and varying complexity.

\textbf{Initial Images Statistics Summary}. A total of 820,816 images were collected and annotated from a substantial collection of financial documents. The initial dataset's distributions across content themes, image types, compliance levels, and complexity degrees are presented in Table~\ref{tab:initial_distributions_summary}.

\begin{table}[htbp]
    \centering
    
    \begin{subfigure}[b]{0.48\textwidth}
        \centering
        \resizebox{\textwidth}{!}{%
        \begin{tabular}{@{}p{2.8cm}lrr@{}}
        \toprule
        \textbf{Category} & \textbf{Classification} & \textbf{Count} & \textbf{Percentage} \\
        \midrule
        \multirow{17}{*}{\textbf{Content Theme}} 
         & Macroeconomic Indicators & -- & 2.66\% \\
         & Financial Markets \& Products & -- & 63.20\% \\
         & Commodities \& Real Estate & -- & 0.99\% \\
         & Bonds \& Fixed Income & -- & 0.73\% \\
         & Monetary \& Fiscal Policy & -- & 0.88\% \\
         & International Trade & -- & 0.88\% \\
         & Corporate Finance \& Valuation & -- & 1.39\% \\
         & Industry Analysis & -- & 2.98\% \\
         & Investment Theory \& PM & -- & 13.42\% \\
         & Risk Models \& Management & -- & 1.13\% \\
         & Economic Cycles \& Theories & -- & 3.06\% \\
         & Microeconomic Principles & -- & 0.43\% \\
         & Demographics \& Socioeconomics & -- & 3.57\% \\
         & Financial Systems & -- & 1.17\% \\
         & Organization \& Regulation & -- & 1.35\% \\
         & Geospatial Economic Data & -- & 1.62\% \\
         & Financial History & -- & 0.53\% \\
        \cline{2-4}
        \addlinespace[0.3em]
         & \textbf{Total} & -- & 100.00\% \\
        \bottomrule
        \end{tabular}%
        }
        \caption{Content Theme Distribution}
        \vspace{0.5em}
        \label{subfig:theme}
    \end{subfigure}

    \begin{subfigure}[b]{0.48\textwidth}
        \centering
        \resizebox{\textwidth}{!}{%
        \begin{tabular}{@{}p{2.8cm}lrr@{}}
        \toprule
        \textbf{Category} & \textbf{Classification} & \textbf{Count} & \textbf{Percentage} \\
        \midrule
        \multirow{11}{*}{\textbf{Chart Type}} 
         & Line Chart & -- & 11.82\% \\
         & Bar/Column Chart & -- & 5.52\% \\
         & Pie/Donut Chart & -- & 0.43\% \\
         & Scatter/Bubble Chart & -- & 0.45\% \\
         & Table & -- & 2.50\% \\
         & Diagram/Schematic & -- & 17.48\% \\
         & Radar Chart & -- & 0.07\% \\
         & Heatmap & -- & 0.05\% \\
         & Candlestick/OHLC & -- & 46.43\% \\
         & Photograph & -- & 3.99\% \\
         & Infographic & -- & 11.27\% \\
        \cline{2-4}
        \addlinespace[0.3em]
         & \textbf{Total} & -- & 100.00\% \\
        \bottomrule
        \end{tabular}%
        }
        \caption{Chart Type Distribution}
        \vspace{0.5em}
        \label{subfig:chart}
    \end{subfigure}

    \begin{subfigure}[b]{0.48\textwidth}
        \centering
        \resizebox{\textwidth}{!}{%
        \begin{tabular}{@{}p{2.8cm}lrr@{}}
        \toprule
        \textbf{Category} & \textbf{Classification} & \textbf{Count} & \textbf{Percentage} \\
        \midrule
        \multirow{10}{*}{\textbf{Compliance Level}} 
         & 1 & 194,325 & 23.67\% \\
         & 2 & 3,703 & 0.45\% \\
         & 3 & 995 & 0.12\% \\
         & 4 & 9 & 0.00\% \\
         & 5 & 117 & 0.01\% \\
         & 6 & 614 & 0.07\% \\
         & 7 & 34,455 & 4.20\% \\
         & 8 & 561,368 & 68.39\% \\
         & 9 & 23,206 & 2.83\% \\
         & 10 & 2,024 & 0.25\% \\
        \cline{2-4}
        \addlinespace[0.3em]
         & \textbf{Total} & 820,816 & 100.00\% \\
        \bottomrule
        \end{tabular}%
        }
        \caption{Compliance Level Distribution}
        \vspace{0.5em}
        \label{subfig:compliance}
    \end{subfigure}

    \begin{subfigure}[b]{0.48\textwidth}
        \centering
        \resizebox{\textwidth}{!}{%
        \begin{tabular}{@{}p{2.8cm}lrr@{}}
        \toprule
        \textbf{Category} & \textbf{Classification} & \textbf{Count} & \textbf{Percentage} \\
        \midrule
        \multirow{10}{*}{\textbf{Complexity Degree}} 
         & 0 & 35 & 0.00\% \\
         & 1 & 52,841 & 6.44\% \\
         & 2 & 49,852 & 6.07\% \\
         & 3 & 49,937 & 6.08\% \\
         & 4 & 49,743 & 6.06\% \\
         & 5 & 82,867 & 10.10\% \\
         & 6 & 408,909 & 49.82\% \\
         & 7 & 126,258 & 15.38\% \\
         & 8 & 296 & 0.04\% \\
         & 9 & 78 & 0.01\% \\
        \cline{2-4}
        \addlinespace[0.3em]
         & \textbf{Total} & 820,816 & 100.00\% \\
        \bottomrule
        \end{tabular}%
        }
        \caption{Complexity Degree Distribution}
        \vspace{-0.5em}
        \label{subfig:complexity}
    \end{subfigure}
    \caption{Initial Image Distributions}
    \label{tab:initial_distributions_summary}
\end{table}

\textbf{Sampling Images Statistics Summary}. A total of 3,092 images were selected from the initial dataset for analysis. The distribution across image types and themes is visualized in Figure~\ref{fig:introduction} (c) and (d), respectively. Regarding the compliance level of the final dataset, 2,642 images (85.45\%) were rated at level 9, while 450 images (14.55\%) were rated at level 10. For complexity degree, the majority of images, 2,741 (88.65\%), were rated at level 7, followed by 188 (6.08\%) at level 8, 140 (4.53\%) at level 6, 17 (0.55\%) at level 9, and 6 (0.19\%) at level 5. 

\section{Limitations}
\label{sec:app-limitations}

We identify several key limitations in our current work:

\noindent \textbf{Potential Data Contamination.} Since \textbf{\emph{PyFi}-600K} is derived from publicly available financial reports and books, there is a risk that these documents were included in the pre-training corpora of the evaluated VLMs. To address this, we implemented rigorous data-level filtering. We conducted "blind tests" to determine if questions could be answered correctly without visual input and "completion tests" to check if models could reconstruct content from text alone. Samples where models succeeded without visual context were flagged as potential leakage and excluded from our core evaluation set to ensure that performance gains reflect genuine multimodal reasoning. Although it may always be beneficial to include documents published after the release of large language models, doing so can significantly reduce the scalability and size of the dataset, introduce manual effort, and prevent the data from being updated continuously.

\noindent \textbf{Computational Constraints on Scale.} Although \textbf{\emph{PyFi}-600K} contains 600k samples, computational limits restricted us from scaling further. However, the pipeline is designed to be extensible and can leverage the vast array of financial datasets available (e.g., on Hugging Face) to construct billion-scale datasets in the future.

\noindent \textbf{Reward Modeling Noise.} The reward scores used during the adversarial interactions and MCTS may contain noise. Accurately evaluating open-ended financial reasoning is inherently challenging, and minor inaccuracies in the reward model can affect the optimization process.

\noindent \textbf{Adversarial Framework Complexity.} The \textit{PyFi-adv} mechanism, while effective for data synthesis, introduces complexity that may limit its deployment in resource-constrained environments. Furthermore, the stochastic nature of VLMs can occasionally lead to inconsistent chain lengths or hallucinations, which requires careful post-hoc filtering.

\noindent \textbf{Incomplete Process Supervision Analysis.} Due to time and resource limitations, we have not yet fully explored the upper bounds of process supervision provided by our framework, although our initial experiments demonstrate its significant value.

\section{Discussion on Data Leakage}
\label{sec:dataleakage}

In constructing \textbf{\emph{PyFi}-600K}, we acknowledge the potential risk of \textbf{data leakage}, particularly due to the use of publicly available financial documents that may have been included in the pretraining corpora of some VLMs. To ensure a fair and reliable evaluation, we conduct a \textbf{systematic and stringent} analysis to identify and mitigate such leakage.

Existing research employs various methods to detect data leakage. \citet{GenOrMem-arxiv24} identify leakage by measuring the peakedness of the model's output distribution, where high similarity among sampled outputs indicates memorization. \citet{ReasonOrMem-arxiv25} assess leakage by the model's ability to reconstruct or answer problems based only on partial prompt prefixes. \citet{TrainOnBench-aaai25} detect leakage in multiple-choice benchmarks by identifying if the log-probability of the original option order is a significant outlier among permutations. However, these approaches are often indirect, model-specific, or limited to particular benchmark formats, making them less applicable to financial image understanding tasks where the interplay between image and text is essential.

We define a sample as exhibiting data leakage if \textbf{more than half of the evaluated models} can answer the question correctly \textbf{without access to the image}. This criterion is based on the assumption that if models can consistently answer a question without visual input, the answer is likely encoded in the textual content or has been memorized during pretraining, rather than being inferred from the image.  
\textbf{Our method is more direct, task-agnostic, and interpretable}: by explicitly ablating the visual modality and measuring performance across diverse models, we capture leakage that genuinely compromises multimodal evaluation fairness—making it the most suitable strategy for benchmarks like \emph{PyFi}-600K that hinge on visual grounding.

To quantify this, we conduct a comprehensive study where we evaluate all 15 VLMs on the original 1,000-sample test set under two conditions: with and without the image input. Samples that meet the leakage criterion are subsequently excluded from the evaluation set. As a result, the test set is reduced from 1,000 to \textbf{301 samples}, forming the \textbf{Evaluation Dataset after Exclusion}. \textbf{All results reported in the main paper are derived exclusively from this rigorously filtered set.}

This filtered dataset is used for all final evaluations and fine-tuning experiments reported in the main paper. The significant reduction in sample size underscores the prevalence of potential leakage in financial Q\&A datasets and highlights the importance of this filtering step.

This approach is both \textbf{reasonable and effective} in mitigating the influence of data leakage. By removing samples that can be answered without visual reasoning, we ensure that the remaining questions require genuine visual understanding and financial reasoning, aligning with the goals of \textbf{\emph{PyFi}-600K}. Furthermore, this method is model-agnostic and does not rely on external knowledge of pretraining data, making it broadly applicable.

\textbf{In summary, our data leakage mitigation strategy enhances the validity of the evaluation results and supports the development of more robust and generalizable financial VLMs.}

\section{Discussion on Pyramid-like Structure}
\label{sec:discussion-pyramid}

Financial image analysis requires more than object detection or simple classification; true understanding is a multi-stage cognitive process from visual perception to strategic decision-making. To address this, we propose \textbf{\emph{PyFi}-600K}, a pyramid-like hierarchy structured across six levels: \textbf{Perception}, \textbf{Data Extraction}, \textbf{Calculation Analysis}, \textbf{Pattern Recognition}, \textbf{Logical Reasoning}, and \textbf{Decision Support}.

\noindent \textbf{Level 1: Perception}.  This is the foundational base of the pyramid. At this stage, the model performs basic visual recognition. It identifies fundamental elements within the image: "What objects are present?" or "What is the color of that line?" This involves detecting axes, labels, data points, lines, bars, pie segments, and textual annotations. The questions and answers are a structured inventory of visual components.

\noindent \textbf{Level 2: Data Extraction}. Building on the perceived components, this level translates visual information into machine-readable quantitative or categorical data. For example, it converts the height of a bar in a chart into a numerical value based on the Y-axis scale, or it extracts the exact numerical percentage from a pie chart label. The output is a clean dataset ready for computation.

\noindent \textbf{Level 3: Calculation Analysis}. Here, the extracted data undergoes initial processing. This involves performing basic arithmetic, such as calculating percentage changes, growth rates or variances, and descriptive statistics. This level answers "What are the basic quantitative relationships?"

\noindent \textbf{Level 4: Pattern Recognition}. This level moves from discrete calculations to identifying broader trends and shapes within the data. It answers "What is happening over time or across categories?" Examples include recognizing an upward or downward trend in a stock price chart, identifying a period of high volatility, or spotting seasonal patterns in sales data.

\noindent \textbf{Level 5: Logical Reasoning}. This is a critical step towards higher-order cognition. The model integrates the identified patterns with external or common-sense financial knowledge to answer "Why is this happening, or what does this imply?" For instance, reasoning that a sharp drop in a company's stock price coinciding with a negative earnings report indicates a causal relationship. It involves drawing inferences and establishing cause-and-effect links.

\noindent \textbf{Level 6: Decision Support}. The apex of the pyramid leverages all previous levels to provide actionable insights. It answers "What should be done?" Based on the reasoned analysis and the options or answers might include recommendations.

\textbf{\emph{PyFi}-600K} provides a robust and principled framework for building a multi-modal financial image understanding dataset. By decomposing the task into six hierarchically organized cognitive levels for automated, reward-driven generation of logical chains, we create a resource that promotes deep, structured, and explainable reasoning. 
\section{Discussion on Question Chain}
\label{sec:questionchain}

\begin{figure}[!h]
    \centering
    \includegraphics[width=\columnwidth]{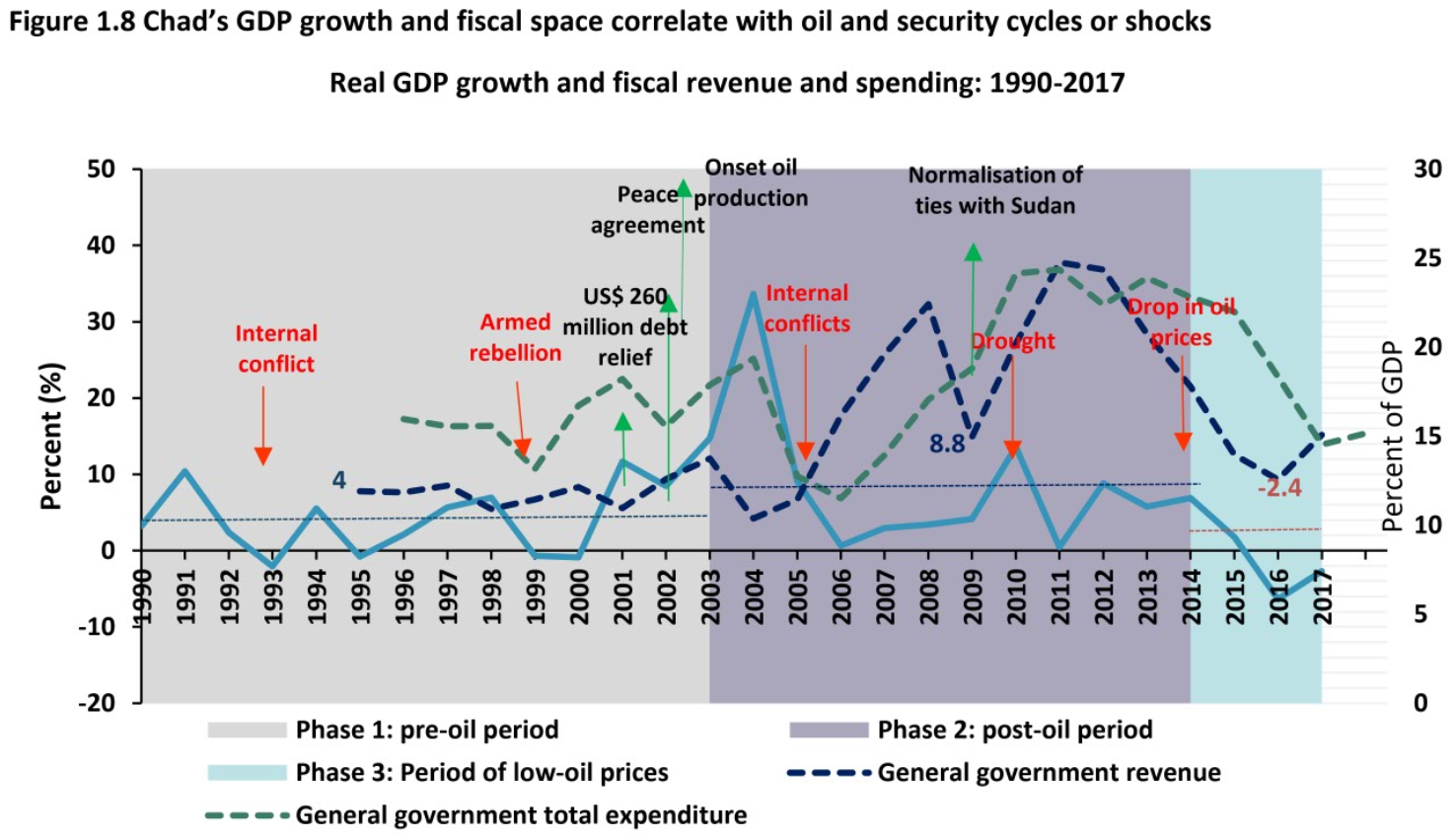}
    \caption{Illustration of the complete question chain presented in the Figure~\ref{fig:framework} of the main paper.}
    \label{fig:questionchain}
\end{figure}

A complete question chain can refer to the Figure~\ref{fig:questionchain}, Figure~\ref{fig:questionchain_example}, and Figure~\ref{fig:questionchain_example_cont}.

\begin{samepage}
\vspace{-12pt}

\begin{figure*}
\begin{tcolorbox}[
    title=Complete Example of Question Chain,
    colback=white,
    colframe=gray!20,
    coltitle=black,
    fonttitle=\bfseries\small,
    sharp corners,
    boxrule=0.8pt,
    left=4pt, right=4pt, top=4pt, bottom=4pt,
    before skip=8pt, after skip=8pt,
    enhanced,
    breakable=false,
    overlay first={
        \draw[gray!30!white, line width=0.8pt] (frame.north west) -- (frame.north east);
    }
]

\noindent\textbf{1.} \textit{Question:} Which shaded region in the chart represents the 'Period of low-oil prices'?\\
\textit{Options:}\\
\hspace*{5pt}A. Gray (Phase 1: pre-oil period)\\
\hspace*{5pt}B. Purple (Phase 2: post-oil period)\\
\hspace*{5pt}C. Light blue (Phase 3: Period of low-oil prices)\\
\hspace*{5pt}D. Green (not a shaded region)\\
\textit{Answer:} \textbf{C}

\vspace{3pt}
\noindent\textbf{2.} \textit{Question:} What is the color of the shaded region labeled 'Phase 2: post-oil period' in the chart?\\
\textit{Options:}\\
\hspace*{5pt}A. Light gray\\
\hspace*{5pt}B. Dark gray\\
\hspace*{5pt}C. Light blue\\
\hspace*{5pt}D. Cyan\\
\textit{Answer:} \textbf{B}

\vspace{3pt}
\noindent\textbf{3.} \textit{Question:} During Phase 3: Period of low-oil prices, what is the approximate value of general government revenue (blue dashed line) in Chad, as shown in the chart?\\
\textit{Options:}\\
\hspace*{5pt}A. 5\% of GDP\\
\hspace*{5pt}B. -2.4\% of GDP\\
\hspace*{5pt}C. 10\% of GDP\\
\hspace*{5pt}D. 20\% of GDP\\
\textit{Answer:} \textbf{B}

\vspace{3pt}
\noindent\textbf{4.} \textit{Question:} Which event is associated with the decline in general government revenue (blue dashed line) during Phase 3 (Period of low-oil prices) in Chad, as shown in the chart?\\
\textit{Options:}\\
\hspace*{5pt}A. Internal conflict\\
\hspace*{5pt}B. Drought\\
\hspace*{5pt}C. Drop in oil prices\\
\hspace*{5pt}D. Normalisation of ties with Sudan\\
\textit{Answer:} \textbf{C}

\vspace{3pt}
\noindent\textbf{5.} \textit{Question:} In 2016, during the Period of low-oil prices (Phase 3), what is the approximate value of (general government total expenditure - general government revenue) in Chad?\\
\textit{Options:}\\
\hspace*{5pt}A. ~17.4\%\\
\hspace*{5pt}B. ~-17.4\%\\
\hspace*{5pt}C. ~5\%\\
\hspace*{5pt}D. ~-5\%\\
\textit{Answer:} \textbf{A}

\vspace{3pt}
\noindent\textbf{6.} \textit{Question:} What pattern is observed between general government revenue and total expenditure during the 'Onset oil production' period (marked in the chart)?\\
\textit{Options:}\\
\hspace*{5pt}A. Both revenue and expenditure increase significantly\\
\hspace*{5pt}B. Revenue increases while expenditure decreases\\
\hspace*{5pt}C. Revenue decreases while expenditure increases\\
\hspace*{5pt}D. Both revenue and expenditure remain stable\\
\textit{Answer:} \textbf{A}
\end{tcolorbox}
\caption{Complete example of question chain.}
\label{fig:questionchain_example}
\end{figure*}
\end{samepage}

\begin{figure*}
\begin{tcolorbox}[
    title=Complete Example of Question Chain (continued),
    colback=white,
    colframe=gray!20,
    coltitle=black,
    fonttitle=\bfseries\small,
    sharp corners,
    boxrule=0.8pt,
    left=4pt, right=4pt, top=4pt, bottom=4pt,
    before skip=8pt, after skip=8pt,
    breakable,
    enhanced,
    overlay first={
        \draw[gray!30!white, line width=0.8pt] (frame.north west) -- (frame.north east);
    }
]

\vspace{3pt}
\noindent\textbf{7.} \textit{Question:} What pattern does the general government total expenditure (green dashed line) follow during periods of internal conflict (marked by red arrows) in Chad from 1990 to 2017?\\
\textit{Options:}\\
\hspace*{5pt}A. It decreases consistently\\
\hspace*{5pt}B. It increases significantly\\
\hspace*{5pt}C. It remains stable\\
\hspace*{5pt}D. It fluctuates randomly\\
\textit{Answer:} \textbf{B}

\vspace{3pt}
\noindent\textbf{8.} \textit{Question:} How does the volatility of general government revenue in Chad during the Period of low-oil prices (Phase 3) contribute to economic vulnerability?\\
\textit{Options:}\\
\hspace*{5pt}A. It increases military spending, leading to internal conflict.\\
\hspace*{5pt}B. It reduces the government's ability to withstand shocks and invest in stability.\\
\hspace*{5pt}C. It causes oil prices to drop further, worsening revenue.\\
\hspace*{5pt}D. It eliminates the need for non-oil revenue sources.\\
\textit{Answer:} \textbf{B}

\vspace{3pt}
\noindent\textbf{9.} \textit{Question:} Based on the image illustrating Chad's economic volatility driven by oil price cycles and security shocks, how would establishing a sovereign wealth fund help reduce revenue vulnerability during low-oil price periods?\\
\textit{Options:}\\
\hspace*{5pt}A. By increasing military spending to preempt conflict and stabilize revenue\\
\hspace*{5pt}B. By investing oil revenues during high-price periods to fund expenditures during low-price periods\\
\hspace*{5pt}C. By reducing reliance on non-oil revenue sources, such as IMF-ECF support\\
\hspace*{5pt}D. By increasing general government revenue through borrowing from the World Bank during booms\\
\textit{Answer:} \textbf{B}

\vspace{3pt}
\noindent\textbf{10.} \textit{Question:} Which policy would most effectively address both Chad's oil revenue volatility (from external shocks) and vulnerability to internal conflict?\\
\textit{Options:}\\
\hspace*{5pt}A. Increase military spending to preempt conflict\\
\hspace*{5pt}B. Establish a sovereign wealth fund to stabilize revenues and investments\\
\hspace*{5pt}C. Reduce all non-oil revenue sources to avoid over-reliance on oil\\
\hspace*{5pt}D. Borrow heavily from the World Bank during low-oil price periods\\
\textit{Answer:} \textbf{B}

\vspace{3pt}
\noindent\textbf{11.} \textit{Final Question:} Based on the image illustrating Chad's fiscal volatility driven by oil cycles and security shocks, which policy would most effectively reduce economic vulnerability to future external shocks and internal conflict?\\
\textit{Options:}\\
\hspace*{5pt}A. Increase military spending to preempt conflict\\
\hspace*{5pt}B. Establish a sovereign wealth fund to stabilize revenues and investments\\
\hspace*{5pt}C. Reduce all non-oil revenue sources to avoid over-reliance on oil\\
\hspace*{5pt}D. Borrow heavily from the World Bank during low-oil price periods\\
\textit{Answer:} \textbf{B}

\end{tcolorbox}
\caption{Complete example of question chain (continued).}
\label{fig:questionchain_example_cont}
\end{figure*}

\section{Model Performance with and without Fine-Tuning}
\label{sec:comparisonfinetuning}

\begin{figure*}[p]
    \centering
    \includegraphics[width=\textwidth, height=\textheight, keepaspectratio]{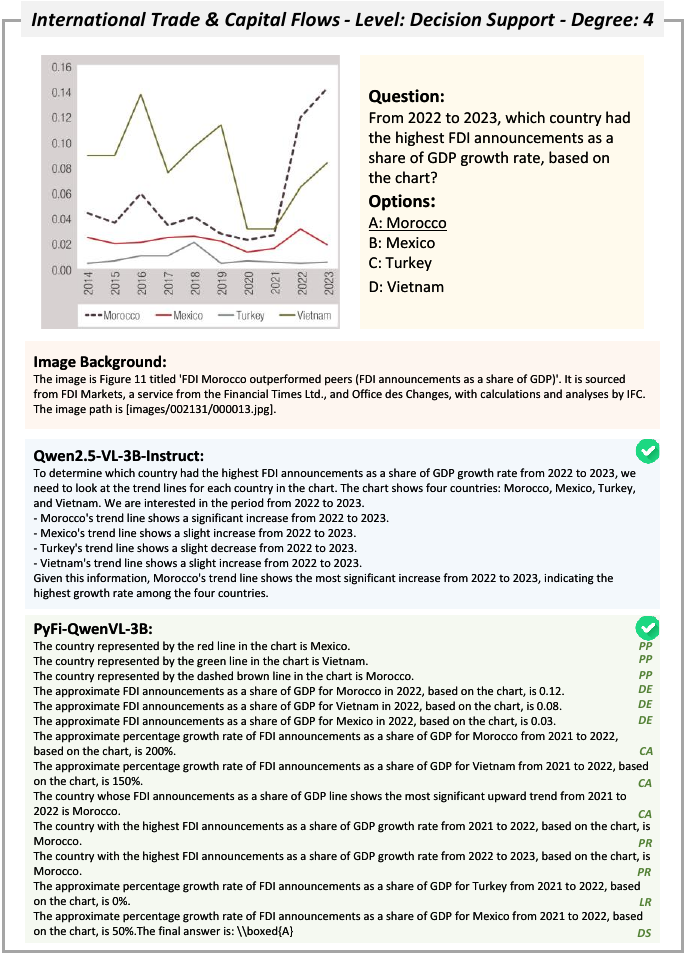}
    \caption{Model performance before and after fine-tuning (Qwen2.5-VL-3B-Instruct)}
    \label{fig:comparison_finetuning_3B_1}
\end{figure*}

\begin{figure*}[p]
    \centering
    \includegraphics[width=\textwidth, height=\textheight, keepaspectratio]{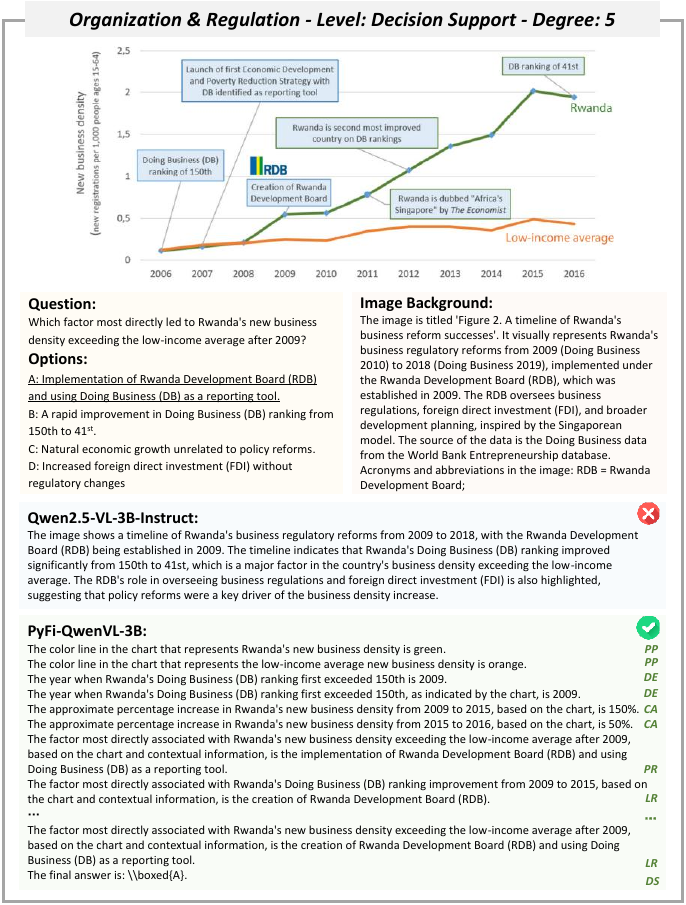}
    \caption{Model performance before and after fine-tuning (Qwen2.5-VL-3B-Instruct)}
    \label{fig:comparison_finetuning_3B_2}
\end{figure*}

\begin{figure*}[p]
    \centering
    \includegraphics[width=\textwidth, height=\textheight, keepaspectratio]{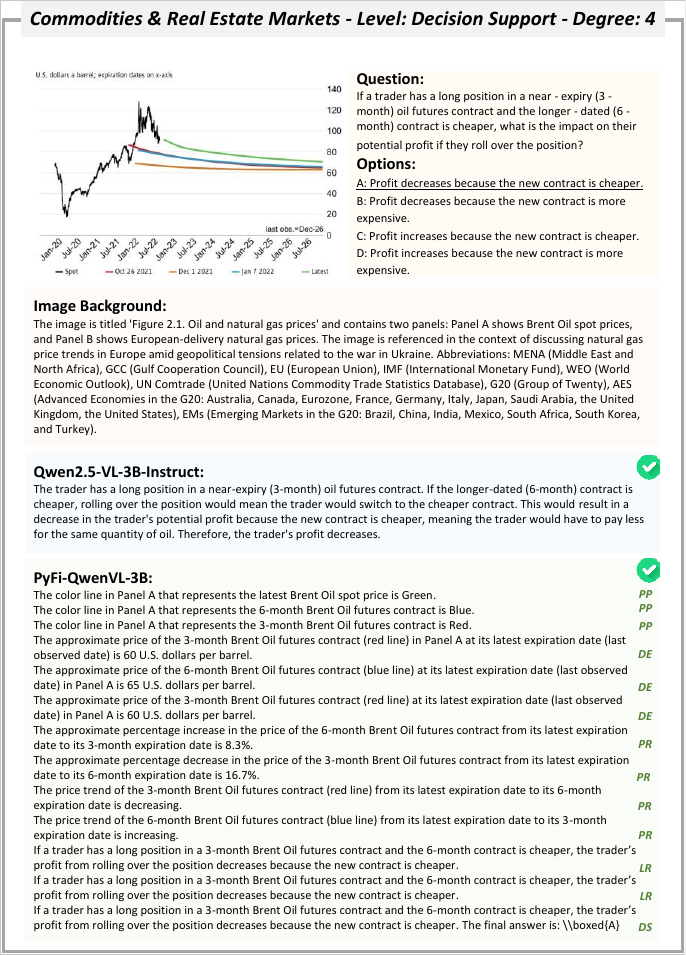}
    \caption{Model performance before and after fine-tuning (Qwen2.5-VL-3B-Instruct)}
    \label{fig:comparison_finetuning_3B_4}
\end{figure*}

\begin{figure*}[p]
    \centering
    \includegraphics[width=\textwidth, height=\textheight, keepaspectratio]{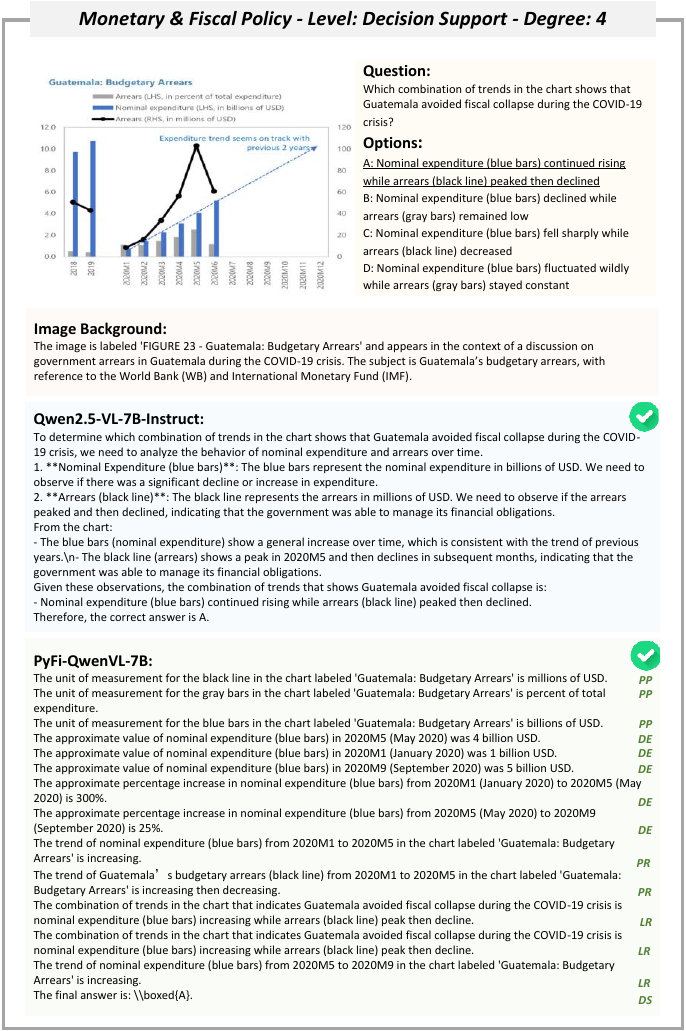}
    \caption{Model performance before and after fine-tuning (Qwen2.5-VL-7B-Instruct)}
    \label{fig:comparison_finetuning_7B_1}
\end{figure*}

\begin{figure*}[p]
    \centering
    \includegraphics[width=\textwidth, height=\textheight, keepaspectratio]{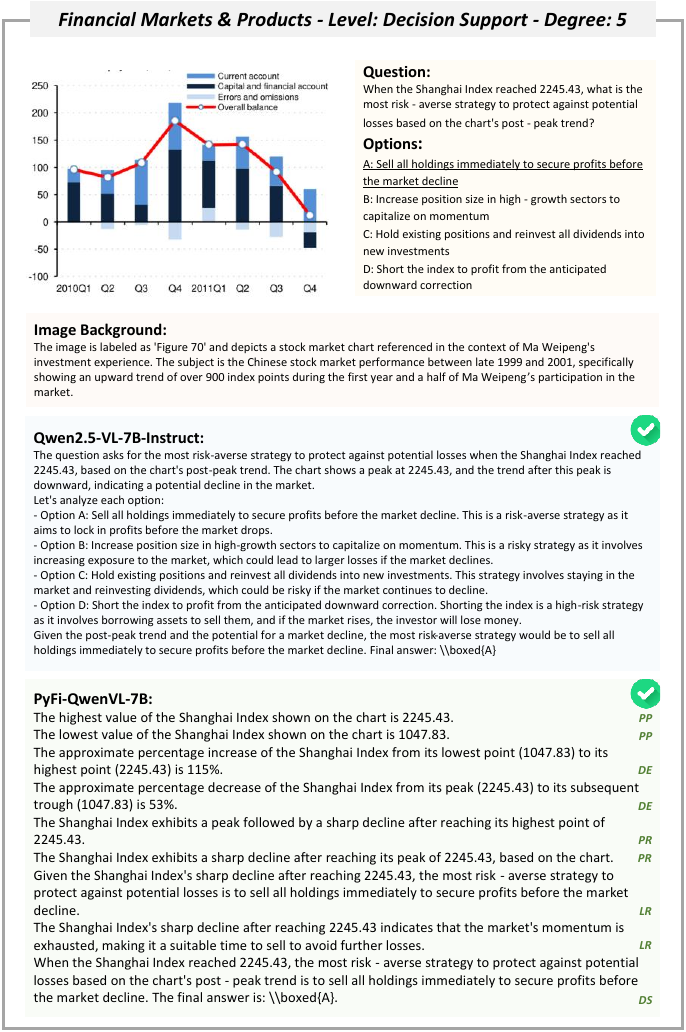}
    \caption{Model performance before and after fine-tuning (Qwen2.5-VL-7B-Instruct)}
    \label{fig:comparison_finetuning_7B_2}
\end{figure*}

\begin{figure*}[p]
    \centering
    \includegraphics[width=\textwidth, height=\textheight, keepaspectratio]{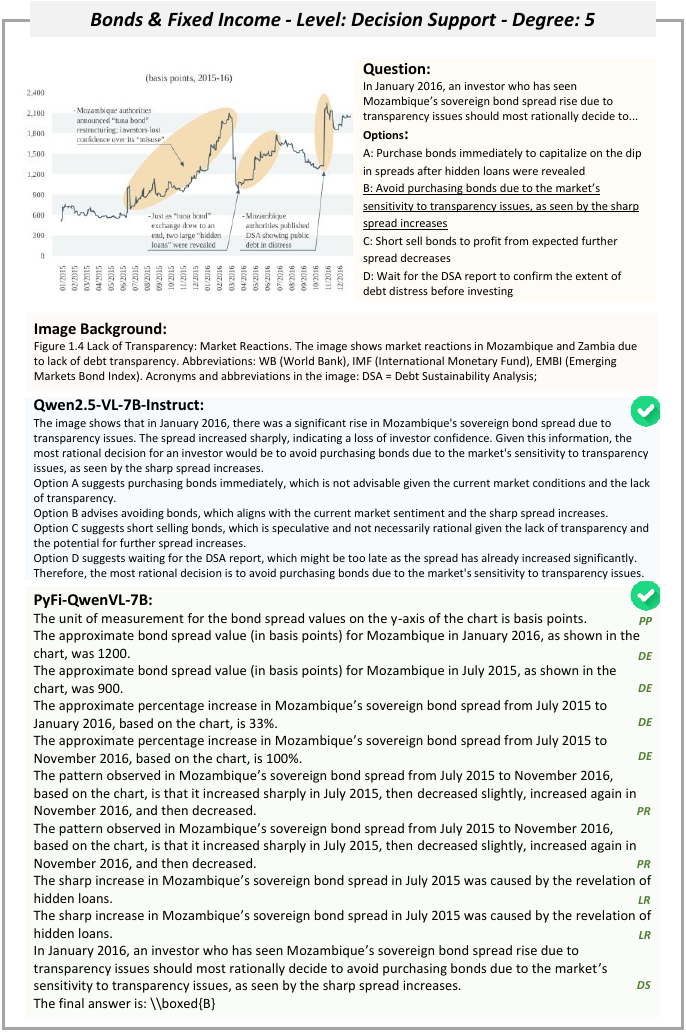}
    \caption{Model performance before and after fine-tuning (Qwen2.5-VL-7B-Instruct)}
    \label{fig:comparison_finetuning_7B_3}
\end{figure*}

\begin{figure*}[p]
    \centering
    \includegraphics[width=\textwidth, height=\textheight, keepaspectratio]{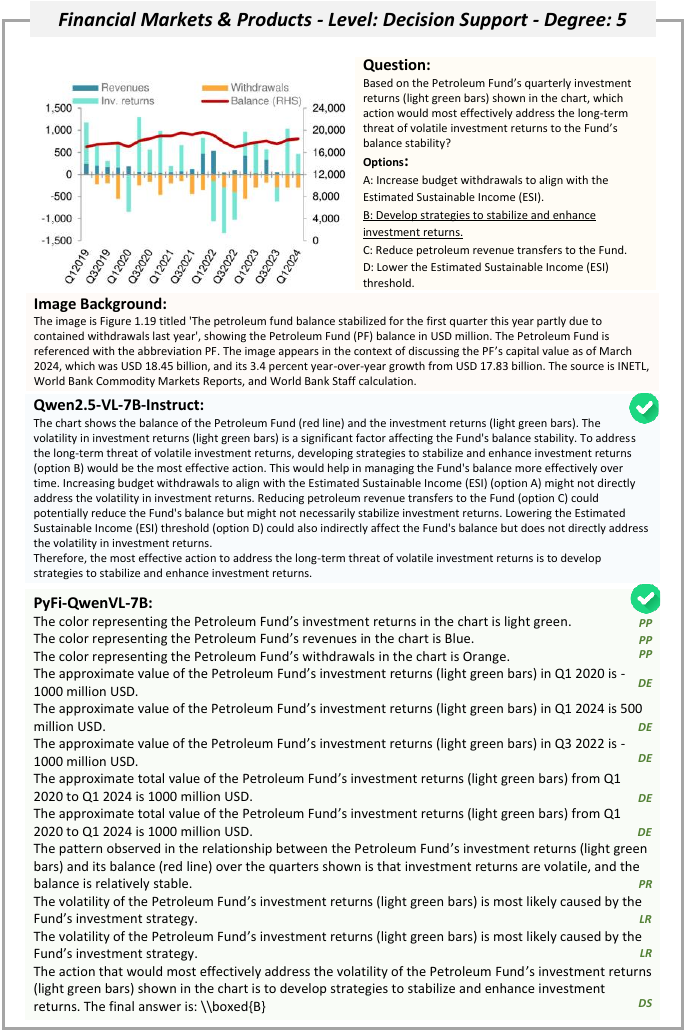}
    \caption{Model performance before and after fine-tuning (Qwen2.5-VL-7B-Instruct)}
    \label{fig:comparison_finetuning_7B_4}
\end{figure*}

The examples of Qwen2.5-VL-3B-Instruct performance before and after fine-tuning can refer to the Figures~\ref{fig:comparison_finetuning_3B_1}, \ref{fig:comparison_finetuning_3B_2}, and \ref{fig:comparison_finetuning_3B_4}, and the examples of Qwen2.5-VL-7B-Instruct can refer to the Figures~\ref{fig:comparison_finetuning_7B_1}, \ref{fig:comparison_finetuning_7B_2}, \ref{fig:comparison_finetuning_7B_3} and \ref{fig:comparison_finetuning_7B_4}.

\clearpage
\clearpage

\clearpage

\clearpage

\clearpage

\clearpage

\clearpage

\clearpage

\newpage
\section{Detailed Results}
\label{sec:statistics}

\subsection{Experiments Consideration and Ablation Studies}
\label{subsec:ablation-studies}

%%%%%%%
%%% Temporarily placed here, it still needs to be adjusted and modified.
%%%%%%%

In order to comprehensively evaluate the performance of various VLMs on our samples from \textbf{\emph{PyFi}-600K} and eliminate potential biases, we consider the following factors and conduct comparative experiments and ablation studies accordingly.

\textbf{The Impact of Contextual Information}. To investigate the impact of image-related contextual information on model performance, we provide the target model with an image and a question, followed by additional textual information in one of four conditions: (1) providing image background information only, (2) providing image contextual analysis only, (3) providing both background information and contextual analysis, and (4) providing no additional text. The background information includes the image's title, topic, and the full forms of all abbreviations, offering necessary context for answering the question without revealing the answer itself. The contextual analysis, generated by a large language model, summarizes relevant contextual information present in the image and offers direct guidance for solving the problem. This experimental design aims to determine whether the model genuinely understands the visual content or relies on external textual information to infer the answer.

\textbf{Risk of Data Leakage}. The VLMs evaluated in this study were pretrained on massive corpora. Since our data primarily consists of publicly available financial documents sourced from the internet and predominantly dated before 2025, there is a potential risk of data leakage. To address this concern, we additionally evaluated the VLMs in a setting where images were not provided. By comparing model accuracy between the two settings, with and without images, we can assess whether the pretraining corpus might already contain images or questions from our dataset. As for the formal statistics, we exclude those samples that can be correctly answered by more than half of the evaluated models even without image input, and the remaining samples are defined as Evaluation Dataset after Exclusion  while the initial ones are defined as Original Evaluation Dataset.

\textbf{Impact of Option Distribution}. Our analysis of the ground-truth labels reveals severe class imbalance. Specifically, 43.39\% of the samples designate option A as the correct answer, 34.20\% select option B, 15.81\% choose option C, and only 6.08\% pick option D. All remaining options combined account for less than 1\% of the total answers. Details can be found in the appendix~\ref{tab:answer_distribution}. Such extreme imbalance may introduce bias during both training and evaluation, potentially undermining model performance. To address this issue and obtain more robust and reliable evaluation results, we rebalance the sample options so that, in a set of one thousand samples, each of the four options A, B, C, and D constitutes approximately 25\%. We then compare model performance before and after this rebalancing.

All experimental results are presented in the following sections. The meanings of key abbreviations and the group configurations are defined in the tables below.

\begin{table}[h!]
\centering
\caption{Meaning of Abbreviations}
\label{tab:abbreviations}
\begin{adjustbox}{max width=\linewidth}
\begin{tabular}{lr}
\hline
\textbf{Abbreviation} & \textbf{Meaning} \\
\hline
PP & Perception \\
DE & Data Extraction \\
CA & Calculation Analysis \\
PR & Pattern Recognition \\
LR & Logical Reasoning \\
DS & Decision Support \\
\hline
\end{tabular}
\end{adjustbox}
\end{table}

\begin{table}[h!]
\centering
\caption{Experimental Group Configurations}
\label{tab:groups}
\begin{adjustbox}{max width=\linewidth}
\begin{tabular}{lr}
\hline
\textbf{Group} & \textbf{Input Configuration} \\
\hline
I & Image and Image Background \\
II & Image and Analysis Information \\
III & Image, Image Background and Analysis Information \\
IV & Image Only \\
\hline
\end{tabular}
\end{adjustbox}
\end{table}

%%%%%%%%%%%%%%%%%%%%%%
%%%%%%%%%%%%%%%%%%%%%%

\subsection{Detailed Experiment Results on Original Evaluation Dataset with Image}

\begin{table*}[t!]
\centering
\caption{Model Performance Comparison By Group On Original Evaluation Dataset (Samples=1000, Groups with Image)}
\label{tab:model_comparison_original_imgonly_overall_group}
\begin{adjustbox}{max width=\linewidth}
\begin{tabular}{l *{7}{>{\centering\arraybackslash}p{1.2cm}}}
\toprule
\textbf{Model} &
\textbf{Overall} &
\textbf{TC} &
\textbf{TQ} &
\textbf{I} &
\textbf{II} &
\textbf{III} &
\textbf{IV} \\
\midrule
\hline
\multicolumn{8}{c}{\textbf{Pre-trained VLMs}} \\
\hline
GPT-4.1 & 76.08 & 3043 & 4000 & 75.4 & 75.0 & 77.5 & 76.4 \\
InternVL3-38B & 73.72 & 2949 & 4000 & 73.9 & 74.6 & 73.7 & 72.7 \\
Claude-opus-4-1-20250805 & 79.88 & 3195 & 4000 & 80.8 & 79.5 & 80.6 & 78.6 \\
GLM-4.5V & 85.22 & 3409 & 4000 & 85.8 & 85.3 & 85.5 & 84.3 \\
DeepSeek-VL2 & 63.20 & 2528 & 4000 & 61.8 & 63.3 & 65.6 & 62.1 \\
Hunyuan-Large-Vision & 76.60 & 3064 & 4000 & 77.3 & 76.4 & 77.5 & 75.2 \\
ERNIE-4.5-turbo-vl & 68.83 & 2753 & 4000 & 68.0 & 70.0 & 71.1 & 66.2 \\
Moonshot-V1-128k-Vision-Preview & 76.40 & 3056 & 4000 & 75.3 & 77.4 & 78.2 & 74.7 \\
Moonshot-V1-32k-Vision-Preview & 76.22 & 3049 & 4000 & 75.1 & 77.1 & 77.8 & 74.9 \\
Moonshot-V1-8k-Vision-Preview  & 76.53 & 3061 & 4000 & 75.0 & 77.5 & 77.8 & 75.8 \\
Qwen3-VL-Plus & 76.70 & 3068 & 4000 & 76.5 & 76.6 & 76.8 & 76.9 \\
Qwen2.5-VL-72B-Instruct & 74.28 & 2971 & 4000 & 72.9 & 75.4 & 75.5 & 73.3 \\
Qwen2.5-VL-32B-Instruct & 71.78 & 2871 & 4000 & 71.2 & 72.4 & 73.0 & 70.5 \\
Qwen2.5-VL-7B-Instruct & 67.55 & 2702 & 4000 & 65.0 & 68.4 & 69.7 & 67.1 \\
Qwen2.5-VL-3B-Instruct & 59.00 & 2360 & 4000 & 57.3 & 59.7 & 62.1 & 56.9 \\

\midrule
\multicolumn{8}{c}{\textbf{SFT VLMs, Trained with 500 Sample Chains}} \\
\hline
Qwen2.5-VL-3B-Instruct-L& 58.80 & 2352 & 4000 & 58.4 & 60.3 & 60.0 & 56.5 \\
PyFi-QwenVL-3B-500 & 61.52 & 2461 & 4000 & 60.6 & 62.8 & 63.6 & 59.1 \\
PyFi-QwenVL-3B-COT-500& 62.32 & 2493 & 4000 & 61.8 & 63.9 & 63.3 & 60.3 \\
Qwen2.5-VL-7B-Instruct-L & 66.60 & 2664 & 4000 & 65.5 & 67.5 & 68.4 & 65.0 \\
PyFi-QwenVL-7B-500 & 66.27 & 2651 & 4000 & 62.3 & 67.1 & 67.2 & 68.5 \\
PyFi-QwenVL-7B-COT-500 & 65.72 & 2629 & 4000 & 64.4 & 66.4 & 66.0 & 66.1 \\

\midrule
\multicolumn{8}{c}{\textbf{SFT VLMs, Trained with 47K Sample Chains}} \\
\hline
Qwen2.5-VL-3B-Instruct-L& 58.80 & 2352 & 4000 & 58.4 & 60.3 & 60.0 & 56.5 \\
PyFi-QwenVL-3B-47K & 61.85 & 2474 & 4000 & 61.6 & 63.0 & 62.5 & 60.3 \\
PyFi-QwenVL-3B-COT-47K & 64.40 & 2576 & 4000 & 64.2 & 63.6 & 66.6 & 63.2 \\
Qwen2.5-VL-7B-Instruct-L & 66.60 & 2664 & 4000 & 65.5 & 67.5 & 68.4 & 65.0 \\
PyFi-QwenVL-7B-47K & 66.35 & 2654 & 4000 & 65.7 & 67.1 & 67.2 & 65.4 \\
PyFi-QwenVL-7B-COT-47K & 67.42 & 2697 & 4000 & 67.8 & 67.6 & 68.0 & 66.3 \\

\bottomrule
\end{tabular}
\end{adjustbox}
\end{table*}

\textbf{Table~\ref{tab:model_comparison_original_imgonly_overall_group}} presents the overall accuracy and performance across groups for all models on the original evaluation dataset (1000 samples). GLM-4.5V achieves the highest overall accuracy (85.22\%) among all pre-trained models, significantly outperforming others. Claude-opus follows as the second-best pre-trained model (79.88\%). Among the Qwen series, the larger models generally perform better, with Qwen3-VL-Plus (76.70\%) leading, showing a clear scaling effect from 3B (59.00\%) to 72B (74.28\%). For SFT models, fine-tuning with 500 or 47K samples shows modest but inconsistent improvements over their base Instruct-L versions. For instance, PyFi-QwenVL-3B-COT-47K reaches 64.40\%, a noticeable gain over the base Qwen2.5-VL-3B-Instruct (59.00\%). However, for the 7B models, gains from SFT are marginal, with some variants even underperforming the base model. This indicates that the effectiveness of SFT may be model-size and data-dependent.

\vspace{1em}

The following tables present performance on the Evaluation Dataset after Exclusion, which removes samples that more than half the models could answer correctly without the image. This subset (301 samples) is designed to mitigate data leakage concerns and focus on problems that genuinely require visual understanding.

\begin{table*}[t!]
\centering
\caption{Model Performance Comparison By Group On Evaluation Dataset After Exclusion (Samples=301, Groups with Image)}
\label{tab:model_comparison_exclusion_imgonly_overall_group}
\begin{adjustbox}{max width=\linewidth}
\begin{tabular}{l *{7}{>{\centering\arraybackslash}p{1.2cm}}}
\toprule
\textbf{Model} &
\textbf{Overall} &
\textbf{TC} &
\textbf{TQ} &
\textbf{I} &
\textbf{II} &
\textbf{III} &
\textbf{IV} \\
\midrule
\hline
\multicolumn{8}{c}{\textbf{Pre-trained VLMs}} \\
\hline
GPT-4.1 & 52.99 & 638 & 1204 & 51.5 & 52.16 & 52.82 & 55.48 \\
InternVL3-38B & 52.91 & 637 & 1204 & 55.81 & 52.16 & 50.17 & 53.49 \\
Claude-opus-4-1-20250805 & 64.70 & 779 & 1204 & 65.78 & 62.46 & 63.12 & 67.44 \\
GLM-4.5V & 74.75 & 900 & 1204 & 74.75 & 75.08 & 73.75 & 75.42 \\
DeepSeek-VL2 & 45.18 & 544 & 1204 & 45.85 & 42.86 & 45.18 & 46.84 \\
Hunyuan-Large-Vision & 59.72 & 719 & 1204 & 60.13 & 60.8 & 59.47 & 58.47 \\
ERNIE-4.5-turbo-vl & 34.47 & 415 & 1204 & 32.56 & 34.55 & 34.88 & 35.88 \\
Moonshot-V1-128k-Vision-Preview & 54.57 & 657 & 1204 & 53.82 & 54.15 & 54.15 & 56.15 \\
Moonshot-V1-32k-Vision-Preview & 54.40 & 655 & 1204 & 54.82 & 53.82 & 53.16 & 55.81 \\
Moonshot-V1-8k-Vision-Preview  & 54.90 & 661 & 1204 & 53.82 & 54.15 & 54.15 & 57.48 \\
Qwen3-VL-Plus & 51.00 & 614 & 1204 & 52.16 & 49.17 & 48.17 & 54.49 \\
Qwen2.5-VL-72B-Instruct & 48.84 & 588 & 1204 & 47.18 & 47.51 & 46.51 & 54.15 \\
Qwen2.5-VL-32B-Instruct & 43.19 & 520 & 1204 & 40.86 & 42.19 & 42.19 & 47.51 \\
Qwen2.5-VL-7B-Instruct & 37.87 & 456 & 1204 & 36.21 & 36.21 & 36.54 & 42.52 \\
Qwen2.5-VL-3B-Instruct & 20.51 & 247 & 1204 & 18.6 & 18.94 & 21.26 & 23.26 \\

\midrule
\multicolumn{8}{c}{\textbf{SFT VLMs, Trained with 500 Sample Chains}} \\
\hline
Qwen2.5-VL-3B-Instruct-L& 20.85 & 251 & 1204 & 20.6 & 21.93 & 17.28 & 23.59 \\
PyFi-QwenVL-3B-500 & 26.74 & 322 & 1204 & 28.57 & 25.25 & 23.92 & 29.24 \\
PyFi-QwenVL-3B-COT-500& 27.66 & 333 & 1204 & 29.57 & 25.58 & 21.59 & 33.89 \\
Qwen2.5-VL-7B-Instruct-L & 34.55 & 416 & 1204 & 34.55 & 32.89 & 32.23 & 38.54 \\
PyFi-QwenVL-7B-500 & 43.44 & 523 & 1204 & 39.53 & 41.53 & 40.2 & 52.49 \\
PyFi-QwenVL-7B-COT-500 & 39.53 & 476 & 1204 & 40.86 & 38.87 & 34.88 & 43.52 \\

\midrule
\multicolumn{8}{c}{\textbf{SFT VLMs, Trained with 47K Sample Chains}} \\
\hline
Qwen2.5-VL-3B-Instruct-L& 20.85 & 251 & 1204 & 20.6 & 21.93 & 17.28 & 23.59 \\
PyFi-QwenVL-3B-47K & 25.25 & 304 & 1204 & 25.58 & 24.25 & 22.92 & 28.24 \\
PyFi-QwenVL-3B-COT-47K & 40.37 & 486 & 1204 & 40.2 & 37.21 & 38.87 & 45.18 \\
Qwen2.5-VL-7B-Instruct-L & 34.55 & 416 & 1204 & 34.55 & 32.89 & 32.23 & 38.54 \\
PyFi-QwenVL-7B-47K & 27.08 & 326 & 1204 & 26.58 & 25.91 & 24.92 & 30.9 \\
PyFi-QwenVL-7B-COT-47K & 42.61 & 513 & 1204 & 44.19 & 38.87 & 42.52 & 44.85 \\

\bottomrule
\end{tabular}
\end{adjustbox}
\end{table*}

\textbf{Table~\ref{tab:model_comparison_exclusion_imgonly_overall_group}} provides an overall summary on the excluded dataset, including total correct (TC) out of total questions (TQ = 4 groups * 301 samples = 1204). The ranking of pre-trained models remains consistent. It also shows performance per group (I, II, III, IV). The top models like GLM-4.5V and Claude-opus maintain high and consistent accuracy across all four groups, demonstrating robustness. The SFT results reinforce previous observations: The 47K COT training for the 3B model (PyFi-QwenVL-3B-COT-47K) yields the most dramatic improvement, boosting overall accuracy from 20.85\% to 40.37\% and showing gains across all groups. For the 7B models, the 500-sample SFT (PyFi-QwenVL-7B-500) provides a strong boost, but the 47K COT variant also delivers competitive performance. Notably, the non-COT 47K SFT for the 7B model (PyFi-QwenVL-7B-47K) underperforms significantly, highlighting the critical importance of the Chain-of-Thought training methodology alongside data volume for achieving gains on this challenging visual reasoning benchmark.

\vspace{1em}

\begin{table*}[t!]
\centering
\caption{Model Performance Comparison By Capability On Evaluation Dataset After Exclusion (Samples=301, Groups with Image)}
\label{tab:model_comparison_exclusion_imgonly_capability}
\begin{adjustbox}{max width=\linewidth}
\begin{tabular}{l *{7}{>{\centering\arraybackslash}p{1.2cm}}}
\toprule
\textbf{Model} &
\textbf{Overall} &
\textbf{PP} &
\textbf{DE} &
\textbf{CA} &
\textbf{PR} &
\textbf{LR} &
\textbf{DS} \\
\midrule
\hline
\multicolumn{8}{c}{\textbf{Pre-trained VLMs}} \\
\hline
GPT-4.1 & 52.99 & 86.84 & 47.64 & 40.62 & 57.14 & 54.84 & 38.46 \\
InternVL3-38B & 52.91 & 76.32 & 56.37 & 44.53 & 43.88 & 53.23 & 30.77 \\
Claude-opus-4-1-20250805 & 64.70 & 80.92 & 62.74 & 66.02 & 58.16 & 72.58 & 32.69 \\
GLM-4.5V & 74.75 & 89.47 & 78.07 & 65.23 & 75.51 & 75.81 & 46.15 \\
DeepSeek-VL2 & 45.18 & 76.32 & 44.58 & 36.33 & 42.35 & 42.74 & 19.23 \\
Hunyuan-Large-Vision & 59.72 & 90.79 & 58.96 & 36.72 & 63.27 & 69.35 & 51.92 \\
ERNIE-4.5-turbo-vl & 34.47 & 34.87 & 33.96 & 36.72 & 25.00 & 45.16 & 36.54 \\
Moonshot-V1-128k-Vision-Preview & 54.57 & 80.92 & 61.08 & 39.06 & 41.84 & 58.87 & 38.46 \\
Moonshot-V1-32k-Vision-Preview & 54.40 & 80.26 & 61.56 & 37.11 & 40.31 & 62.90 & 38.46 \\
Moonshot-V1-8k-Vision-Preview  & 54.90 & 80.26 & 61.79 & 39.84 & 42.35 & 58.06 & 38.46 \\
Qwen3-VL-Plus & 51.00 & 77.63 & 54.25 & 37.50 & 41.84 & 57.26 & 32.69 \\
Qwen2.5-VL-72B-Instruct & 48.84 & 70.39 & 59.91 & 33.20 & 35.71 & 46.77 & 26.92 \\
Qwen2.5-VL-32B-Instruct & 43.19 & 60.53 & 45.28 & 37.50 & 36.73 & 42.74 & 28.85 \\
Qwen2.5-VL-7B-Instruct & 37.87 & 54.61 & 49.53 & 34.77 & 18.37 & 24.19 & 15.38 \\
Qwen2.5-VL-3B-Instruct & 20.51 & 36.84 & 15.57 & 21.48 & 16.84 & 21.77 & 19.23 \\

\midrule
\multicolumn{8}{c}{\textbf{SFT VLMs, Trained with 500 Sample Chains}} \\
\hline
Qwen2.5-VL-3B-Instruct-L & 20.85 & 35.53 & 17.22 & 21.09 & 15.82 & 20.97 & 25.00 \\
PyFi-QwenVL-3B-500 & 26.74 & 34.87 & 28.77 & 28.52 & 16.84 & 24.19 & 21.15 \\
PyFi-QwenVL-3B-COT-500 & 27.66 & 37.50 & 28.77 & 26.17 & 22.45 & 25.81 & 21.15 \\
Qwen2.5-VL-7B-Instruct-L & 34.55 & 52.63 & 42.92 & 31.64 & 18.37 & 25.00 & 11.54 \\
PyFi-QwenVL-7B-500 & 43.44 & 71.05 & 44.58 & 43.36 & 30.10 & 37.90 & 17.31 \\
PyFi-QwenVL-7B-COT-500  & 39.53 & 61.84 & 41.51 & 40.23 & 26.02 & 34.68 & 17.31 \\

\midrule
\multicolumn{8}{c}{\textbf{SFT VLMs, Trained with 47K Sample Chains}} \\
\hline
Qwen2.5-VL-3B-Instruct-L & 20.85 & 35.53 & 17.22 & 21.09 & 15.82 & 20.97 & 25.00 \\
PyFi-QwenVL-3B-47K & 25.25 & 42.76 & 19.81 & 18.75 & 29.59 & 29.03 & 25.00 \\
PyFi-QwenVL-3B-COT-47K & 40.37 & 62.50 & 36.56 & 29.69 & 41.84 & 48.39 & 34.62 \\
Qwen2.5-VL-7B-Instruct-L & 34.55 & 52.63 & 42.92 & 31.64 & 18.37 & 25.00 & 11.54 \\
PyFi-QwenVL-7B-47K & 27.08 & 36.84 & 23.35 & 30.47 & 17.35 & 37.10 & 25.00 \\
PyFi-QwenVL-7B-COT-47K & 42.61 & 68.42 & 41.75 & 34.77 & 32.65 & 49.19 & 34.62 \\

\bottomrule
\end{tabular}
\end{adjustbox}
\end{table*}

\textbf{Table~\ref{tab:model_comparison_exclusion_imgonly_capability}} breaks down model performance by required capability on the challenging 301-sample subset. The overall accuracy drops dramatically for all models compared to the full dataset, confirming the difficulty of these leakage-filtered samples. GLM-4.5V again dominates with 74.75\% overall, excelling particularly in Perception (PP, 89.47\%) and Data Extraction (DE, 78.07\%). Claude-opus (64.70\%) shows strong performance in Calculation Analysis (CA, 66.02\%) and Logical Reasoning (LR, 72.58\%). Hunyuan-Large-Vision is notable for its high Perception (90.79\%) and relatively strong Decision Support (DS, 51.92\%). A clear capability-wise difficulty trend emerges: Perception is generally easiest for top models, while Calculation Analysis and, especially, Decision Support are most challenging. Among SFT models, PyFi-QwenVL-3B-COT-47K shows a remarkable leap to 40.37\%, more than doubling the performance of its base model (20.85\%) and significantly outperforming its 500-sample counterpart. This suggests that Chain-of-Thought (COT) training on a larger, high-quality dataset (47K) is highly effective for smaller models, particularly improving Pattern Recognition (PR) and Logical Reasoning (LR). For the 7B models, PyFi-QwenVL-7B-500 achieves the best overall (43.44\%), but the 47K COT variant also shows strong gains in LR.

\vspace{1em}

\begin{table*}[t!]
\centering
\caption{Model Performance Comparison By Complexity Degree On Evaluation Dataset After Exclusion (Samples=301, Groups with Image)}
\label{tab:model_comparison_exclusion_imgonly_complexity}
\begin{adjustbox}{max width=\linewidth}
\begin{tabular}{l *{6}{>{\centering\arraybackslash}p{1.2cm}}}
\toprule
\textbf{Model} &
\textbf{Overall} &
\textbf{1} &
\textbf{2} &
\textbf{3} &
\textbf{4} &
\textbf{5} \\
\midrule
\hline
\multicolumn{7}{c}{\textbf{Pre-trained VLMs}} \\
\hline
GPT-4.1 & 52.99 & 74.51 & 46.12 & 52.63 & 46.55 & 53.85 \\
InternVL3-38B & 52.91 & 75.00 & 52.91 & 41.78 & 44.83 & 67.31 \\
Claude-opus-4-1-20250805 & 64.70 & 76.47 & 61.89 & 66.12 & 55.60 & 73.08 \\
GLM-4.5V & 74.75 & 87.75 & 76.21 & 72.37 & 64.22 & 73.08 \\
DeepSeek-VL2 & 45.18 & 67.65 & 43.45 & 34.54 & 42.24 & 46.15 \\
Hunyuan-Large-Vision & 59.72 & 80.88 & 54.85 & 50.00 & 57.76 & 80.77 \\
ERNIE-4.5-turbo-vl & 34.47 & 32.84 & 33.25 & 38.82 & 25.43 & 65.38 \\
Moonshot-V1-128k-Vision-Preview & 54.57 & 74.02 & 56.31 & 46.71 & 43.97 & 57.69 \\
Moonshot-V1-32k-Vision-Preview & 54.40 & 73.53 & 56.31 & 45.72 & 43.97 & 61.54 \\
Moonshot-V1-8k-Vision-Preview  & 54.90 & 73.53 & 56.80 & 48.03 & 43.97 & 55.77 \\
Qwen3-VL-Plus & 51.00 & 72.55 & 50.00 & 45.72 & 37.50 & 65.38 \\
Qwen2.5-VL-72B-Instruct & 48.84 & 67.16 & 57.77 & 36.51 & 31.03 & 57.69 \\
Qwen2.5-VL-32B-Instruct & 43.19 & 58.33 & 43.69 & 39.14 & 30.60 & 59.62 \\
Qwen2.5-VL-7B-Instruct & 37.87 & 55.39 & 44.90 & 31.91 & 17.67 & 38.46 \\
Qwen2.5-VL-3B-Instruct & 20.51 & 31.86 & 12.62 & 23.03 & 18.97 & 30.77 \\

\midrule
\multicolumn{7}{c}{\textbf{SFT VLMs, Trained with 500 Sample Chains}} \\
\hline
Qwen2.5-VL-3B-Instruct-L & 20.85 & 33.33 & 13.11 & 21.38 & 21.12 & 28.85 \\
PyFi-QwenVL-3B & 26.74 & 40.69 & 20.87 & 28.29 & 20.69 & 36.54 \\
PyFi-QwenVL-3B-COT & 27.66 & 39.22 & 22.57 & 27.96 & 24.57 & 34.62 \\
\hline
Qwen2.5-VL-7B-Instruct-L & 34.55 & 55.88 & 37.86 & 30.26 & 15.52 & 34.62 \\
PyFi-QwenVL-7B & 43.44 & 66.18 & 44.17 & 39.47 & 28.45 & 38.46 \\
PyFi-QwenVL-7B-COT & 39.53 & 58.82 & 39.32 & 36.51 & 23.71 & 53.85 \\

\midrule
\multicolumn{7}{c}{\textbf{SFT VLMs, Trained with 47K Sample Chains}} \\
\hline
Qwen2.5-VL-3B-Instruct-L & 20.85 & 33.33 & 13.11 & 21.38 & 21.12 & 28.85 \\
PyFi-QwenVL-3B & 25.25 & 37.25 & 18.93 & 19.74 & 31.47 & 32.69 \\
PyFi-QwenVL-3B-COT & 40.37 & 59.80 & 32.77 & 36.51 & 40.09 & 48.08 \\
\hline
Qwen2.5-VL-7B-Instruct-L & 34.55 & 55.88 & 37.86 & 30.26 & 15.52 & 34.62 \\
PyFi-QwenVL-7B & 27.08 & 35.29 & 21.60 & 26.64 & 25.86 & 46.15 \\
PyFi-QwenVL-7B-COT & 42.61 & 62.25 & 38.35 & 39.80 & 32.76 & 59.62 \\

\bottomrule
\end{tabular}
\end{adjustbox}
\end{table*}

\textbf{Table~\ref{tab:model_comparison_exclusion_imgonly_complexity}} analyzes performance stratified by problem complexity degree (1=Least, 5=Most Complex). A strong negative correlation between model performance and complexity degree is evident for most models. For example, GLM-4.5V scores 87.75\% on degree-1 problems but drops to 64.22\% on degree-4. Interestingly, performance on degree-5 (the most complex) does not always follow this monotonic decline; for models like Claude-opus, Hunyuan, and some Qwen variants, degree-5 accuracy is higher than degree-4. This may indicate that the highest complexity problems, while difficult, sometimes involve different (e.g., more structured) reasoning that some models can handle. The SFT models show interesting patterns: PyFi-QwenVL-3B-COT-47K demonstrates substantial improvements across all complexity levels, most notably on degrees 4 and 5 (40.09\% and 48.08\%, up from ~21\% and ~29\% in the base model). This suggests that the COT training on extensive data particularly enhances the model's ability to tackle high-complexity, multi-step reasoning problems. The PyFi-QwenVL-7B-COT-47K model also shows strong gains on degree-5 problems (59.62\%).

\vspace{1em}

\noindent\textbf{Summary of Key Findings}

\noindent \textbf{Performance Gap}. A significant performance gap exists between top-tier proprietary models (GLM-4.5V, Claude-opus) and open-source models, especially on the challenging leakage-filtered subset.

\noindent \textbf{Scaling Laws}. For the Qwen family, model size strongly correlates with performance on this financial visual QA task.

\noindent \textbf{Effectiveness of SFT}: Supervised Fine-Tuning can substantially improve smaller models (3B), particularly when using a large (47K), high-quality dataset with Chain-of-Thought reasoning steps. Gains for larger models (7B) are more modest and sensitive to the training recipe.

\noindent \textbf{Capability \& Complexity}.  Perception and Data Extraction are relative strengths for leading models, while Calculation Analysis and Decision Support remain challenging. Problem complexity is inversely correlated with accuracy, but advanced training (COT) can specifically improve performance on high-complexity problems.

\section{Algorithms}
\label{sec:algorithms}

The proposed adversarial agent dataset construction method is outlined in Algorithm~\ref{alg:agent_debate}. Its runtime is primarily dictated by the base VLM's inference latency. Our implementation, using Doubao-seed-1-6-flash-250828, completes approximately 100 adversarial rounds (100 samples) in about 60 minutes. While the MCTS process for a single image runs sequentially, each adversarial step can generate multiple branches in parallel for internal acceleration. The framework also enables concurrent processing of multiple images across servers and processes. Consequently, the approach is readily extensible to large-scale synthesis with manageable time costs.

\newpage
\begin{algorithm*}[t!]
\caption{Dataset Construction via Adversarial Agents under the MCTS Paradigm}
\label{alg:agent_debate}
\SetAlgoLined
\KwIn{Challenger Agent \(\psi\), Solver Agent \(\phi\),  Image $I$ with background $b_I$ and analysis info $f_I$, capability levels $\mathcal{L} = \{l_1, \dots, l_6\}$, complexity levels $\mathcal{C} = \{c_1, \dots, c_5\}$, MCTS parameters $\alpha, \beta$, max chains $T_{\max}$}
\KwOut{Pyramid-structured dataset $\mathcal{S}$ with question-answer chains}

Generate final question $Q_F$ using LLMs with: 
 $b_I$, $f_I$, $\text{\(l\)}(Q_F) = l_6$,\;
Determine ground truth answer $A_F$ for $Q_F$ via multi-LLM consensus\;

Initialize MCTS tree $\mathcal{T}$ with root $Q_F$\;
$t \leftarrow 0$\;
\While{$t < T_{\max}$ and $\mathcal{T}$ not terminated}{
    ${\psi}, {\phi} \leftarrow$ root of $\mathcal{T}$\;
    \While{${\psi}$ and ${\phi}$ ready and debate not terminated}{
        \tcbox[colback=gray!30, colframe=gray!30, sharp corners, boxsep=1pt, left=2pt, right=2pt, top=0pt, bottom=0pt, on line]{\textbf{Compete (Challenger Agent \(\psi\) Turn)}}\;
        $N_{\text{child}} \leftarrow$ children of $q_{\psi}$\;
        \eIf{$|N_{\text{child}}| = 0$}{
            Determine current and next capability level $l_{\text{curr}}$, $l_{\text{next}}$ based on pyramid progression\;
            Generate new question $q_{\text{new}}$ with:
            $b_I$,
            $l_{\text{curr}} = l_{\text{next}}$, 
            $\text{complexity} \in \mathcal{C}$,
            $\text{visit\_count} = 1$, 
            $\text{victory\_count} = 0$\;
            $q_{\psi} \leftarrow q_{\text{new}}$\;
        }{
            $p_{\text{gen}} \leftarrow \sigmoid(\alpha \cdot (|N_{\text{child}}| - \beta))$\;
            Sample $z \sim \text{Bernoulli}(p_{\text{gen}})$\;
            \eIf{$z = 1$}{
                Determine current and next capability level $l_{\text{curr}}$, $l_{\text{next}}$ based on pyramid progression\;
                Generate new question $q_{\text{new}}$ with:
                $b_I$,
                $\text{capability} = l_{\text{next}}$, $\text{complexity} \in \mathcal{C}$,
                $\text{visit\_count} = 1$, 
                $\text{victory\_count} = 0$\;
                $q_{\psi} \leftarrow q_{\text{new}}$\;
            }{
                $q_{\psi} \leftarrow \arg\max_{n' \in N_{\text{child}}} \text{UCT}(n')$\;
                Increment $\text{visit\_count}(q_{\psi})$\;
            }
        }

    \tcbox[colback=gray!30, colframe=gray!30, sharp corners, boxsep=1pt, left=2pt, right=2pt, top=0pt, bottom=0pt, on line]{\textbf{Compete (Solver Agent \(\phi\) Turn)}}\;
    Generate answer $a_{\phi}$ to $q_{\psi}$\;
    \eIf{$a_{\phi} \in$ existing actions}{
        Increment $\text{visit\_count}(a_{\phi})$\;
    }{
        Create new $a_{\phi}$ with $\text{visit\_count} \leftarrow 1$, $\text{victory\_count} \leftarrow 0$\;
    }
    Add edge $(q_{\psi}, a_{\phi})$ to $\mathcal{T}$\;
    
    \If{$l_{\text{curr}}$ = 6 and LLM judges chain can answer $Q_F$}{
        Generate final answer $a_{\text{final}}$\;
        \If{$a_{\text{final}} = A_F$}{
            Backpropagate and increment $\text{victory\_count}$ for all nodes\;
        }
        Break\;
    }
    }

    $t \leftarrow t + 1$\;
}
Add successful chains to $\mathcal{S}$\;

\end{algorithm*}

\end{appendices}

\end{document}